\magnification=\magstep1

\input amstex
\documentstyle{myamsppt}




\input eplain
%
%
%






\def\setRuledStrut{\relax}

\catcode`@=11                                   
\catcode`\|=12                                  
\catcode`\&=4                                   

\newcount\ncols         \ncols=\z@              
\newcount\nrows         \nrows=\z@              
\newcount\curcol        \curcol=\z@             
     
\newdimen\thinsize      \thinsize=0.6pt         
\newdimen\thicksize     \thicksize=1.5pt        
\newdimen\framesize     \framesize=1.5pt        
\newdimen\tablewidth    \tablewidth=-\maxdimen  
\newdimen\parasize      \parasize=4in           

\newif\iftableinfo      \tableinfofalse         
\newif\ifcentertables   \centertablestrue       
%
%
     
\let\plaincr=\cr                        
\let\plainspan=\span                    
\let\plaintab=&                         
\let\lparen=(                           
\let\NX=\noexpand                       

     
\def\ruledtable{\relax                          
    \@BeginRuledTable                     
        \@RuledTable}


\def\@BeginRuledTable{
   \ncols=0\nrows=0                             
   \begingroup                                  
    \offinterlineskip                           
    \def~{\phantom{0}}
    \def\span{\plainspan\omit\relax\colcount\plainspan}
    \let\cr=\crrule                             
    \let\CR=\crthick                            
    \let\nr=\crnorule                           
    \let\|=\Vb                                  
    \def\hfill{\hskip0pt plus1fill\hbox{}}
%
%
    \ifx\tablestrut\undefined\relax             
    \else\let\tstrut=\tablestrut\fi             
    \catcode`\|=13 \catcode`\&=13\relax         
    \TableActive                                
    \curcol=1                                   
%
%
    \ifdim\tablewidth>-\maxdimen\relax          %
      \edef\@Halign{\NX\halign to \NX\tablewidth\NX\bgroup\TablePreamble}%
      \tabskip=0pt plus 1fil                    
    \else                                       %
      \edef\@Halign{\NX\halign\NX\bgroup\TablePreamble}%
      \tabskip=0pt                              
    \fi                                         %
%
%
    \ifcentertables                             
       \ifhmode\vskip 0pt\fi                    
       \line\bgroup\hss                         
    \else\hbox\bgroup                           
    \fi}


\long\def\@RuledTable#1\endruledtable{
   \vrule width\framesize                       
     \vbox{\@Halign                             
       \framerule                               
       #1\killspace                             
       \tstrut                                  
       \linecount                               
       \plaincr\framerule                       
     \egroup}
   \vrule width\framesize                       
   \ifcentertables\hss\fi\egroup                
  \endgroup                                     
  \global\tablewidth=-\maxdimen                 
  \iftableinfo                                  
      \immediate\write16{[Nrows=\the\nrows, Ncols=\the\ncols]}%
   \fi}
     

\def\TablePreamble{
   \TableItem{####}
   \plaintab\plaintab                   
   \TableItem{####}
   \plaincr}


\def\@TableItem#1{
   \hfil\tablespace                     
   #1\killspace
   \tablespace\hfil                     
    }%

\def\@tableright#1{
   \hfil\tablespace\relax               
   #1\killspace
   \tablespace\relax}

\def\@tableleft#1{
   \tablespace\relax                    
   #1\killspace
   \tablespace\hfil}

\let\TableItem=\@TableItem              
     
\def\RightJustifyTables{\let\TableItem=\@tableright}
\def\LeftJustifyTables{\let\TableItem=\@tableleft}
\def\NoJustifyTables{\let\TableItem=\@TableItem}


\def\LooseTables{\let\tablespace=\quad}
\def\TightTables{\let\tablespace=\space}
\LooseTables                                    


\def\TrailingSpaces{\let\killspace=\relax}      
\def\NoTrailingSpaces{\let\killspace=\unskip}   
\TrailingSpaces                                 

%


\def\setRuledStrut{
   \dimen@=\baselineskip                        
   \advance\dimen@ by-\normalbaselineskip       
   \ifdim\dimen@<.5ex \dimen@=.5ex\fi           
   \setbox0=\hbox{\lparen}
   \dimen1=\dimen@ \advance\dimen1 by \ht0      
   \dimen2=\dimen@ \advance\dimen2 by \dp0      
   \def\tstrut{\vrule height\dimen1 depth\dimen2 width\z@}%
   }%

\def\tstrut{\vrule height 3.1ex depth 1.2ex width 0pt}


\def\bigitem#1{
   \dimen@=\baselineskip                        
   \advance\dimen@ by-\normalbaselineskip       
   \ifdim\dimen@<.5ex \dimen@=.5ex\fi           
   \setbox0=\hbox{#1}
   \dimen1=\dimen@ \advance\dimen1 by \ht0      
   \dimen2=\dimen@ \advance\dimen2 by \dp0      
   \vrule height\dimen1 depth\dimen2 width\z@   
   \copy0}

     
%

     
\def\nextcolumn#1{
   \plaintab\omit#1\relax\colcount              
   \plaintab}
     
\def\tab{
   \nextcolumn{\relax}}


\def\vb{
   \nextcolumn{\vrule width\thinsize}}

\def\Vb{
   \nextcolumn{\vrule width\thicksize}}


     
{\catcode`\|=13 \let|0
 \catcode`\&=13 \let&0
 \gdef\TableActive{\let|=\vb \let&=\tab}%
}


\def\crrule{\killspace                  
   \tstrut                              
   \linecount                           
   \plaincr\tablerule                   
  }%

\def\crthick{\killspace                 
   \tstrut                              
   \linecount                           
   \plaincr\thickrule                   
  }%
     
\def\crnorule{\killspace                
   \tstrut                              
   \linecount                           
   \plaincr                             
   }%
   

     
\def\tablerule{\noalign{\hrule height\thinsize depth 0pt}}%
\def\thickrule{\noalign{\hrule height\thicksize depth 0pt}}%
\def\framerule{\noalign{\hrule height\framesize depth 0pt}}


%
%
%
     

\def\linecount{
   \global\advance\nrows by1
   \ifnum\ncols>0
      \ifnum\curcol=\ncols\relax\else           
      \immediate\write16
      {\NX\ruledtable warning: Ncols=\the\curcol\space for Nrow=\the\nrows}%
      \fi\fi                                    
   \global\ncols=\curcol                        
   \global\curcol=1}                            

\def\colcount{\relax                            %
   \global\advance\curcol by 1\relax}


%

%

\def\begintable{\relax                          
    \@BeginRuledTable                           
    \@begintable}

\long\def\@begintable#1\endtable{
   \@RuledTable#1\endruledtable}



\catcode`@=12                                   


\let\ref=\eplainref  

\TagsOnRight
\parskip 1ex

\hsize=6.5truein
\vsize=9truein

\loadbold
\loadeurm

\font\bbf=cmbx12
\font\bbigbf=cmbx12 scaled\magstep1
\font\bi=cmbxti10	
\font\bigbf=cmbx12
\font\bigbi=cmbxti10 scaled\magstep1	

\font\sc=cmcsc10
\font\smallrm=cmr9
\font\smallit=cmti9
\font\smallbf=cmbx9
\font\smallbi=cmbxti10 at 8pt
\font\sf=cmss10

\font\smalltt=cmtt8
\font\ssf=cmss8 
\font\ssmallrm=cmr8

\document

\define\de{\define}
\de\AAA{{\tx{${\scr A}$}}}
\de\AAa{{\tx{${\Bbb A}$}}}
\de\ABBB{{\tx{$A^\BBB$}}}
\de\ABN{{\tx{$A^{B,N}$}}}
\de\AN{{\tx{$A^N$}}}
\de\Ab{{\bf A}}
\de\Abi{{\bi A}}
\de\Abar{{\tx{$\bar A$}}}
\de\Abars{\tx{$\Abar_s$}}
\de\Abaru{\tx{$\Abar^u$}}
\de\AinKK{\tx{$A \in \KK$}}
\de\AinAlgSig{\tx{$A \in \AlgSig$}}
\de\AinNStdAlgSig{\tx{$A \in \NStdAlgSig$}}
\de\AinStdAlgSig{\tx{$A \in \StdAlgSig$}}
\de\Akvec{\tx{$A[\kvec]$}}
\de\Akxvec{\tx{$A[\kxvec]$}}
\de\Aklxvec{\tx{$A[\kvec,\lxvec]$}}
\de\Alg{\tx{\bi Alg}}
\de\AlgSig{\Alg\,(\Sig)}
\de\Alvec{\tx{$A[\lvec\,]$}}
\de\Alxvec{\tx{$A[\lxvec]$}}
\de\Aps{\tx{\sf Ap}}
\de\As{\tx{$A_s$}}
\de\Asbar{\tx{$\Abar_s$}}
\de\Asx{\tx{$A_s^*$}}
\de\AssLang{\tx{\bi Ass}\Lang}
\de\Assumption{\pr{Assumption}\sl}
\de\Assumptionn#1{\pr{Assumption #1}\sl}
\de\Assumptions{\pr{Assumptions}\sl}
\de\Asuu{\tx{$A^\uu_s$}}
\de\At{\tx{\bi At}}
\de\AtSig{\At(\Sig)}
\de\AtSt{\tx{\bi AtSt}}
\de\AtStSig{\tx{$\AtSt(\Sig)$}}
\de\Atil{{\tx{$\til A$}}}
\de\Au{\tx{$A^u$}}
\de\Aul{{\tx{$\ul A$}}}
\de\Auu{\tx{$A^\uu$}}
\de\AuuN{\tx{$A^{\uu,N}$}}
\de\Av{\tx{$A^v$}}
\de\Aw{\tx{$A^w$}}
\de\Ax{{\tx{$A^*$}}}
\de\Axs{\tx{$A^*_s$}}
\de\Axx{(\Ax)\str}
\de\Bbi{\tx{\bi B}}
\de\BB{{\tx{${\Bbb B}$}}}
\de\BBB{{\tx{${\scr B}$}}}
\de\BBbar{\tx{$\ol \BB$}}
\de\BBuu{\tx{${\Bbb B}^\uu$}}
\de\Bb{\tx{\bf B}}
\de\Bool{\tx{\bi Bool}}
\de\BoolSig{\tx{$\Bool(\Sig)$}}
\de\Bs{\tx{\sf B}}
\de\Bss{{\tx{\ssf B}}}
\de\Btil{\tx{$\til B$}}
\de\Bul{\tx{$\ul B$}}
\de\CC{\tx{${\Bbb C}$}}
\de\CCn{\tx{$\CC^n$}}
\de\CCC{\tx{$\scr C$}}
\de\CCCL{\tx{${\CCC}^<$}}
\de\CR{\tx{\bi CR}}
\de\CRSig{\tx{$\CR(\Sig)$}}
\de\CT{{\tx{\bi CT}}}
\de\CTSig{\CT(\Sig)}
\de\CTsmall{{\tx{\smallbi CT}}}
\de\Cbar{\tx{$\ol C$}}
\de\Cf{{\it Cf\.}}
\de\Chead#1{\bn\cbb{#1}\bn}
\de\Comp{\tx{\bi Comp}}
\de\CompA{\tx{$\Comp^A$}}
\de\CompLengthA{\tx{$\Comp\Length^A$}}
\de\CompSeqA#1{\tx{$\Comp\Seq^A(#1)$}}
\de\Compu{\tx{\bi Compu}}
\de\Con{\tx{\bi Con}}
\de\Cond{\tx{\bi Cond}}
\de\CondSig{\tx{$\Cond(\Sig)$}}
\de\Constrn#1{\pr{Construction \ #1}\rm}
\de\Cont{\tx{\bi Cont}}
\de\ContA{\tx{$\Cont(A)$}}
\de\Conventionn#1{\pr{Convention \ #1}\sl}
\de\Cor{\pr{Corollary}\sl}
\de\Corn#1{\prn{Corollary #1}\sl}
\de\Cors{\pr{Corollaries}\sl}
\de\Cs{\tx{\sf C}}
\de\Cul{\tx{$\ul C$}}
\de\DD{{\tx{${\Bbb D}$}}}
\de\DDD{{\tx{$\scr D$}}}
\de\Dbar{\tx{$\ol{D}$}}
\de\Dbars{\tx{$\ol{D}_s$}}
\de\Def{\pr{Definition}\rm}
\de\Defn#1{\prn{Definition #1}\rm}
\de\Defs{\pr{Definitions}\rm}
\de\Defsn#1{\prn{Definitions #1}\rm}
\de\Del{\tx{$\Delta$}}
\de\Discussion{\pr{Discussion}\rm}
\de\Discussionn#1{\pr{Discussion #1}\rm}
\de\Ds{\tx{\sf D}}
\de\Dss{{\tx{\ssf D}}}
\de\Ebar{\tx{$\ol E$}}
\de\Eg{{\it E.g.}}
\de\ElemInd{\tx{\sf ElemInd}}
\de\ElemIndSig{\tx{$\ElemInd(\Sig)$}}
\de\EqSort{\tx{\bi EqSort\/}}
\de\EqSortSig{\EqSort(\Sig)}
\de\Eqb{\tx{\bi Eq}}
\de\Eqs{\tx{\sf Eq}}
\de\Equiv{\tx{$\pmb{Equiv}$}}
\de\Etil{\tx{$\til{E}$}}
\de\Example{\pr{Example}}
\de\Examplen#1{\prn{Example {#1}}}
\de\Examples{\pr{Examples}}
\de\Examplesn#1{\prn{Examples {#1}}}
\de\Exercise{\pr{Exercise}}
\de\Exercisen#1{\prn{Exercise {#1}}}
\de\Exercises{\pr{Exercises}}
\de\Exercisesn#1{\prn{Exercises {#1}}}
\de\F{\tx{\bi F}}
\de\FA{\tx{$F^A$}}
\de\FAuu{\tx{$F^{A,\uu}$}}
\de\FFF{\tx{$\scr F$}}
\de\FFt{\tx{\tt F}}
\de\FinFuncSig{\tx{$F \in \FuncSig$}}
\de\First{\tx{\bi First}}
\de\For{\tx{\bi For}}
\de\ForA{\tx{$\For(A)$}}
\de\ForN{\tx{$\For^N$}}
\de\ForNA{\tx{$\ForN(A)$}}
\de\ForNSig{\tx{$\ForN(\Sig)$}}
\de\ForSig{\tx{$\For(\Sig)$}}
\de\ForSigN{\tx{$\For(\SigN)$}}
\de\ForSigx{\tx{$\For(\Sigx)$}}
\de\Form{\tx{\bi Form}}
\de\FormSig{\tx{$\Form(\Sig)$}}
\de\Foro{\tx{$\For_0$}}
\de\Forolam{\tx{$\For_0^{\lam}$}}
\de\Forx{\tx{$\For^\starb$}}
\de\ForxA{\tx{$\Forx(A)$}}
\de\ForxSig{\tx{$\Forx(\Sig)$}}
\de\Fs{{\tx{\smallbi F}}}
\de\Ftil{\tx{$\til{F}$}}
\de\Func{\tx{\bi Func}}
\de\FuncSig{\Func\,(\Sig)}
\de\FuncSigp{\tx{$\Func(\Sigp)$}}
\de\FuncSigus{\tx{$\FuncSig_\utos$}}
\de\GG{\tx{${\Bbb G}$}}
\de\GGG{\tx{$\scr G$}}
\de\Gtt{\tx{\tt G}}
\de\Gam{\tx{$\Gamma$}}
\de\Gbar{\tx{$\ol G$}}
\de\Gs{{\tx{\sf G}}}
\de\Halt{\tx{\bi Halt}}
\de\HaltTests{\tx{\sf HaltTest}}
\de\HaltA#1{\tx{$\Halt^A\,(#1)$}}
\de\Hbar{\tx{$\bar H$}}
\de\Hh{\tx{\bi H}}
\de\Hs{{\tx{\sf H}}}
\de\Hss{{\tx{\ssf H}}}
\de\Htt{\tx{\tt H}}
\de\Ib{\tx{{\bf I}}}
\de\II{{\tx{${\Bbb I}$}}}
\de\III{\tx{${\scr I}$}}
\de\IIIN{{\tx{$\III^N$}}}
\de\IL{\tx{\bi IL}}
\de\IO{\tx{\bi IO}}
\de\Ibar{{\tx{$\bar I$}}}
\de\Ie{{\it I.e.}}
\de\Ind{\tx{\sf Ind}}
\de\IndSig{\tx{$\Ind(\Sig)$}}
\de\Indi{\tx{$\Ind_i$}}
\de\IndiSig{\Indi(\Sig)}
\de\Init{\tx{\bi Init}}
\de\Is{\tx{\sf I}}
\de\Iss{{\tx{\ssf I}}}
\de\JJJ{\tx{${\scr J}$}}
\de\Jbar{{\tx{$\bar J$}}}
\de\KK{{\tx{${\Bbb K}$}}}
\de\KKbar{\tx{$\ol{\KK}$}}
\de\KKN{\tx{${\Bbb K}^N$}}
\de\KKuu{\tx{${\Bbb K}^\uu$}}
\de\KKx{\tx{${\Bbb K}^*$}}
\de\LL{\tx{${\Bbb L}$}}
\de\LLL{\tx{${\scr L}$}}
\de\Lam{\tx{$\Lambda$}}
\de\Lang{\tx{\bi Lang}}
\de\LangSig{\tx{$\Lang(\Sig)$}}
\de\LangSigx{\tx{$\Lang(\Sigx)$}}
\de\Langx{\tx{$\Lang^*$}}
\de\LangxSig{\tx{$\Langx(\Sig)$}}
\de\Lemma{\pr{Lemma}\sl}
\de\Lemman#1{\prn{Lemma #1}\sl}
\de\Lemmas{\pr{Lemmas}\sl}
\de\Lemmasn#1{\prn{Lemmas #1}\sl}
\de\Length{\tx{\bi Length}}
\de\Lgths{\tx{\sf Lgth}}
\de\Lss{{\tx{\ssf L}}}
\de\MMM{\tx{${\scr M}$}}
\de\Min{\tx{\bi Min}}
\de\MinNStdAlg{\Min\NStdAlg}
\de\Mod{\tx{\bi Mod}}
\de\NN{{\tx{${\Bbb N}$}}}
\de\NNN{{\tx{${\scr N}$}}}
\de\NNNB{{\tx{${\NNN^B}$}}}
\de\NNNo{{\tx{$\NNN_0$}}}
\de\NNu{\tx{${\Bbb N}^\uu$}}
\de\NStdAlg{\tx{\bi NStdAlg\/}}
\de\NStdAlgSig{\tx{$\NStdAlg\,(\Sig)$}}
\de\Nbd{\tx{\bi Nbd}}
\de\Nbi{{\tx{\bi N}}}
\de\Nbismall{{\tx{\smallbi N}}}
\de\Newlengths{\tx{\sf Newlength}}
\de\Notation{\pr{Notation}\rm}
\de\Notationn#1{\pr{Notation \ #1}\rm}
\de\Note{\pr{Note}\rm}
\de\Noten#1{\pr{Note #1}\rm}
\de\Notes{\pr{Notes}\rm}
\de\Notesn#1{\pr{Notes #1}\rm}
\de\Ns{\tx{\sf N}}
\de\Nss{{\tx{\ssf N}}}
\de\Nnu{\tx{$\Nn^\uu$}}
\de\Nulls{\tx{\sf Null}}
\de\Nullvec{\tx{$\vec{\Nulls}$}}
\de\Om{{\tx{$\Omega$}}}
\de\Pp{\tx{\bk P}}
\de\PA{\tx{$P^A$}}
\de\PE{\tx{\bi PE}}
\de\PL#1{\tx{$\ProgLang_{#1}$}}
\de\PPP{\tx{$\scr P$}}
\de\PR{\tx{PR}}
\de\PRA{\tx{$\PR(A)$}}
\de\PRAutos{\tx{$\PRA_\utos$}}
\de\PRSig{\tx{$\PR(\Sig)$}}
\de\PRSigb{\tx{$\PRb(\Sigb)$}}
\de\PRSigutos{\tx{$\PRSig_\utos$}}
\de\PRb{\tx{\bf PR}}
\de\PRmin{\tx{$\PR_-$}}
\de\PRo{\tx{$\PR_0$}}
\de\PRolam{\tx{$\PR_0^{\lam}$}}
\de\PRutos{\tx{$\PR_\utos$}}
\de\PRx{\PR\str}
\de\PRxA{\tx{$\PRx(A)$}}
\de\PRxSig{\tx{\PRx(\Sig)}}
\de\PRxSigb{\tx{\PRxb(\Sigb)}}
\de\PRxb{\PRb\strb}
\de\PTE{\tx{\bi PTE}}
\de\PTEA{\tx{$\PTE^A$}}
\de\PTerm{\tx{\bi PTerm}}
\de\PTermSig{\tx{$\PTerm(\Sig)$}}
\de\Pbi{\tx{\bi P}}
\de\Pf{\n{\bf Proof:\ \,}}
\de\Ph{\tx{$\Phi$}}
\de\PreStr{\tx{\bi PreStr}}
\de\PreStrSig{\PreStr\,(\Sig)}
\de\Problem{\pr{Problem}}
\de\Problemn#1{\pr{Problem {#1}}}
\de\Problems{\pr{Problems}}
\de\Problemsn#1{\pr{Problems {#1}}}
\de\Propsn#1{\pr{Propositions {#1}}}
\de\Proc{\tx{\bi Proc}}
\de\ProcN{\tx{$\Proc^N$}}
\de\ProcNSig{\ProcN(\Sig)}
\de\ProcSig{\Proc(\Sig)}
\de\ProcSigN{\Proc(\SigN)}
\de\ProcSiguv{\tx{$\Proc(\Sig)_\utov$}}
\de\ProcSigx{\tx{$\Proc(\Sigx)$}}
\de\Procuv{\tx{$\Proc_\utov$}}
\de\ProcuvSig{\tx{$\Procuv(\Sig)$}}
\de\Procx{\Proc\str}
\de\ProcxSig{\Procx(\Sig)}
\de\ProcxSiguv{\tx{$\ProcxSig_\utov$}}
\de\Procxuv{\tx{$\Proc^*_\utov$}}
\de\Prod{\tx{\bi Prod}}
\de\ProdType{\tx{\bi ProdType}}
\de\ProdTypeSig{\tx{$\ProdType(\Sig)$}}
\de\Prog{\tx{\bi Prog}}
\de\ProgLang{\Prog\Lang}
\de\ProgTerm{\Prog\Term}
\de\ProgTermSig{\tx{$\ProgTerm(\Sig)$}}
\de\Prop{\pr{Proposition}\sl}
\de\Propn#1{\prn{Proposition #1}\sl}
\de\Propp{\tx{\bi Prop}}
\de\Props{\pr{Propositions}\sl}
\de\Ps{\tx{$\Psi$}}
\de\Q{{\bf Q}}
\de\QQ{\tx{${\Bbb Q}$}}
\de\Qs{\tx{\sf Q}}
\de\Question{\pr{Question}\sl}
\de\RR{{\tx{${\Bbb R}$}}}
\de\RRR{{\tx{$\scr R$}}}
\de\RRRB{\tx{$\RRR^B$}}
\de\RRRL{\tx{${\RRR}^<$}}
\de\RRRN{\tx{$\RRR^N$}}
\de\RRn{\tx{$\RR^n$}}
\de\RRx{\tx{$\RR\str$}}
\de\RRq{\tx{$\RR^q$}}
\de\Ref{\tx{\bf Ref:\/}}
\de\Rem{\tx{\bi Rem}}
\de\RemA{\tx{$\Rem^A$}}
\de\RemSeqA#1{\tx{$\Rem\Seq^A(#1)$}}
\de\RemSet#1{\tx{\bi RemSet\,$(#1)$}}
\de\Remark{\pr{Remark}\rm}
\de\Remarkn#1{\prn{Remark #1}\rm}
\de\Remarks{\pr{Remarks}\rm}
\de\Remarksn#1{\prn{Remarks #1}\rm}
\de\Rep{\tx{\bi Rep}}
\de\Rest{\tx{\bi Rest}}
\de\Rs{\tx{\sf R}}
\de\Rss{{\tx{\ssf R}}}
\de\SA{\tx{$\bb{S}^A$}}
\de\SS{{\tx{${\Bbb S}$}}}
\de\SSS{{\tx{${\scr S}$}}}
\de\Sarrow{\tx{$S^{\to}$}}
\de\Sat{\tx{$S_{\tx{\ssf at}}$}}
\de\Sbar{\tx{$\ol \Sbi$}}
\de\Sbi{\tx{\bi S}}
\de\Seq{\tx{\bi Seq}}
\de\Shead#1#2{\bn{\bigbf {#1}\ \ \ {#2}}\sn}
\de\Sheads#1#2{\bn{\bigbf {#1}\ \ \ {#2}}}
\de\Sig{{\tx{$\varSigma$}}}
\de\SigA{\tx{$\Sig(A)$}}
\de\SigBBB{\tx{$\Sig^\BBB$}}
\de\SigBN{\tx{$\Sig^{B,N}$}}
\de\SigN{\tx{$\Sig^N$}}
\de\SigT{\tx{$(\Sig,T)$}}
\de\Sigb{\tx{$\bs\varSigma$}}
\de\Sigbar{\tx{$\ol \Sig$}}
\de\Sigo{\tx{$\varSigma_0$}}
\de\Sigox{\tx{$\varSigma_0^*$}}
\de\Sigone{\tx{$\Sigb_1$}}
\de\SigoneInd{\tx{$\Sigb_{\pmb1}$-$\pmb{Ind}\,$}}
\de\Sigonex{\tx{$\Sigb_1^*$}}
\de\SigonexInd{\tx{$\Sigb_{\pmb1}^{\pmb*}$-$\pmb{Ind}\,$}}
\de\SigonexIndl{\tx{$\Sigb_{\pmb1}^{\pmb*}$-$\pmb{Ind_\ell}\,$}}
\de\SigonexIndc{\tx{$\Sigb_{\pmb1}^{\pmb*}$-$\pmb{Ind_c}\,$}}
\de\SigonexIndi{\tx{$\Sigb_{\pmb1}^{\pmb*}$-$\pmb{Ind_i}\,$}}
\de\Sigp{{\tx{$\Sig'$}}}
\de\SigpTp{\tx{$(\Sig',T')$}}
\de\Sigpp{{\tx{$\Sig''$}}}
\de\SigppTpp{\tx{$(\Sig'',T'')$}}
\de\Sigstk{\tx{$\Sig^\stkss$}}
\de\Siguu{\tx{$\varSigma^\uu$}}
\de\SiguuN{\tx{$\varSigma^{\uu,N}$}}
\de\Sigx{\tx{$\varSigma^*$}}
\de\Snap{\tx{\bi Snap}}
\de\SnapA{\tx{$\Snap^A$}}
\de\SnapSeqA#1{\tx{$\Snap\Seq^A(#1)$}}
\de\Sort{\tx{\bi Sort\/}}
\de\SortSig{\Sort(\Sig)}
\de\SortSigsmall{\Sortsmall\,(\Sig)}
\de\SortSigp{\tx{$\Sort(\Sigp)$}}
\de\SortSigx{\tx{$\Sort(\Sigx)$}}
\de\Sortsmall{\tx{\smallbi Sort}}
\de\Spec{\tx{\sf Spec}}
\de\Ss{\tx{\sf S}}
\de\Sss{{\tx{\ssf S}}}
\de\State{\tx{\bi State}}
\de\StateA{\tx{$\State(A)$}}
\de\Step{\tx{\bi Step}}
\de\Stmt{\tx{\bi Stmt}}
\de\StmtSig{\tx{$\Stmt(\Sig)$}}
\de\StmtStepA{\tx{$\Stmt\Step^A$}}
\de\Std{\tx{\bi Std\/}}
\de\StdAlg{\Std\Alg}
\de\StdAlgSig{\tx{$\StdAlg\,(\Sig)$}}
\de\StdAlgSigx{\tx{$\StdAlg\,(\Sigx)$}}
\de\StdMod{\Std\Mod}
\de\Str{\tx{\bi Str}}
\de\StrSig{\Str\,(\Sig)}
\de\SubalgA#1{\tx{${\bi \ Subalg}_A(#1)$}}
\de\Tb{\tx{\bf T}}
\de\Tbi{\tx{\bi T}}
\de\TComp{\tx{\bi TComp}}
\de\TE{\tx{\bi TE}}
\de\TEA{\tx{$\TE^A$}}
\de\TOL{\TagsOnLeft}
\de\TOR{\TagsOnRight}
\de\TSig{\Tbi(\Sig)}
\de\TT{\tx{${\Bbb T}$}}
\de\TTT{\tx{$\scr T$}}
\de\Tab{\tx{\bi Tab}}
\de\TabSig{\Tab(\Sig)}
\de\Tbar{\tx{$\ol T$}}
\de\Term{\tx{\bi Term}}
\de\TermSig{\tx{$\Term(\Sig)$}}
\de\TermTup{\Term\Tup}
\de\TermTupSig{\tx{$\TermTup(\Sig)$}}
\de\Terminology{\pr{Terminology}\rm}
\de\Th{\tx{\bi Th}}
\de\Thesisn#1{\pr{Thesis \ #1}\sl}
\de\Thm{\pr{Theorem}\sl}
\de\Thmn#1{\prn{Theorem #1}\sl}
\de\Thms{\pr{Theorems}\sl}
\de\Tm{\tx{\bi Tm}}
\de\TmSig{\tx{$\Tm(\Sig)$}}
\de\Tms{\tx{$\Tm_s$}}
\de\TmsSig{\tx{$\Tms(\Sig)$}}
\de\Ts{\tx{$\Tbi_s$}}
\de\TsSig{\tx{$\Ts(\Sig)$}}
\de\Tup{\tx{\bi Tup}}
\de\Ub{\tx{\bf U}}
\de\Ubi{\tx{\bi U}}
\de\Univ{\tx{\bi Univ}}
\de\Univs{\tx{\sf Univ}}
\de\Univuv{\tx{$\Univs_\utov$}}
\de\UnivuvA{\tx{$\Univs_\utov^A$}}
\de\Univuvx{\tx{$\Univs_\utov^*$}}
\de\UnivuvxA{\tx{$\Univs_\utov^{*,A}$}}
\de\Univus{\tx{$\Univ_\utos$}}
\de\UnivusA{\tx{$\Univ_\utos^A$}}
\de\Unspecs{\tx{\sf Unspec}}
\de\Updates{\tx{\sf Update}}
\de\UpdateD{\tx{$\Updates^D$}}
\de\Us{\tx{\sf U}}
\de\VVV{\tx{$\scr V$}}
\de\Val{\tx{\bi Val}}
\de\Var{\tx{\bi Var}}
\de\VarSig{\tx{$\Var(\Sig)$}}
\de\VarTup{\Var\Tup}
\de\VarTupSig{\tx{$\VarTup(\Sig)$}}
\de\Vars{\tx{$\Var_s$}}
\de\VarsSig{\tx{$\Vars(\Sig)$}}
\de\Vbb{\tx{\bf V}}
\de\WWW{\tx{${\scr W}$}}
\de\Wb{\tx{\bf W}}
\de\While{\tx{\bi While}}
\de\WhileA{\tx{$\While(A)$}}
\de\WhileN{\tx{$\While^\Nbismall$}}
\de\WhileNA{\tx{$\WhileN(A)$}}
\de\WhileNSig{\tx{$\WhileN(\Sig)$}}
\de\WhileSig{\tx{$\While(\Sig)$}}
\de\WhileSigN{\tx{$\While(\SigN)$}}
\de\WhileSigx{\tx{$\While(\Sigx)$}}
\de\Whilebig{\tx{\bigbi While}}
\de\WhilebigSig{\tx{$\Whilebig(\Sigb)$}}
\de\WhilebigN{\tx{$\Whilebig^\Nbi$}}
\de\Whilebigx{\tx{$\Whilebig^\starb$}}
\de\Whilex{\tx{$\While^\starb$}}
\de\WhilexA{\Whilex(A)}
\de\WhilexSig{\Whilex(\Sig)}
\de\WhilexSigN{\Whilex(\SigN)}
\de\Whilexbig{\tx{$\Whilebig^{\dz{\starb}}$}}
\de\XXX{{\tx{${\scr X}$}}}
\de\ZZ{\tx{${\Bbb Z}$}}
\de\ZZZ{{\tx{${\scr Z}$}}}
\de\Zbi{\tx{\bi Z}}
\de\Zs{\tx{\sf Z}}
\de\Zss{{\tx{\ssf Z}}}
\de\Zz{\tx{\sf Z}}
\de\abi{\tx{\bi a}}
\de\aas{\tx{\sf a}}
\de\abar{\tx{$\bar a$}}
\de\adt{{\bi adt}}
\de\aee{\tx{\bi ae}}
\de\ainAu{\tx{$a \in \Au$}}
\de\ainAv{\tx{$a \in \Av$}}
\de\ainAs{\tx{$a \in \As$}}
\de\al{{\tx{$\alpha$}}}
\de\alA{\tx{$\alpha^A$}}
\de\alb{\tx{$\bs\alpha$}}
\de\albar{{\tx{$\ol \alpha$}}}
\de\algebras{\tx{\sf algebra}}
\de\alhat{{\tx{$\hat{\alpha}$}}}
\de\aliass{\tx{\sf alias}}
\de\aliasss{{\tx{\ssf alias}}}
\de\all{\forall}
\de\alvec{\tx{$\vec{\alpha}$}}
\de\ands{\tx{\sf and}}
\de\ang#1{\tx{$\langle #1 \rangle$}}
\de\angg#1{\tx{$\langg #1 \rangg$}}
\de\arity#1{\tx{\rm arity$(#1)$}}
\de\arr{\arrow <0.1in> [0.2,0.5]}
\de\att{{\tx{\tt a}}}
\de\atts{{\tx{\smalltt a}}}
\de\attx{\att\str}
\de\auxs{\tx{\sf aux}}
\de\avec{\tx{$\vec{a}$}}
\de\ax{\tx{$a^*$}}
\de\bA{\tx{$\bb{b}^A$}}
\de\bB{\bk{B}}
\de\bFor{\tx{\bigbi For}}
\de\bForN{\tx{$\bFor^\Nbi$}}
\de\bForx{\tx{$\bFor^\starb$}}
\de\bProc{\tx{\bigbi Proc}}
\de\bb#1{\tx{$\lbb #1 \rbb$}}
\de\bbar{[\ns]}
\de\bbi{\tx{\bi b}}
\de\bbs{\tx{\sf b}}
\de\bdot{{\bs\cdot}}
\de\be{{\tx{$\beta$}}}
\de\beA{\tx{$\beta^A$}}
\de\beb{\tx{$\bs\beta$}}
\de\begins{\tx{\sf begin}}
\de\bigcaplim#1{\underset#1\to\bigcap}
\de\bigcon{\bigwedge}
\de\bigdis{\bigvee}
\de\bigskipn{\bigskip\nin}
\de\biu{\tx{\bi u}}
\de\bk{\boldkey}
\de\bn{\bigskip\nin}
\de\bools{\tx{\sf bool}}
\de\boolss{{\tx{\ssf bool}}}
\de\br{\ |\ }
\de\bs{\boldsymbol}
\de\btt{{\tx{\tt b}}}
\de\btts{{\tx{\smalltt b}}}
\de\bttx{\btt\str}
\de\bu{$\bullet$}
\de\bul{\item{\bu}}
\de\bull{\itemm{\bu}}
\de\bulll{\itemmm{\bu}}
\de\bx{\tx{$b^*$}}
\de\capt#1#2{\ce{{\sc #1}. \ \ #2}}
\de\card#1{\tx{\bi card\,$(#1)$}}
\de\carrierss{\tx{\sf carriers}}
\de\cart#1{\tx{\bi cart\, $(#1)$}}
\de\cb#1{\ce{\bf #1}}
\de\cbb#1{\ce{\bbf #1}}
\de\cbbb#1{\ce{\bbigbf #1}}
\de\cbi#1{\ce{\bi #1}}
\de\cbbi#1{\ce{\bigbi #1}}
\de\ccc{{\tx{\bi c}}}
\de\ce{\centerline}
\de\cf{{\it cf\.}}
\de\chead#1{\bn\cb{#1}\mn}
\de\chooses{\tx{\sf choose}}
\de\cnr#1{\tx{$\ulc #1 \urc$}}
\de\comlabs{\tx{\com\bf-\labs}}
\de\comlam{\tx{$\com^\lam$}}
\de\compb{\tx{\bi comp}}
\de\compl#1{\tx{\bi compl\,$(#1)$}}
\de\complexs{\tx{\sf complex}}
\de\complexss{{\tx{\ssf complex}}}
\de\comps{\tx{\sf comp}}
\de\con{\land}
\de\concon{{\bigcon\nss\nss\nss\bigcon}}
\de\conconlim#1{\underset#1\to\concon}
\de\conconlims#1#2{\overset{#2}\to{\conconlim{#1}}}
\de\cons{\tx{\sf con}}
\de\conss{{\tx{\ssf con}}}
\de\constantss{\tx{\sf constants}}
\de\consts{\tx{\sf const}}
\de\constss{{\tx{\ssf const}}}
\de\cpxs{\tx{\sf complex}}
\de\ccs{\tx{\sf c}}
\de\css{{\tx{\ssf c}}}
\de\ctt{{\tx{\tt c}}}
\de\ctts{{\tx{\smalltt c}}}
\de\cttx{\ctt\str}
\de\curl#1{\tx{$\{#1\}$}}
\de\curly#1{\tx{$\{\,#1\,\}$}}
\de\cvals{\tx{\sf cval}}
\de\cvec{\tx{$\vec{c}$}}
\de\cx{\tx{$c^*$}}
\de\da{\downarrow}
\de\dash{\item{---}}
\de\dashh{\itemm{---}}
\de\dashhh{\itemmm{---}}
\de\datas{{\tx{\sf data}}}
\de\datass{{\tx{\ssf data}}}
\de\dcs{\tx{\sf dc}}
\de\ddis{\lor\!\!\!\!\!\lor}
\de\defaults{\tx{\sf default}}
\de\degg#1{\tx{\bi deg$(#1)$}}
\de\del{{\tx{$\delta$}}}
\de\delb{\tx{$\bs\delta$}}
\de\delhat{\hat{\delta}}
\de\depth#1{\tx{\bi depth$(#1)$}}
\de\dis{\lor}
\de\disdis{{\bigdis\nss\nss\nss\bigdis}}
\de\disdislim#1{\underset#1\to\disdis}
\de\disdislims#1#2{\overset{#2}\to{\disdislim{#1}}}
\de\displayquote{\narrower\narrower\smallrm\nin}
\de\displaytext{\narrower\narrower\sl\nin}
\de\diss{\tx{\sf dis}}
\de\divN{\tx{$\divs_{\natss}$}}
\de\divNR{\tx{$\divs_\natss^\RRR$}}
\de\divs{\tx{\sf div}}
\de\dos{\tx{\sf do}}
\de\dom#1{\tx{\bi dom$(#1)$}}
\de\down{\tx{$^\vee$}}
\de\ds{{\tx{\sf d}}}
\de\dt{\tx{\bi dt}}
\de\dtt{{\tx{\tt d}}}
\de\dz{\dsize}
\de\ebar{{\tx{$\bar e$}}}
\de\eg{{\it e.g.}}
\de\ehat{tx{$\hat e$}}
\de\el{\tx{$\ell$}}
\de\elses{\tx{\sf else}}
\de\endpf{\qed\smskip}
\de\endpr{\rm\smskip}
\de\ends{\tx{\sf end}}
\de\enum{\tx{\bi enum}}
\de\eps{{\tx{$\epsilon$}}}
\de\epsb{{\tx{$\bs\epsilon$}}}
\de\eqbool{\tx{$\eqs_{\boolss}$}}
\de\eqdata{\tx{$\eqs_\datass$}}
\de\eqdf{=_{df}}
\de\eqint{\tx{$\eqs_{\intss}$}}
\de\eqintvl{\tx{$\eqs_{\intvlss}$}}
\de\eqnat{\tx{$\eqs_{\natss}$}}
\de\eqnatN{\tx{$\eqs_{\natss}^\NNN$}}
\de\eqreal{\tx{$\eqs_{\realss}$}}
\de\eqrealR{\tx{$\eqs_{\realss}^\RRR$}}
\de\eqs{\tx{\sf eq}}
\de\eqss{\tx{$\eqs_s$}}
\de\equal{\ &= \ }
\de\es{\tx{\sf e}}
\de\etal{{\it et al.}}
\de\ett{{\tx{\tt e}}}
\de\eval{\tx{\bi eval}}
\de\evals{\tx{\sf eval}}
\de\evec{\tx{$\vec{e}$}}
\de\ex{\exists}
\de\expands{\tx{\sf expand}}
\de\fA{\tx{$f^A$}}
\de\falses{\tx{\sf false}}
\de\fbi{\tx{\bi f}}
\de\fff{\tx{{\sf f}\!{\sf f}}}
\de\ffrom{\tx{$\Leftarrow$}}
\de\fcheck{\tx{$\check f$}}
\de\fhat{\tx{$\hat f$}}
\de\first{\tx{\bi first}}
\de\fis{\tx{\sf fi}}
\de\floor#1{\tx{$\llc #1 \lrc$}}
\de\fn{\footnote}
\de\fors{\tx{\sf for}}
\de\free{\tx{\bi free}}
\de\from{\leftarrow}
\de\fs{{\tx{\sf f}}}
\de\fss{{\tx{\ssf f}}}
\de\ftil{\tx{$\til f$}}
\de\funcs{\tx{\sf func}}
\de\functionss{\tx{\sf functions}}
\de\fuu{\tx{$f^\uu$}}
\de\fvec{\tx{$\vec{f}$}}
\de\gam{{\tx{$\gamma$}}}
\de\gamA{\tx{$\gam^A$}}
\de\gamb{\tx{$\bs\gamma$}}
\de\gamx{\tx{$\gam^*$}}
\de\ghat{\tx{$\hat g$}}
\de\gtil{\tx{$\til g$}}
\de\gn{\tx{\bi gn}}
\de\gotos{\tx{\sf goto}}
\de\graph{\tx{\bi graph}}
\de\gs{{\tx{\sf g}}}
\de\gtt{{\tx{\tt g}}}
\de\gvec{\tx{$\vec{g}$}}
\de\halt{\tx{\bi halt}}
\de\hs{{\tx{\sf h}}}
\de\htil{\tx{$\til h$}}
\de\htt{{\tx{\tt h}}}
\de\ibar{{\tx{$\bar{\imath}$}}}
\de\id{\tx{\bi id}}
\de\ident{\equiv}
\de\idi{\tx{$\id_i$}}
\de\ie{{\it i.e.}}
\de\ifbool{\tx{$\ifs_{\boolss}$}}
\de\ifdata{\tx{$\ifs_{\datass}$}}
\de\ifff{\ \,\llongtofrom\ }
\de\ifffdf{\ \ \tx{$\llongtofrom_{df}$}\ }
\de\ifint{\tx{$\ifs_{\intss}$}}
\de\ifintvl{\tx{$\ifs_{\intvlss}$}}
\de\ifnat{\tx{$\ifs_{\natss}$}}
\de\ifnatN{\tx{$\ifs_{\natss}^\NNN$}}
\de\ifreal{\tx{$\ifs_{\realss}$}}
\de\ifrealR{\tx{$\ifs_\realss^\RRR$}}
\de\ifs{\tx{\sf if}}
\de\ifss{\tx{$\ifs_s$}}
\de\ifstk{\tx{$\ifs_{\stkss}$}}
\de\ift#1{&\tx{if \ {#1}}}
\de\ii{{\tx{\bi i}}}
\de\iii{{\tx{\sf i}}}
\de\iinI{{\tx{$i \in I$}}}
\de\ims{\tx{\sf im}}
\de\imp{\to}
\de\impp{\ \,\llongto\ }
\de\imps{\tx{\sf imp}}
\de\imports{\tx{\sf import}}
\de\inbull{\itemitem{\bu}}
\de\indash{\itemitem{---}}
\de\indentt{\indent\quad}
\de\ins{\tx{\sf in}}
\de\ints{{\tx{\sf int}}}
\de\intss{{\tx{\ssf int}}}
\de\intvls{{\tx{\sf intvl}}}
\de\intvlss{{\tx{\ssf intvl}}}
\de\io{{\tx{$\iota$}}}
\de\is{\tx{\sf i}}
\de\isom{\cong}
\de\itt{{\tx{\tt i}}}
\de\ivec{\tx{$\vec{\imath}$}}
\de\ivecstrut{\vec{\mathstrut i}}
\de\jbar{{\tx{$\bar{\jmath}$}}}
\de\js{\tx{\sf j}}
\de\jvec{\tx{$\vec{\jmath}$}}
\de\jvecstrut{\vec{\mathstrut j}}
\de\kap{{\tx{$\kappa$}}}
\de\kapb{\tx{$\bs\kappa$}}
\de\kbar{\tx{$\bar k$}}
\de\kinNN{\tx{$k\in\NN$}}
\de\ktt{{\tx{\tt k}}}
\de\kvec{\tx{$\vec{k}$}}
\de\kvecstrut{\vec{\mathstrut k}}
\de\kxvec{\tx{$\kvec^*$}}
\de\labs{\lamb\tx{\bi abs}}
\de\lam{{\tx{$\lambda$}}}
\de\lamPR{\lam\PR}
\de\lamPRmin{\lam\PRmin}
\de\lamPRx{\lam\PRx}
\de\lamb{\tx{$\bs\lambda$}}
\de\lammuPR{\lam\muu\PR}
\de\lammuPRmin{\lam\muu\PRmin}
\de\lammuPRx{\lam\muu\PRx}
\de\langg{\langle\!|}
\de\lb{\linebreak}
\de\lbar{\tx{$\ol l$}}
\de\lbb{[\![}
\de\leasts{\tx{\sf least}}
\de\lele#1#2#3{{\tx{$#1 \le #2 \le #3$}}}
\de\lel#1#2#3{{\tx{$#1 \le #2 < #3$}}}
\de\lgth#1{\tx{\bi lgth$(#1)$}}
\de\llc{\llcorner}
\de\lle#1#2#3{{\tx{$#1 < #2 \le #3$}}}
\de\llongfrom{\tx{$\Longleftarrow$}}
\de\llongto{\tx{$\Longrightarrow$}}
\de\llongtofrom{\tx{$\Longleftrightarrow$}}
\de\loccit{{\it loc\. cit.}}
\de\locs{\tx{\sf loc}}
\de\longfrom{\longleftarrow}
\de\longimp{\longrightarrow}
\de\longto{\longrightarrow}
\de\longtofrom{\longleftrightarrow}
\de\loops{\tx{\sf loop}}
\de\lowerbox#1#2{{\lower#1pt\hbox{$#2$}}}
\de\lrc{\lrcorner}
\de\lss{\tx{\sf less}}
\de\lsint{\tx{$\lss_\intss$}}
\de\lsintvl{\tx{$\lss_\intvlss$}}
\de\lsnat{\tx{$\lss_\natss$}}
\de\lsnatN{\tx{$\lss_\natss^\NNN$}}
\de\lsreal{\tx{$\lss_\realss$}}
\de\lsrealR{\tx{$\lss_\realss^\RRR$}}
\de\lvec{\tx{$\vec{l}$}}
\de\lxvec{\tx{$\vec{l}^*$}}
\de\lxx{\lower.2ex}
\de\lxxx{\lower.3ex}
\de\lxxxx{\lower.4ex}
\de\macros{\tx{\sf macro}}
\de\macross{{\tx{\ssf macro}}}
\de\magn#1{\magnification=\magstep#1}
\de\mapdownl#1{\mapdown\lft{\dz{#1}}}
\de\mapdownlr#1#2{\mapdown\lft{\dz{#1}}\rt{\dz{#2}}}
\de\mapdownr#1{\mapdown\rt{\dz{#1}}}
\de\maprightd#1{\mapright_{\dz{#1}}}
\de\maprightdlower#1#2{\mapright_{\lower#1pt\hbox{$\dz{#2}$}}}
\de\maprightu#1{\mapright^{\dz{#1}}}
\de\maprightud#1#2{\mapright^{\dz{#1}}_{\dz{#2}}}
\de\maprighturaise#1#2{\mapright^{\raise#1pt\hbox{$\dz{#2}$}}}
\de\maprighturaised#1#2#3{\mapright^{\raise#1pt\hbox{$\dz{#2}$}}_{\dz{#3}}}
\de\maprightudlower#1#2#3{\mapright^{\dz{#1}}_{\lower#2pt\hbox{$\dz{#3}$}}}
\de\mapupl#1{\mapup\lft{\dz{#1}}}
\de\mapuplr#1#2{\mapup\lft{\dz{#1}}\rt{\dz{#2}}}
\de\mapupr#1{\mapup\rt{\dz{#1}}}
\de\mbar{\tx{$\bar m$}}
\de\medskipn{\medskip\nin}
\de\mg#1{\tx{\bi mg$(#1)$}}
\de\mi{\tx{$^-$}}
\de\mins{\tx{\sf min}}
\de\mn{\medskip\nin}
\de\mtt{{\tx{\tt m}}}
\de\muCR{\mub\CR}
\de\muCRA{\tx{$\muCR(A)$}}
\de\muCRSig{\tx{$\muCR(\Sig)$}}
\de\muPR{\muu\PR}
\de\muPRA{\tx{$\muPR(A)$}}
\de\muPRSig{\tx{$\muPR(\Sig)$}}
\de\muPRSigb{\tx{$\muPRb(\Sigb)$}}
\de\muPRb{\mub\PRb}
\de\muPRmin{\muu\PRmin}
\de\muPRo{\muu\PRo}
\de\muPRolam{\muu\PRolam}
\de\muPRx{\muPR\str}
\de\muPRxb{\mub\PRxb}
\de\muPRxA{\tx{$\muPRx(A)$}}
\de\muPRxSig{\tx{$\muPRx(\Sig)$}}
\de\muPRxSigutos{\tx{$\muPRxSig_\utos$}}
\de\muPRxSigb{\tx{$\muPRxb(\Sigb)$}}
\de\muPRxutos{\tx{$\muPRx_\utos$}}
\de\mub{\tx{$\bs\mu$}}
\de\muu{{\tx{$\mu$}}}
\de\mvec{\tx{$\vec{m}$}}
\de\mxvec{\tx{$\mvec^*$}}
\de\n{\noindent}
\de\names{\tx{\sf name}}
\de\namess{{\tx{\ssf name}}}
\de\nats{\tx{\sf nat}}
\de\natss{{\tx{\ssf nat}}}
\de\nbar{\tx{$\bar n$}}
\de\negs{\tx{\sf neg}}
\de\news{\tx{\sf new}}
\de\newss{{\tx{\ssf new}}}
\de\nil{\tx{$\emptyset$}}
\de\nin{\noindent}
\de\ninNN{\tx{$n\in\NN$}}
\de\nl{\newline}
\de\nneg{\neg\!\!\!\!\!\neg}
\de\notover{\tx{\bi notover}}
\de\notimplies{\tx{$\ \not\nsss\llongto$}}
\de\notimplied{\tx{$\ \not\nsss\llongfrom$}}
\de\nots{\tx{\sf not}}
\de\npn{\NoPageNumbers}
\de\ns{\negthinspace}
\de\nss{\negthickspace}
\de\nsss{\nss\nss}
\de\nssss{\nsss\nsss}
\de\ntt{{\tx{\tt n}}}
\de\nub{\tx{$\bs\nu$}}
\de\nx{\tx{$n^*$}}
\de\ol{\overline}
\de\opcit{{\it op\. cit\.}}
\de\ods{\tx{\sf od}}
\de\om{{\tx{$\omega$}}}
\de\omb{{\tx{$\bs\omega$}}}
\de\onto{\twoheadrightarrow}
\de\ors{\tx{\sf or}}
\de\outs{\tx{\sf out}}
\de\ow{&\tx{otherwise}}
\de\pa{{\it p.a.}}
\de\pairs{\tx{\sf pair}}
\de\pe{\tx{\bi pe}}
\de\ph{{\tx{$\varphi$}}}
\de\pie{\tx{$\pi$}}
\de\pmapsto{{\,\overset\ {\cdot} \to\mapsto\,}}
\de\pr#1{\mn{\bf #1}. \,}
\de\prb#1{\bn{\bbf #1}.\sn}
\de\prc#1{\sn{\sc #1}.\ \ }
\de\prn#1{\mn{\bf #1}.\,}
\de\prims{\tx{\sf prim}}
\de\procs{\tx{\sf proc}}
\de\procss{{\tx{\ssf proc}}}
\de\proj{\tx{\bi proj}}
\de\projs{\tx{\sf proj}}
\de\ps{{\tx{$\psi$}}}
\de\pte{\tx{\bi pte}}
\de\pto{{\,\overset\ {\cdot} \to\longto\,}}
\de\ptt{{\tx{\tt p}}}
\de\pttx{\ptt\str}
\de\px{\tx{$p^*$}}
\de\qqquad{\qquad \qquad}
\de\qqqquad{\qqquad \qqquad}
\de\qfors{`\,\fors'}
\de\qforxs{`$\,\fors^*$'}
\de\qtt{{\tx{\tt q}}}
\de\qwhiles{`\,\whiles'}
\de\qwhilex{`$\,\whiles^*$'}
\de\rR{\tx{$\bk R$}}
\de\raisebox#1#2{{\raise#1pt\hbox{$#2$}}}
\de\ran#1{\tx{\bi ran$(#1)$}}
\de\rand{\ \tx{\rm and}\ }
\de\rangg{|\!\rangle}
\de\rats{\tx{\sf rat}}
\de\ratss{{\tx{\ssf rat}}}
\de\rbb{]\!]}
\de\reals{\tx{\sf real}}
\de\realss{{\tx{\ssf real}}}
\de\rec{\tx{\bi rec}}
\de\redSig{\,|\,_{\Sig}}
\de\reffs{\tx{\sf ref}}
\de\refss{{\tx{\ssf ref}}}
\de\rel#1{\tx{\rm (rel $#1$)}}
\de\rem{\tx{\bi rem}}
\de\res{\tx{\sf re}}
\de\rest{\restriction}
\de\restt{\tx{\bi rest}}
\de\rh{\tx{$\rho$}}
\de\rtt{{\tx{\tt r}}}
\de\rvec{\tx{$\vec{r}$}}
\de\rxx{\raise.2ex}
\de\rxxx{\raise.3ex}
\de\rxxxx{\raise.4ex}
\de\sbar{{\tx{$\bar s$}}}
\de\scr{\Cal}
\de\se{\tx{\bi se}}
\de\seq{\tx{\bi seq}}
\de\seqt{\longmapsto}
\de\setss{{\tx{\ssf set}}}
\de\shead#1#2{\mn{\bf {#1}\ \ \ {#2}}\sn}
\de\sheadb#1#2{\mn{\bf {#1}\ \ \ {#2}}\nl}
\de\sheadrun#1#2{\mn{\bf {#1}\ \ \ {#2}.\ \ }}
\de\sheads#1#2{\mn{\bf {#1}\ \ \ {#2}}}
\de\sig{{\tx{$\sigma$}}}
\de\signatures{\tx{\sf signature}}
\de\sigo{\tx{$\sig_0$}}
\de\sigp{\tx{$\sig'$}}
\de\sinSortSig{{\tx{$s \in \SortSig$}}}
\de\sinSortSigx{\tx{$s \in \SortSigx$}}
\de\skips{\tx{\sf skip}}
\de\smskip{\smallskip}
\de\smskipn{\smallskip\nin}
\de\sn{\smallskip\nin}
\de\snap{\tx{\bi snap}}
\de\sort#1{\tx{\bi sort\,{$(#1)$}}}
\de\sortss{\tx{\sf sorts}}
\de\sq{\simeq}
\de\sqsseq{\sqsubseteq}
\de\srchs{\tx{\sf srch}}
\de\ssA{{\ssz A}}
\de\ssB{{\ssz B}}
\de\ssbe{{\ssz \beta}}
\de\sset{\subset}
\de\sseq{\subseteq}
\de\sshead#1#2{{\bf {#1}\ \ \ {#2}}\sn\!}
\de\ssheadrun#1#2{{\bf {#1}\ \ \ {#2}.\ \ }}
\de\ssheads#1#2{{\bf {#1}\ \ \ {#2}}}
\de\sss{\tx{\sf s}}
\de\sz{\ssize}
\de\ssz{\sssize}
\de\starb{{\tx{$\bk *$}}}
\de\state{\tx{\bi state}}
\de\stks{{\tx{\sf stk}}}
\de\stkss{{\tx{\ssf stk}}}
\de\str{\tx{$^*$}}
\de\stt{{\tx{\tt s}}}
\de\strb{\tx{$^{\bk*}$}}
\de\subex{\tx{\bi subex}}
\de\subst{\tx{\bi subst}}
\de\sucs{\tx{\sf suc}}
\de\suu{{\tx{$s^\uu$}}}
\de\sx{{\tx{$s^*$}}}
\de\tA{\tx{$\bb{t}^A$}}
\de\ta{\tx{$\tau$}}
\de\tagx{\tag{$*$}}
\de\tagxx{\tag{$**$}}
\de\tagxxx{\tag{${*}{*}{*}$}}
\de\tagxxxx{\tag{${*}{*}{*}{*}$}}
\de\tagxxxxx{\tag{${*}{*}{*}{*}{*}$}}
\de\tagxxxxxx{\tag{${*}{*}{*}{*}{*}{*}$}}
\de\tbar{\tx{$\bar t$}}
\de\te{\tx{\bi te}}
\de\term{\tx{\bi term}}
\de\that{\tx{$\hat t$}}
\de\thens{\tx{\sf then}}
\de\tif#1{&\tx{if \ {#1}}}
\de\til{\tilde}
\de\tiltil#1{\tilde{\tilde#1}}
\de\tinT{\tx{$t\in\TT$}}
\de\th{\tx{$\theta$}}
\de\tofrom{\leftrightarrow}
\de\tos{\tx{\sf to}}
\de\toto{\rightrightarrows}
\de\trues{\tx{\sf true}}
\de\tstile{\vdash}
\de\tto{\tx{$\Rightarrow$}}
\de\ttofrom{\tx{$\Leftrightarrow$}}
\de\ttstile{\models}
\de\ttt{\tx{{\sf t}\!{\sf t}}}
\de\tup#1#2#3{\tx{$#1_{#2},\dots,#1_{#3}$}}
\de\tuptimes#1#2#3{\tx{$#1_{#2} \times \dots \times #1_{#3}$}}
\de\tvec{\tx{$\vec{t}$}}
\de\tvecstrut{\vec{\mathstrut t}}
\de\twiddle{{\tx{$\sim$}}}
\de\twiddlek{\tx{$\sim_k$}}
\de\tx{\text}
\de\txvec{\tx{$\vec{t^*}$}}
\de\type#1{\tx{\bi type$(#1)$}}
\de\ua{{\tx{$\uparrow$}}}
\de\ub#1{\tx{\underbar{#1}}}
\de\ubi{{\tx{\bi u}}}
\de\ubrivec{\tx{$u | \ivec$}}
\de\uinProdTypeSig{\tx{$u \in \ProdTypeSig$}}
\de\ul{\underline}
\de\ulc{\ulcorner}
\de\up{\tx{$^\wedge$}}
\de\urc{\urcorner}
\de\unspecs{\tx{\sf unspec}}
\de\unspecN{\tx{$\unspecs_\natss$}}
\de\unspecB{\tx{$\unspecs_\boolss$}}
\de\ups{{\tx{$\upsilon$}}}
\de\utos{{\tx{$u \to s$}}}
\de\utov{{\tx{$u \to v$}}}
\de\uu{{\tx{\ssf u}}}
\de\uuu{\tx{{\sf u}\!\i}}
\de\uuus{\tx{$\uuu_s$}}
\de\uvec{\tx{$\vec{u}$}}
\de\ux{{\tx{$u^*$}}}
\de\val{\tx{\bi val}}
\de\vals{\tx{\sf val}}
\de\valss{{\tx{\ssf val}}}
\de\var#1{\tx{\bi var$(#1)$}}
\de\vars{\tx{\sf var}}
\de\vects{\tx{\sf vect}}
\de\vinProdTypeSig{\tx{$v \in \ProdTypeSig$}}
\de\vvec{\tx{$\vec{v}$}}
\de\vvecstrut{\vec{\mathstrut v}}
\de\vx{{\tx{$v^*$}}}
\de\vxvec{\tx{$\vec{v}\,^*$}}
\de\whiles{\tx{\sf while}}
\de\wrt{w.r.t\.}
\de\wvec{\tx{$\vec{w}$}}
\de\wx{{\tx{$w^*$}}}
\de\x{\tx{\bi x}}
\de\xbar{\tx{$\bar x$}}
\de\xib{\tx{$\bs\xi$}}
\de\xii{\tx{$\xi$}}
\de\xinAbaru{\tx{$x \in \Abaru$}}
\de\xinAu{\tx{$x \in \Au$}}
\de\xinAw{\tx{$x \in \Aw$}}
\de\xtt{{\tx{\tt x}}}
\de\xttx{\tx{$\xtt^*$}}
\de\xtts{{\tx{\smalltt x}}}
\de\xttvec{\tx{$\vec{\xtt}$}}
\de\xvec{\tx{$\vec{x}$}}
\de\xx{\tx{$x^*$}}
\de\xxvec{\tx{$\xvec^*$}}
\de\xxx{\tx{${*}{*}{*}$}}
\de\xxxx{\tx{${*}{*}{*}{*}$}}
\de\xxxxx{\tx{${*}{*}{*}{*}{*}$}}
\de\xxxxxx{\tx{${*}{*}{*}{*}{*}{*}$}}
\de\yinAs{\tx{$y \in \As$}}
\de\yinAv{\tx{$y \in \Av$}}
\de\yvec{\tx{$\vec{y}$}}
\de\yx{\tx{$y^*$}}
\de\yxvec{\tx{$\yvec\,^*$}}
\de\ytt{{\tx{\tt y}}}
\de\yttx{\ytt\str}
\de\ztt{{\tx{\tt z}}}
\de\zttx{\ztt\str}
\de\zvec{\tx{$\vec{z}$}}
\de\zx{\tx{$z^*$}}
\de\zxvec{\tx{$\zvec\,^*$}}


\newcount\mmm
\de\itemn#1{\mmm=#1\par\indentn\mmm \advance\mmm by 1
      \hangindent\mmm \parindent \textindent
 }

\newcount\nnn

\global\de\indentn#1{
\nnn=#1
\loop \ifnum\nnn>0 
      \indent
      \advance\nnn by -1
\repeat
}

\de\itemm{\itemn1}
\de\itemmm{\itemn2}

\newcount\ppp

\global\de\quadn#1{
\n=#1
\loop \ifnum\ppp>0
        \quad
        \advance\ppp by -1
\repeat
}

\firstfoot{{\ssmallrm ACM Transactions on Computational Logic,
           Vol.\ TBD, No.\ TBD, TBD TBD, Pages TBD.}}
\runningfoot{{\ssmallrm ACM Transactions on Computational Logic,
           Vol.\ TBD, No.\ TBD, TBD TBD.}}

\de\AExA{\tx{$\AEE\,_\xtt^A$}}
\de\APhi{\tx{$A^\Phi$}}
\de\APsi{\tx{$A^\Psi$}}
\de\ATTT{\tx{$(A,\TTT)$}}
\de\AbarS{\tx{$\Abar^{\Sbi}$}}
\de\AbsComp{\tx{\bi AbsComp}}
\de\AbsCompA{\tx{$\AbsComp(A)$}}
\de\Ad{\tx{$(A,d)$}}
\de\Af{\tx{$(A,f)$}}
\de\AfalA{\tx{$(A,\,\falA)$}}
\de\AgalAfalA{\tx{$(A,\,\galA,\,\falA)$}}
\de\AinMinNStdAlgSigE{\tx{$A\in\MinNStdAlgSigE$}}
\de\AinMinNStdAlgSigT{\tx{$A\in\MinNStdAlgSigT$}}
\de\AinNStdAlgSigE{\tx{$A\in\NStdAlgSigE$}}
\de\AinNStdAlgSigT{\tx{$A\in\NStdAlgSigT$}}
\de\AlgSigE{\Alg\SigE}
\de\AlgSigF{\Alg\SigF}
\de\AlgSigT{\Alg\SigT}
\de\Approx{\tx{\bi Approx}}
\de\ArrAx{\tx{\sf ArrAx}}
\de\ArrAxSig{\tx{$\ArrAx(\Sig)$}}
\de\Asprod{\tx{$A_{s_1}\times\dots\times A_{s_n}$}}
\de\Aspx{\tx{$A_{s'}^*$}}
\de\Ass{\tx{$A_{s_1},\dots,, A_{s_n}$}}
\de\AtStx{\tx{$\AtSt\,_\xtt$}}
\de\Axbar{\tx{$\ol{\Ax}$}}
\de\AxfalA{\tx{$(\Ax,\,\falA)$}}
\de\AxgalAfalA{\tx{$(\Ax,\,\galA,\,\falA)$}}
\de\Axp{\tx{$\Ax'$}}
\de\angte{\ang{\teasA \mid \sinSortSig}}
\de\BBBbar{\tx{$\ol \BBB$}}
\de\BUEq{\tx{\bi BUEq}}
\de\BUEqSig{\BUEq(\Sig)}
\de\BddAx{\tx{\sf BddAx}}
\de\BddAxSig{\tx{$\BddAx(\Sig)$}}
\de\BddAxo{\tx{$\BddAx^0$}}
\de\BddAxoSig{\tx{$\BddAxo(\Sig)$}}
\de\CA{\tx{$C(A)$}}
\de\CCCInd{\tx{\CCC-\Ind}}
\de\CCCIndSig{\tx{\CCC-\IndSig}}
\de\CCCIndi{\tx{\CCC-\Indi}}
\de\CCCIndiSig{\tx{\CCC-\IndiSig}}
\de\CCCN{\tx{$\CCC^N$}}
\de\CCCNL{\tx{$\CCC^{N,<}$}}
\de\CCCo{\tx{$\CCC_0$}}
\de\CompI{\tx{$\Comp_1$}}
\de\CompIA{\tx{$\Comp_1^A$}}
\de\CompN{\tx{$\Comp(\NN)$}}
\de\CompR{\tx{$\Comp(\RR)$}}
\de\CompalA{\tx{$\Comp_\al(A)$}}
\de\CompxA{\tx{$\Comp\,_\xtt^A$}}
\de\ConcRep{\tx{\bi ConcRep}}
\de\ConcRepA{\tx{$\ConcRep(A)$}}
\de\CondBUEq{\tx{\sf CondBUEq}}
\de\CondBUEqSig{\tx{$\CondBUEq(\Sig)$}}
\de\CondEq{\tx{\sf CondEq}}
\de\CondEqSig{\tx{$\CondEq(\Sig)$}}
\de\CondEqom{\tx{$\CondEq_\omb$}}
\de\CondEqomSig{\tx{$\CondEqom(\Sig)$}}
\de\CondSUEq{\tx{\sf CondSUEq}}
\de\CondSUEqom{\tx{$\CondSUEq_\om$}}
\de\CondSUEqomSig{\tx{$\CondSUEqom(\Sig)$}}
\de\EU{\tx{$E^U$}}
\de\EUus{\tx{$E^U_{u,s}$}}
\de\EVm{\tx{$E^V_m$}}
\de\EVus{\tx{$E^V_{u,s}$}}
\de\EWm{\tx{$E^W_m$}}
\de\EWn{\tx{$E^W_n$}}
\de\Emz{\tx{$E_m(\ztt)$}}
\de\Eal{\tx{$E_\al$}}
\de\Ealx{\tx{$E_\al^*$}}
\de\Ee{\tx{\sf E}}
\de\Ef{\tx{$E^f$}}
\de\Egame{\tx{$E_{\game}$}}
\de\Ekbar{\tx{$E(\kbar)$}}
\de\Enbar{\tx{$E(\nbar)$}}
\de\EqSig{\Eqb(\Sig)}
\de\Eqn{\tx{\bi Eqn}}
\de\EqnA{\tx{$\Eqn(A)$}}
\de\Eqom{\tx{$\Eqb_\omb$}}
\de\EqomSig{\Eqom(\Sig)}
\de\EtilWn{\tx{$\til{E}^W_n$}}
\de\Eval{\tx{\bi Eval}}
\de\EvalAx{\tx{\sf EvalAx}}
\de\EvalAxSig{\tx{$\EvalAx(\Sig)$}}
\de\Ez{\tx{$E(\ztt)$}}
\de\FAP{\tx{\bi FAP}}
\de\FAPA{\tx{$\FAP(A)$}}
\de\FAPC{\tx{\bi FAPC}}
\de\FAPCA{\tx{$\FAPC(A)$}}
\de\FAPCS{\tx{\bi FAPCS}}
\de\FAPCSA{\tx{$\FAPCS(A)$}}
\de\FAPS{\tx{\bi FAPS}}
\de\FAPSA{\tx{$\FAPS(A)$}}
\de\FOL{\tx{\sf FOL}}
\de\FOLSig{\tx{$\FOL(\Sig)$}}
\de\FOLom{\tx{$\FOL_\omb$}}
\de\FOLomSig{\tx{$\FOLom(\Sig)$}}
\de\FOLomo{\tx{$\FOL^{\bk 0}_\omb$}}
\de\FOLomoSig{\tx{$\FOLomo(\Sig)$}}
\de\FPD{\tx{\bi FPD}}
\de\FPDA{\tx{$\FPD(A)$}}
\de\FU{\tx{$F^U$}}
\de\FUus{\tx{$F^U_{u,s}$}}
\de\FVus{\tx{$F^V_{u,s}$}}
\de\Fal{\tx{$F_\al$}}
\de\Falhat{\tx{$F_\alhat$}}
\de\Falx{\tx{$F_\al^*$}}
\de\Falminx{\tx{$F_\almin^*$}}
\de\Ff{\tx{$F^f$}}
\de\Fgame{\tx{$F_{\gamma,e}$}}
\de\Fgamex{\tx{$F_{\gamma,e}^*$}}
\de\Firstx{\tx{$\First\,_\xtt$}}
\de\Fo{{\tx{$F^\circ$}}}
\de\Fotil{\tx{$\til Fo$}}
\de\ForProcSig{\tx{\For\Proc(\Sig)}}
\de\Fmu{\tx{$F_\mu$}}
\de\ForSigbar{\tx{$\For\,(\Sigbar)$}}
\de\ForoAbar{\tx{$\Foro(\Abar)$}}
\de\ForoSigbar{\tx{$\Foro(\Sigbar)$}}
\de\ForolamSigbar{\tx{$\Forolam(\Sigbar)$}}
\de\ForolamAbar{\tx{$\Forolam(\Abar)$}}
\de\Forxp{\tx{$\Forx'$}}
\de\GGGN{\tx{$\GGG^N$}}
\de\GGGo{\tx{$\GGG_0$}}
\de\GamDel{\tx{$\Gam \seqt \Del$}}
\de\GamP{\tx{$\Gam \seqt P$}}
\de\GamQ{\tx{$\Gam \seqt Q$}}
\de\GamR{\tx{$\Gam \seqt R$}}
\de\GL{\tx{\rm GL}}
\de\GLmI{\tx{$\GL_m(I)$}}
\de\GLtmI{\tx{$\GL^{\ssz\tx{\rm T}}_m(I)$}}
\de\HF{\tx{\bi HF}}
\de\HFA{\tx{$\HF(A)$}}
\de\HaltAP{\HaltA{P}}
\de\HaltTestu{\tx{$\HaltTests_u$}}
\de\HaltTestuA{\tx{$\HaltTests_u^A$}}
\de\IIId{\tx{$\III^d$}}
\de\IIIp{\tx{$\III_p$}}
\de\IIItN{\tx{$\III_t^N$}}
\de\INS{\Init\NStdAlg}
\de\INSSigE{\tx{$\INS\SigE$}}
\de\INSSigF{\tx{$\INS\SigF$}}
\de\INSSigT{\tx{$\INS\SigT$}}
\de\IOP{\tx{$\IO_P$}}
\de\ISigE{\Init\SigE}
\de\ISigKK{\Init\SigKK}
\de\ISigT{\Init\SigT}
\de\In{\tx{$I^n$}}
\de\Iq{\tx{$I^q$}}
\de\JSig{\tx{$J\redSig$}}
\de\KKPhi{\tx{$\Bbb K ^{\Phi}$}}
\de\KKbarS{\tx{$\KKbar^{\Sbi}$}}
\de\KKp{\tx{$\Bbb K '$}}
\de\LJ{\tx{\sf LJ}}
\de\LJe{\tx{$\LJ_{\es}$}}
\de\LJeSig{\LJe(\Sig)}
\de\LK{\tx{\sf LK}}
\de\LKe{\tx{$\LK_{\es}$}}
\de\LKeSig{\LKe(\Sig)}
\de\Lomiom{\tx{$L_{\om_1\om}$}}
\de\MA{\tx{$\MMM_A$}}
\de\MCA{\tx{$\MC(A)$}}
\de\MCAbar{\tx{$\MC(\Abar)$}}
\de\NNmin{\tx{$\NN^-$}}
\de\NNp{\tx{$\NN'$}}
\de\NNtoA{\tx{$[\NN\to \!A]$}}
\de\NNtoAs{\tx{$[\NN\to \!A_s]$}}
\de\NNtoBB{\tx{$[\NN\to \BB]$}}
\de\NNtoRR{\tx{$[\NN\to \RR]$}}
\de\NNNbar{\tx{$\ol \NNN$}}
\de\NNNmin{\tx{$\NNN^-$}}
\de\NNNmino{\tx{$\NNN_0^-$}}
\de\NNbar{\tx{$\ol{\NN}$}}
\de\NNtoNN{\tx{$[\NN\to\NN]$}}
\de\MinNStdAlgSigE{\tx{$\MinNStdAlg\SigE$}}
\de\MinNStdAlgSigT{\tx{$\MinNStdAlg\SigT$}}
\de\NStdAlgSigE{\tx{$\NStdAlg\SigE$}}
\de\NStdAlgSigF{\tx{$\NStdAlg\SigF$}}
\de\NStdAlgSigG{\tx{$\NStdAlg\SigG$}}
\de\NStdAlgSigT{\tx{$\NStdAlg\SigT$}}
\de\NStdAlgSigp{\NStdAlg(\Sigp)}
\de\NStdAlgSigx{\NStdAlg(\Sigx)}
\de\NStdAx{\tx{\sf NStdAx}}
\de\NStdAxo{\tx{$\NStdAx_0$}}
\de\NStdAxSig{\tx{$\NStdAx(\Sig)$}}
\de\NStdAxoSig{\tx{$\NStdAxo(\Sig)$}}
\de\NtoN{\tx{$\Ns\to\Ns$}}
\de\Omal{\tx{$\Om_\al$}}
\de\PAN{\tx{$P^{A^N}$}}
\de\PEabc{\tx{$\PE_\abct$}}
\de\PEabcA{\tx{$\PE\,_\abct^A$}}
\de\PRAuv{\tx{$\PRA_{\utov}$}}
\de\PRAbar{\tx{$\PR(\Abar)$}}
\de\PRN{\tx{$\PR(\NNN)$}}
\de\PRNbar{\tx{$\PR(\NNNbar)$}}
\de\PRSchemeSig{\PR\Scheme(\Sig)}
\de\PRSigbar{\tx{$\PR(\Sigbar)$}}
\de\PRSiguv{\tx{$\PRSig_\utov$}}
\de\PRSigbaruv{\tx{$\PR(\Sigbar)_\utov$}}
\de\PRetc{PR (etc.)}
\de\PRminAbar{\tx{$\PRmin(\Abar)$}}
\de\PRminSigbar{\tx{$\PRmin(\Sigbar)$}}
\de\PRoAbar{\tx{$\PRo(\Abar)$}}
\de\PRoSigbar{\tx{$\PRo(\Sigbar)$}}
\de\PRolamAbar{\tx{$\PRolam(\Abar)$}}
\de\PRolamSigbar{\tx{$\PRolam(\Sigbar)$}}
\de\PRor{\PR\ (or \PRx, or \muPR, or \muPRx)}
\de\PRxAbar{\tx{$\PRx(\Abar)$}}
\de\PRxSigbar{\tx{$\PRx(\Sigbar)$}}
\de\PRxSiguv{\tx{$\PRxSig_\utov$}}
\de\Per#1{\tx{\bi Per\,$(#1)$}}
\de\PinProcuv{\tx{$P\in\Procuv$}}
\de\Procabc{\tx{$\Proc\,_\abct$}}
\de\Psrch{\tx{$P_{srch}$}}
\de\QF{\tx{\sf QF}}
\de\RAB{\tx{$R_A(B)$}}
\de\RABs{\tx{$\RAB_s$}}
\de\REC{\tx{REC}}
\de\RRBBfield{\tx{$\RR,\,\BB;\,0,\,1,\,x+y,\,-x,\,x\cdot y,\,x^{-1}$}}
\de\RRRLE{\tx{$\RRR^{<,E}$}}
\de\RRRLEN{\tx{$\RRR^{<,E,N}$}}
\de\RRRLN{\tx{$\RRR^{<,N}$}}
\de\RRRLp{\tx{$\RRR_p^<$}}
\de\RRRNt{\tx{$\RRR^N_t$}}
\de\RRRNIt{\tx{$\RRR^{N,I}_t$}}
\de\RRRd{\tx{$\RRR^d$}}
\de\RRRo{\tx{$\RRR_0$}}
\de\RRRp{\tx{$\RRR_p$}}
\de\RRRtN{\tx{$\RRR_t^N$}}
\de\RRfield{\tx{$\RR;\,0,\,1,\,x+y,\,-x,\,x\cdot y,\,x^{-1}$}}
\de\RRtwon{\tx{$\RR^{2n}$}}
\de\Rc{\tx{$R^\css$}}
\de\Rec{\tx{\bi Rec}}
\de\RemxA{\tx{$\Rem\,_\xtt^A$}}
\de\RepxA{\tx{$\Rep\,_\xtt^A$}}
\de\RepxxA{\tx{$\Rep\,_{\xtt \,*}^A$}}
\de\RestA{\tx{$\Rest\,^A$}}
\de\RestxA{\tx{$\Rest\,_\xtt^A$}}
\de\SAsig{\tx{$\SA\sig$}}
\de\SExA{\tx{$\SE\,_\xtt^A$}}
\de\SR{\tx{\bi SR}}
\de\SRA{\tx{$\SR(A)$}}
\de\SUEq{\tx{\bi SUEq}}
\de\SUEqSig{\SUEq(\Sig)}
\de\SaA#1{\tx{$\ang{#1}^A$}}
\de\San{\cS,a,n}
\de\Sadeln{\cS,\,\adel,\,n}
\de\SasA#1{\tx{$\ang{#1}_s^A$}}
\de\SasAx#1{\tx{$\ang{#1}_s^\Ax$}}
\de\SasnA#1#2{\tx{$\ang{#1}_{s,#2}^A$}}
\de\SauA#1{\tx{$\ang{#1}_u^A$}}
\de\SavA#1{\tx{$\ang{#1}_v^A$}}
\de\Se{\tx{$\Sbi_e$}}
\de\SigE{\tx{$(\Sig,E)$}}
\de\SigF{\tx{$(\Sig,F)$}}
\de\SigG{\tx{$(\Sig,G)$}}
\de\SigIIId{\tx{$\Sig(\IIId)$}}
\de\SigKK{\tx{$(\Sig,\KK)$}}
\de\SigTF{\tx{$(\Sig,T,\FFF)$}}
\de\SigXTF{\tx{$(\Sig,X,T,\FFF)$}}
\de\Sigal{\tx{$\Sig_\al$}}
\de\SigalEal{\tx{$(\Sigal,\Eal)$}}
\de\Sigalx{\tx{$\Sig_\al^*$}}
\de\SigalxEal{\tx{$(\Sigalx,\Eal)$}}
\de\SigalxFal{\tx{$(\Sigalx,\Fal)$}}
\de\SigbarS{\tx{$\Sigbar^{\Sbi}$}}
\de\SigIIIdx{\tx{$\SigIIId^*$}}
\de\Sigf{\tx{$\Sig_\fss$}}
\de\SigpSig{{\tx{$\Sigp/\Sig$}}}
\de\Sigtilxn{\tx{$\til{\Sig}^*_n$}}
\de\Sigxgame{\tx{$\Sig_{\gam,e}^*$}}
\de\Sigxm{\tx{$\Sig_m^*$}}
\de\Sigxn{\tx{$\Sig_n^*$}}
\de\Sigxus{\tx{$\Sig_{u,s}^*$}}
\de\Sigxusp{\tx{$\Sig_{u,s}^{*\prime}$}}
\de\Sigxp{\tx{$\Sigx'$}}
\de\Sigxpm{\tx{$\Sig^{*\prime}_m$}}
\de\Sigxpn{\tx{$\Sig^{*\prime}_n$}}
\de\Sinit{\tx{$S_{init}$}}
\de\SinStmtx{\tx{$S\in\Stmtx$}}
\de\SnapxA{\tx{$\Snap_\xtt^A$}}
\de\Splus{\Stimes}
\de\StateAcupx{\tx{$\StateA\cup\{*\}$}}
\de\StatesA{\tx{$\State_s(A)$}}
\de\StdAlgSigE{\tx{$\StdAlg\SigE$}}
\de\StdAlgSigF{\tx{$\StdAlg\SigF$}}
\de\StdAlgSigT{\tx{$\StdAlg\SigT$}}
\de\StdAlgSigbar{\tx{$\StdAlg(\Sigbar)$}}
\de\StdAlgSigbarS{\tx{$\StdAlg(\SigbarS)$}}
\de\StdAlgSigp{\tx{$\StdAlg(\Sigp)$}}
\de\Stmtx{\tx{$\Stmt_{\,\xtt}$}}
\de\SubAlgStage{\tx{\bi SubAlgStage}}
\de\SubAlgStagexus{\tx{$\SubAlgStage\,_{\xtt,u,s}$}}
\de\Sx{\tx{$[S,\xtt]$}}
\de\TExsA{\tx{$\TE\,_{\xtt,s}^A$}}
\de\TP{\tx{$\TTT(P)$}}
\de\TSigKK{\Tbi\SigKK}
\de\TSigTF{\Tbi\SigTF}
\de\TSigX{\tx{$\Tbi(\Sig,X)$}}
\de\TSigXTF{\Tbi\SigXTF}
\de\TSigx{\tx{$\Tbi(\Sig,\xtt)$}}
\de\TSx{\TTT\Sx}
\de\TermTupa{\tx{$\TermTup_\att$}}
\de\TermTupu{\tx{$\TermTup_u$}}
\de\TermTupx{\tx{$\TermTup_\xtt$}}
\de\TermTupxu{\tx{$\TermTup_{\xtt,u}$}}
\de\TermTupxv{\tx{$\TermTup_{\xtt,v}$}}
\de\TermTupxw{\tx{$\TermTup_{\xtt,w}$}}
\de\Terma{\tx{$\Term_{\,\att}$}}
\de\TermaN{\tx{$\Term_{\,\att}^N$}}
\de\Termabool{\tx{$\Term_{\,\att,\boolss}$}}
\de\Termaboolx{\tx{$\Term_{\,\att,\boolss}^*$}}
\de\Termanatx{\tx{$\Term_{\,\att,\natss}^*$}}
\de\Termas{\tx{$\Term_{\,\att,s}$}}
\de\TermasN{\tx{$\Term_{\,\att,s}^N$}}
\de\TermasuuN{\tx{$\Term_{\,\att,s}^{\uu,N}$}}
\de\Termasx{\tx{$\Term_{\,\att,s}^*$}}
\de\TermauuN{\tx{$\Term_{\,\att}^{\uu,N}$}}
\de\Termaw{\tx{$\Term_{\,\att,w}$}}
\de\Termax{\tx{$\Term_{\,\att}^*$}}
\de\Terms{\tx{$\Term_s$}}
\de\Termx{\tx{$\Term_{\,\xtt}$}}
\de\Termxs{\tx{$\Term_{\,\xtt,s}$}}
\de\TTTp{\tx{$\TTT'$}}
\de\TTTs{\tx{$\TTT_s$}}
\de\TsSigx{\tx{$\Tbi_s(\Sig,\xtt)$}}
\de\VarTupu{\tx{$\VarTup_u$}}
\de\WhileAp{\tx{$\While(A')$}}
\de\WhileProc{\While\Proc}
\de\WhileProcSig{\tx{$\WhileProc(\Sig)$}}
\de\WhileSigp{\tx{$\While(\Sigp)$}}
\de\Whilexp{\tx{$\Whilex'$}}
\de\abct{{\tx{$\att,\btt,\ctt$}}}
\de\ack{\tx{\bi ack}}
\de\aeuA{\tx{$\aee\biu^A$}}
\de\aexA{\tx{$\aee\,_\xtt^A$}}
\de\aexSA{\tx{$\aee\,_{\xtt,S}^A$}}
\de\adel{\tx{$(a,\delbA)$}}
\de\alAbar{\tx{$\al^\Abar$}}
\de\allR{\tx{$\all R$}}
\de\allomR{\tx{$\all_\om R$}}
\de\almin{{\tx{$\alpha^-$}}}
\de\approxKK{\tx{$\approx_\KK$}}
\de\approxTF{\tx{$\approx_{T,\FFF}$}}
\de\bSl{\tx{$\bbb_{S,\lam}$}}
\de\bSlk{\tx{$\bbb_{S,\lam_k}$}}
\de\bbSA{\tx{$\bb{S}^A$}}
\de\bbbk{\tx{$\bbb_k$}}
\de\bbbkx{\tx{$\bbb_k^*$}}
\de\bbtA{\tx{$\bb{t}^A$}}
\de\bSk{\tx{$\bbb_{S,k}$}}
\de\cP{\cnr{P}}
\de\cS{\cnr{S}}
\de\compseqxA#1{\tx{$\comp\seq\,_\xtt^A(#1)$}}
\de\compxA{\tx{$\comp\,_\xtt^A$}}
\de\compxSA{\tx{$\comp\,_{\xtt,S}^A$}}
\de\compuaA{\tx{$\comp\biu\,_\att^A$}}
\de\comptree{\tx{\comp\tree}}
\de\concat{^\smallfrown}
\de\cxtt{\cnr{\xtt}}
\de\dI{\tx{$d_\intvlss$}}
\de\dR{\tx{$d_\realss$}}
\de\delbA{\tx{$\delb_A$}}
\de\delbAu{\tx{$\delb_A^u$}}
\de\delbAv{\tx{$\delb_A^v$}}
\de\delbAw{\tx{$\delb_A^w$}}
\de\delbs{\tx{$\delb^s$}}
\de\delbsA{\tx{$\delb^s_A$}}
\de\delbu{\tx{$\delb^u$}}
\de\delbuA{\tx{$\delb^u_A$}}
\de\en{\tx{\bi en}}
\de\enA{\tx{$\en_A$}}
\de\enAs{\tx{$\en_A^s$}}
\de\enAsx{\tx{$\en_A^{s*}$}}
\de\enAx{\tx{$\en_A\str$}}
\de\endd$1{\tx{{\bi end}\,($$1$)}}
\de\eqNm{\tx{$\eqs_{\natss^-}$}}
\de\eqp{\tx{$\eqs_p$}}
\de\evalAs{\tx{$\evals_s^A$}}
\de\exomL{\tx{$\ex_\om L$}}
\de\exps{\tx{\sf exp}}
\de\expss{\tx{\ssf exp}}
\de\fal{\tx{$\fs_\al$}}
\de\falA{\tx{$\fs_\al^A$}}
\de\falhat{\tx{$\fs_\alhat$}}
\de\falhatA{\tx{$\fs_\alhat^A$}}
\de\fmu{\tx{$f_\mu$}}
\de\fps{\tx{\sf fp}}
\de\fsA{\tx{$\fs^A$}}
\de\fsigA{\tx{$\fs_\sig^A$}}
\de\fupsA{\tx{$\fs_\ups^A$}}
\de\gN{\tx{$g^\natss$}}
\de\gal{\tx{$\gs_\al$}}
\de\galA{\tx{$\gs_\al^A$}}
\de\galhat{\tx{$\gs_\alhat$}}
\de\galhatA{\tx{$\gs_\alhat^A$}}
\de\gbeA{\tx{$\gs_\be^A$}}
\de\galiA{\tx{$\gs_{\al,i}^A$}}
\de\game{\tx{$(\gamma,e)$}}
\de\grps{\tx{\sf grp}}
\de\grpss{{\tx{\ssf grp}}}
\de\haltP{\tx{$\halt_P$}}
\de\haltS{\tx{$\halt_S$}}
\de\hN{\tx{$h^\natss$}}
\de\iI{\tx{$\iota_I$}}
\de\ia{\tx{$\io_\att$}}
\de\invs{\tx{\sf inv}}
\de\invexps{\tx{\sf invexp}}
\de\invexpss{\tx{\ssf invexp}}
\de\isint{\tx{\sf is\_int}}
\de\lamMCA{\tx{$\lamMC(A)$}}
\de\lamMCAbar{\tx{$\lamMC(\Abar)$}}
\de\lamPRA{\tx{$\lamPR(A)$}}
\de\lamPRAbar{\tx{$\lamPR(\Abar)$}}
\de\lamPRN{\tx{$\lamPR(\NNN)$}}
\de\lamPRNbar{\tx{$\lamPR(\NNNbar)$}}
\de\lamPRSigbar{\tx{$\lamPR(\Sigbar)$}}
\de\lamPRminAbar{\tx{$\lamPRmin(\Abar)$}}
\de\lamPRminSigbar{\tx{$\lamPRmin(\Sigbar)$}}
\de\lammuMCA{\tx{$\lammuMC(A)$}}
\de\lammuMCAbar{\tx{$\lammuMC(\Abar)$}}
\de\lammuPRA{\tx{$\lammuPR(A)$}}
\de\lammuPRAbar{\tx{$\lammuPR(\Abar)$}}
\de\lammuPRN{\tx{$\lammuPR(\NNN)$}}
\de\lammuPRNbar{\tx{$\lamPR(\NNNbar)$}}
\de\lammuPRSigbar{\tx{$\lammuPR(\Sigbar)$}}
\de\lammuPRminAbar{\tx{$\lammuPRmin(\Abar)$}}
\de\lammuPRminSigbar{\tx{$\lammuPRmin(\Sigbar)$}}
\de\lammuPRxAbar{\tx{$\lammuPRx(\Abar)$}}
\de\lamz{\lam\ztt}
\de\lastA{\tx{$\last^A$}}
\de\lastxA{\tx{$\last_\xtt^A$}}
\de\lsp{\tx{$\lss_p$}}
\de\lseqs{\tx{\sf lseq}}
\de\lseqp{\tx{$\lseqs_p$}}
\de\maxval{\tx{\bi maxval}}
\de\muPRAbar{\tx{$\muPR(\Abar)$}}
\de\muPRSigbar{\tx{$\muPR(\Sigbar)$}}
\de\muPRminAbar{\tx{$\muPRmin(\Abar)$}}
\de\muPRminSigbar{\tx{$\muPRmin(\Sigbar)$}}
\de\muPRoAbar{\tx{$\muPRo(\Abar)$}}
\de\muPRoSigbar{\tx{$\muPRo(\Sigbar)$}}
\de\muPRolamAbar{\tx{$\muPRolam(\Abar)$}}
\de\muPRolamSigbar{\tx{$\muPRolam(\Sigbar)$}}
\de\muPRxAbar{\tx{$\muPRx(\Abar)$}}
\de\muPRxApprox{\tx{\muPRx-\Approx}}
\de\muPRxApproxA{\tx{$\muPRxApprox(A)$}}
\de\muPRxApproxAutos{\tx{$\muPRxApproxA_\utos$}}
\de\muPRxSigbar{\tx{$\muPRx(\Sigbar)$}}
\de\natmin{\tx{$\nats^-$}}
\de\natsp{\tx{$\nats'$}}
\de\natsx{\tx{$\nats^*$}}
\de\natssmin{\tx{$\natss^-$}}
\de\natxutos{\tx{$\nats\times u\to s$}}
\de\natxutoss{{\tx{$\natss\times u\to s$}}}
\de\notoverxA{\tx{$\notover\,_\xtt^A$}}
\de\notoverxSA{\tx{$\notover\,_{\xtt,S}^A$}}
\de\notoveruaA{\tx{$\notover\biu\,_\att^A$}}
\de\ord{\tx{\bi ord}}
\de\ordusA{\tx{$\ord_{u,s}^A$}}
\de\orb#1{\tx{\bi orb\,$(#1)$}}
\de\peabcA{\tx{$\pe\,_\abct^A$}}
\de\peabcPA{\tx{$\pe\,_{\abct,P}^A$}}
\de\per#1{\tx{\bi per\,$(#1)$}}
\de\phipsi{\tx{$(\phi,\psi)$}}
\de\phihat{\tx{$\hat{\phi}$}}
\de\phis{{\tx{$\phi_s$}}}
\de\pib{\tx{$\pi_\btt$}}
\de\plustwo{\tx{\bi plus2}}
\de\posreals{\tx{\sf pos-real}}
\de\prodtt{\tx{\tt prod}}
\de\preds{\tx{\sf pred}}
\de\psis{{\tx{$\psi_s$}}}
\de\qinNN{\tx{$q\in\NN$}}
\de\redSigf{\,|\,_{\Sigf}}
\de\redSigp{\,|\,_{\Sigp}}
\de\remseqxA#1{\tx{$\rem\seq\,_\xtt^A(#1)$}}
\de\remxA{\tx{$\rem\,_\xtt^A$}}
\de\remxSA{\tx{$\rem\,_{\xtt,S}^A$}}
\de\remuaA{\tx{$\rem\biu\,_\att^A$}}
\de\restA{\tx{$\restt^A$}}
\de\restuaA{\tx{$\restt\biu\,_\att^A$}}
\de\restxA{\tx{$\restt\,_\xtt^A$}}
\de\restxSA{\tx{$\restt\,_{\xtt,S}^A$}}
\de\ruu{\tx{$r^\uu$}}
\de\sE{{\tx{\ssf E}}}
\de\saA#1{\tx{$\ang{#1}^A$}}
\de\siginStateA{\tx{$\sig\in\StateA$}}
\de\sigs{\tx{$\sig_s$}}
\de\sinS{\tx{$s \in \Sbi$}}
\de\sinSe{\tx{$s \in \Se$}}
\de\sexA{\tx{$\se\,_\xtt^A$}}
\de\sexSA{\tx{$\se\,_{\xtt,S}^A$}}
\de\sigxtt{\tx{$\sig[\xtt]$}}
\de\snapseqxA#1{\tx{$\snap\seq\,_\xtt^A(#1)$}}
\de\snapuaA{\tx{$\snap\biu\,_{\att}^A$}}
\de\snapxA{\tx{$\snap\,_\xtt^A$}}
\de\snapxSA{\tx{$\snap\,_{\xtt,S}^A$}}
\de\sprod{\tx{${s_1}\times\dots\times {s_n}$}}
\de\stateuaA{\tx{$\state\biu\,_\att^A$}}
\de\statexA{\tx{$\state\,_\xtt^A$}}
\de\statexSA{\tx{$\state\,_{\xtt,S}^A$}}
\de\subalg{\preceq}
\de\tAsig{\tx{$\tA\sig$}}
\de\teA{\tx{$\te^A$}}
\de\teaBA{\tx{$\te\,_{\att,\boolss}^A$}}
\de\teasA{\tx{$\te\,_{\att,s}^A$}}
\de\teauA{\tx{$\te\,_{\att,u}^A$}}
\de\teavA{\tx{$\te\,_{\att,v}^A$}}
\de\teawA{\tx{$\te\,_{\att,w}^A$}}
\de\texsA{\tx{$\te\,_{\xtt,s}^A$}}
\de\texsAx{\tx{$\te\,_{\xtt,s}^\Ax$}}
\de\texstA{\tx{$\te\,_{\xtt,s,t}^A$}}
\de\texuA{\tx{$\te\,_{\xtt,u}^A$}}
\de\texvA{\tx{$\te\,_{\xtt,v}^A$}}
\de\texwA{\tx{$\te\,_{\xtt,w}^A$}}
\de\texyxboolAx{\tx{$\te\,_{\xtt,\yttx,\boolss}^\Ax$}}
\de\tinTSig{\tx{$t\in\TSig$}}
\de\tinTSigx{\tx{$t\in\TSigx$}}
\de\ttil{\tx{$\til t$}}
\de\tree{\tx{\bi tree}}
\de\truncs{\tx{\sf trunc}}
\de\tupprimetimes#1#2#3{\tx{$#1_{#2}' \times \dots \times #1_{#3}'}}
\de\tuu{\tx{$t^\uu$}}
\de\uneqs{\tx{\sf uneq}}
\de\uneqp{\tx{$\uneqs_p$}}
\de\updates{\tx{\sf update}}
\de\us{{\tx{$u,s$}}}
\de\whicht{\tx{\tt which}}
\de\vN{\tx{$v^\natss$}} 
\de\valxaA{\tx{$\val_\xa^A$}}
\de\valxaAt{\tx{$\val_\xa^A(t)$}}
\de\ws{{\tx{$w(s)$}}}
\de\wxx{{\tx{$w^{**}$}}}
\de\xS{\cxtt,\,\cS}
\de\xSan{\xS,\,a,\,n}
\de\xa{{\ang{\xtt/a}}}
\de\xiinNNtoA{\tx{$\xi \in \NNtoA$}} 
\de\xixi{\xi_1,\xi_2}
\de\xttt{\tx{$\ang{\xtt/t}$}}
\de\zxx{\tx{$z^{**}$}}


\cbb{Abstract Computability and Algebraic Specifications}
\bigskip
\bigskip
\ce{{\bf J.V. Tucker}
}
\sn
\ce{\it Department of Computer Science,}
\ce{\it University of Wales, Swansea  SA2 8PP, Wales}
\ce{\tt J.V.Tucker@swansea.ac.uk}
\bn
\ce{{\bf J.I. Zucker}\footnotemark"*"}
\footnotetext"*\ "{
\n The research of the second author was supported 
by a grant from the Natural Sciences
and Engineering Research Council (Canada),
and by a Visiting Fellowship from the Engineering
and Physical Sciences Research Council (U.K.)
}

\ce{\it Department of Computing and Software,}
\ce{\it McMaster University, Hamilton, Ont\.  L8S 4L7, Canada}
\ce{\tt zucker@mcmaster.ca}
\bigskip
\bigskip
\bigskip
\bigskip
\bigskip
\cbb{Abstract}

\bn
{
\smallrm 
Abstract computable functions are defined by
abstract finite deterministic algorithms on many-sorted algebras.
We show that there exist finite universal algebraic specifications
that specify uniquely 
(up to isomorphism) 
\,($i$) all absract computable functions on any many-sorted algebra;
\,and
\,($ii$) all functions effectively approximable by 
abstract computable functions on any metric algebra.
We show that there exist universal algebraic specifications
for all the classically computable functions on the
set \,\RR\ \,of real numbers.
The algebraic specifications used are mainly bounded universal
equations and conditional equations.
We investigate the initial algebra semantics of
these specifications, and derive situations
where algebraic specifications define precisely the computable functions.

\mn
{\smallbf Categories and Subject Descriptors:}
F.1.1 ({\smallbf Computation by Abstract Devices}):
Models of Computation --- {\smallit computability theory\/};
F.4.1 ({\smallbf Mathematical Logic and Formal Languages}):
\,Mathematical Logic --- {\smallit computability theory; \,proof theory\/}

\mn
{\smallbf General Terms:}
\,Abstract Computability,
Algebraic Specification,
Computable Analysis,
Conditional Equations, Equational Logic, Metric Algebras

\mn
{\smallbf Additional Key Words and Phrases:}
\,Birkhoff's theorem, Grzegorczyk-Lacombe computability,
initial  algebras, term models,
many-sorted algebras, Mal'cev's theorem, topological algebras

}

\newpage

\Shead0{Introduction}
Abstract computability theory is the theory of computable
functions and relations over many-sorted algebras. It is a
generalisation of classical recursion theory on the natural
numbers, based on notions of finite deterministic
computation on an arbitrary many-sorted algebra. An
important feature of the theory is its analysis of
computations that are {\it uniform\/} over classes of algebras, and
a natural application of the theory is to analyse the scope and
limits of models of computation and specification over
abstract data types and their implementations. Since the 1960s, many
abstract models of computation have been defined and
classified, starting with the models of 
E. Engeler, Y. Moschovakis, H. Friedman and J.C. Shepherdson,
and generalised Church-Turing Theses for
computation and specification have been formulated and
defended \cite{tz:book,tz:jlp}.
Here we will use the model of computation \,\muPRx\
(a generalised form of Kleene schemes), which involves
{\it simultaneous primitive recursion\/} and {\it least number search\/}
over a many-sorted algebra augmented by the booleans,
natural numbers and finite sequences of every sort. In
\cite{tz:book} the model \,\muPRx\  is shown to be
equivalent to \qwhiles-array programs over these algebras, the
primary mathematical model of imperative programming.

Working with finite computation on any algebra enables us
to develop a number of special computability theories
for algebras, such as {\it rings and fields of real numbers\/} 
\cite{jvt:cas,eng93,bss,bcss-manifesto,bcss}
and {\it topological and metric algebras\/} \cite{tz:top}.
For a comprehensive
introduction to abstract computability, including a survey
of its origins in the 1950s and principal literature, see
our survey \cite{tz:hb}.

In this paper we prove theorems that show that functions
that are abstractly computable over many-sorted algebras,
or have abstractly computable {\it approximations\/} on topological
algebras, can be specified by purely algebraic methods,
but that the converse does not hold in the absence of
certain topological conditions.

Algebraic specification methods characterise functions as
the solutions of systems of algebraic formulae; 
normally, the solutions are
unique. By algebraic formulae, we mean
equations
$$
t(\xtt) \ = \ t'(\xtt)
$$
or conditional equations
$$
t_1(\xtt) = t_1'(\xtt) \,\con \,\dots \,\con \,t_k(\xtt) = t'_k(\xtt) 
\ \ \longto\ \ t(\xtt) = t'(\xtt),
\tagx
$$
or, more generally, conditional formulae 
$$
R_1\con \dots \con R_k \ \longto\ R
\tagxx
$$
where the formulae $R_i$ and $R$ are
generalisations of equations, 
making use of the distinguished sorts \nats\ of naturals and \reals\
of reals
(as we will see below).
To define a unique solution for a system of equations,  in
logic one often thinks of {\it definability up to isomorphism\/},
and in computing one often thinks in terms of {\it initial algebra
semantics\/} (or possibly {\it final algebra semantics\/}). However,
notice that there are many more equational methods, \eg,
for specifying concurrent processes using metric space
methods to solve equations \cite{db-rutten,db-devink}, or
for computing  solutions of differential or integral
equations.

In computation over a many-sorted algebra $A$ we use the
booleans, natural numbers and finite sequences over $A$. 
With regard to algebraic specifications over such structures,
generalising conditional equations
leads to the concept of
{\it conditional bounded universal (BU) equations\/},
in which the formulae $R_i$ and $R$ of ($**$)
may have the form 
$$
t_1 = t_2 \qquad \tx{or} \qquad 
\all \ztt<t \,[\,t_1= t_2\,]
$$
where the variable \ztt\ and term $t$ 
are of sort \nats.
 
Conditional BU equations are new and provide us
with more appropriate axiomatisations for some properties
using the natural number sort; we show they are equivalent
with conditional equations. The main theorems are first
proved for conditional BU equations and the reduction
method applied to obtain conditional equational
specifications.

In the first part of the paper, we begin with the ``simple"
situation where there is a system $E$ of conditional equations
over a signature \Sig, and a \Sig-algebra $A$ such that $E$ has one and only
one solution $f$ on $A$. We call this method of
characterising functions {\it conditional equation definability\/}
on $A$. We address the obvious general question: 

{\displaytext
Does
abstract computability imply conditional equation
definability?

}\n
The answer is yes, and we show that there exist 
{\it universal specifications\/}
that specify {\it all\/} computable functions, as follows
(Section 5, Theorem 4).

\Thmn{A \,(Algebra)}
Given a signature \Sig\ and function type \ta\ over \Sig,
there
exists a finite set of conditional equations \,\Ez\
\,(with a distinguished natural number variable \,\ztt)
over a finite expansion \Sigp\ of \Sig,
\,such that for any
abstract program \al\ over \Sig, if  $A$ is any  \Sig-algebra and
$f$  a total function on $A$ of type \ta\ computed by \al, then $f$ is
defined uniquely on $A$ by \,\Ekbar, \,where \kbar\ is
a numeral instantiating \,\ztt\ which is effectively calculable
from \al. The system \,\Ez\ \,is uniformly computable from \Sig\ and
\ta.
\endpr

Applying our abstract computability theory to {\it metric algebras\/},
we can obtain an important, strictly broader, class of functions:
namely, those {\it uniformly approximable by 
abstractly computable functions\/}.
In metric algebras, approximation is elegantly formulated 
in terms of the distance function,
which uses the sort \reals.
This gives rise to 
a broader class of conditional formulae than ($*$),
called
{\it conditional equations and inequalities\/},
namely formulae ($**$) 
in which the formulae $R_i$ and $R$ 
may have the form
$$
t_1 = t_2 \qquad \tx{or} \qquad 
t_1< t_2
$$
where, in the case of inequality ($t_1<t_2$),
$t_1$ and $t_2$ are of sort \reals.

From Theorem A we then prove (Section 6, Theorem 2):

\Thmn{B \,(Metric algebra)}
Given a signature \Sig\ and function type \ta\ over \Sig,
there exists a finite
set of conditional equations and inequalities 
\,\Ez\ \,(with a distinguished natural number variable \ztt)
over a finite expansion \Sigp\ of \Sig,
\,such that for any
abstract program \al\ over
\Sig, if  $A$ is any  metric \Sig-algebra and $f$  a total function
on $A$ of type \ta, approximable by \al\ in the following sense: 
for all
$a\in A$ and all \,$n$
$$
d(f(a), \,\bb{\al}(n, a)) < 2^{-n},
$$
then $f$ is
defined uniquely on $A$ by \,\Ekbar,
\, where \kbar\ is
a numeral instantiating \,\ztt\ which is effectively calculable
from \al. 
The system \,\Ez\ \,is uniformly computable from
\Sig\ and \ta.
\endpr

Thus, {\sl there is a bound $B(\Sig,\tau)$ on the number of 
conditional equations and inequalities
needed to define all computable or computably
approximable functions, that depends only on the signature \Sig\
and the function type \ta.}

Using Theorem B, we show that all the classically computable functions
of real analysis are unique solutions of finite sets of
conditional equations and inequalities.  
These classically computable functions have several characterisations,
starting with those of Grzegorczyk \cite{grzeg55,grzeg57} 
and Lacombe \cite{lacombe55},
and hence are often called {\it GL-computable\/}.
The familiar functions of analysis, such as 
\,$\sin x$, $e^x$, $\log x$, etc.,
are all GL-computable.
Now there exists a certain simple total metric algebra
\,\IIId\ \,over the real unit interval $I = [0,1]$, such that the 
total functions on $I$ which are uniformly approximable 
by abstractly computable functions on \,\IIId\
\,are precisely the GL-computable functions on $I$
\cite{tz:top,tz:hb}.
We prove the following (Section 6, Theorem 3).

\Thmn{C \,(Metric algebra over a real interval)}
For each positive integer $m$ 
there is a signature \Sigxm\ 
which is an expansion of the signature of \,\IIId\
\,by finitely many function symbols,
\,and a finite system of conditional equations and inequalities
\,\Emz\ 
\,(with a distinguished natural number variable \ztt)
\,over \Sigxm, 
\,such that any total function
\,$f\:[0,1]^m\to\RR$ 
\,that is GL-computable,
is the unique solution of \,\Ekbar\ \,for some substitution of
a numeral \,\kbar\ \,for \,\ztt.  
The specification \,$(\Sigxm,\,\Emz)$ \,is uniformly computable from $m$.
\endpr

Thus
{\sl there is a bound $B(m)$ on the
number of conditional equations and inequalities 
needed to define all $m$-ary GL-computable functions on $[0,1]$.\/}

The signature \Sigxm\ consists of
the sorts of booleans \BB\ and naturals \NN, with their standard operations;
the sort of reals \RR, with its ring operations,
together with division of reals by naturals;
the sort of the unit interval $I$,
with its embedding into \RR;
the sort of finite arrays on \RR\
with their standard operations;
the standard metrics on all these sorts;
a ``universal function" which approximably abstractly computes
all $m$-ary GL-computable total functions on $I$,
together with the auxiliary functions
used in its computation;
the function $2^{-n}$ used for expressing approximations;
and a function for computing bounded quantification over \NN.

This theorem has some interesting consequences,
one of which we illustrate (Section 6, Theorem 4):

\Cor
For each $n>0$,
there is a finite universal algebraic specification,
consisting of conditional equations and inequalities,
for all 
computable finite dimensional dynamical systems
on the unit $n$-cube and over the unit time interval.
\endpr

Next we consider the converse problem:

\Problem\sl
Find (reasonable) conditions under which
algebraic definability implies abstract computability.
\endpr

From Theorem C it follows that
the converse to Theorem A is false,
at least for specifications consisting of
conditional equations and inequalities;
for example, for the sine and cosine functions on the unit interval.

It is an open problem
whether the 
converse of the approximation
result (Theorems B and C) holds.
It seems that some extra {\it topological condition\/}
such as continuity is required for a converse result.
This suggests an
interesting research area;
see the example and discussion in Section 6.3.

In the second part of the paper, we show how the
conditional equational theories, and conditional BU equational
theories, can be used with standard
algebraic specification methods associated with proof
systems, term rewriting and initial algebra semantics.

Now, when using the booleans, natural numbers and finite
sequences, the algebraic specifications and their initial
algebra semantics must define the corresponding standard
models of the booleans, natural numbers and finite arrays. We
develop extensions of the Birkhoff-Mal'cev Completeness
Theorems that underlie the algebraic specification methods,
designed to ensure that these sorts have standard models.
Then we prove (Section 8, Theorems 4 and 5):

\Thmn{D \,(Initial algebra)}
Given a signature \Sig\ and function type \ta\
over \Sig, there exists a finite set of conditional equations
\,\Ez\ \,(with a distinguished natural number variable \ztt)
over a finite expansion \Sigp\ of \Sig,
such that for any abstract program \al\ over \Sig, if \al\
computes a total function $f$ on $A$ 
of type \,\ta, 
\,and $A$ has an initial
algebra specification by a set $E$ of either conditional
equations or conditional BU equations
(with hidden sorts and functions), 
then
\,\Af\ has an initial algebra specification by a set \,$E\cup \Ekbar$,
where \,\Ez\ \,consists of conditional equations and 
\,\kbar\ \,is a numeral instantiating \,\ztt\ \,which is effectively
calculable from \al.  
The system \,\Ez\ \,is uniformly computable
from \Sig\ and \ta.
Furthermore, if the specification $E$ of $A$ has $e$ axioms,
then the specification of \Af\ is finite, with $e + e'$
axioms, where $e'$ is a constant 
computed uniformly from \Sig\ and \ta.
\endpr

This paper is part of our series on abstract computability
theory on many-sorted algebras and its applications,
starting in  \cite{tz:book} and most recently surveyed in \cite{tz:hb}.
Knowledge of computation and our studies of computation
versus specifications 
\cite{tz:jlp,tz:ijfocs}
and verification  
\cite{tz:leeds}
will be helpful, but only our work on topological data types
\cite{tz:top}
is necessary.

The subject of this paper is also a generalisation of the
theory of algebraic specifications for {\it computable,
semicomputable and co-semicomputable} algebras developed by
one of us (JVT) with J.A. Bergstra: see
\cite{bt80a,bt80b,bt82a,bt83b,bt87,bt95}
and the surveys \cite{meseguer-goguen,stolt-jvt95}.
However, at least
initially, the generalised computability raises new
questions concerning topological data types, uniformity and
parameterisation, and standard models. Knowledge of the
theory for computable algebras is not required for this
paper.

In Section 1 we define how to augment structures with the
{\it standard sorts} of the booleans and naturals, and
finite sequences or {\it arrays} over all sorts, together
with the corresponding operations. For the rest of the
paper we consider, without loss of generality, only
N-standard signatures and structures with the booleans and
naturals.

In Section 2 we introduce a number of proof systems, all
based in the calculus of sequents over a many-sorted signature \Sig. These
are systems for ($i$) first order logic over \Sig\ with
equality, ($ii$) conditional equational logic, ($iii$)
conditional bounded universal (BU) equational logic, and
($iv$) conditional standard universal (SU) equational
logic. The systems ($ii$) and ($iii$) are subsystems of the
classical predicate calculus ($i$), and are used in the
following sections, while ($iv$) is an infinitary system
introduced for interest.

In Section 3 we define the basic technical notion of a
theory uniquely specifying a function on an arbitrary algebra with
hidden sorts and functions. This
leads to a simple notion of specifiable parameterisation
which we illustrate by showing how a conditional
equational  (or conditional BU equational) specification of
a standard structure $A$ can be extended to a similar
specification of the array structure \Ax. We also show how
to ``reduce" a conditional BU equational specification  over
\Sig\ to a conditional equational specification over an
expansion of \Sig.

In Section 4 we recall the basic notions of 
{\it computability} of functions, 
including universality of the \,\muPRx\ functions.

In Section 5 we prove Theorem A above,
concerning the conditional equational definability of
computable functions.

In Section 6 we prove Theorems B and C, 
concerning the definability, by conditional equations and inequalities,
of computably approximable functions on metric algebras.

In Section 7 we describe the construction of {\it initial
standard models} for conditional equational and conditional
BU equational theories, and work out the completeness
theorems for the corresponding proof systems in Section 2.
The reduction of a conditional BU equational specification
over \Sig\ to a conditional equational specification over
an expansion of \Sig\ is proved for initial models.

In Section  8  we investigate the relationship between {\it
computability} and {\it algebraic specifiability} of
functions on initial N-standard algebras, and prove Theorem D.
Finally, in Section 9, we consider the converse problem of
finding sufficient conditions for algebraic
specifiability to imply computability on classes of
standard structures. Two equivalence theorems are proved.

We wish to thank an anonymous referee for some
very helpful comments.

\Shead{1}{Many-sorted signatures and algebras}
In this section
we briefly review concepts defined and discussed in
\cite[\S1]{tz:hb},
where more detailed information can be found.
Background information on universal algebra can be found in
\cite{meinke-jvt,ehrig-mahr,wechler}.

\shead{1.1}{Basic definitions}
A {\it signature} \Sig\ (for a many-sorted algebra)
is a pair consisting of
($i$) a finite set \SortSig\ of {\it sorts},
and ($ii$) a finite set \FuncSig\ of {\it (primitive)
function symbols},
each symbol $F$ having a {\it type}  \ $\tuptimes{s}{1}{m}\to s$,
where $s_1 ,\ldots, s_m, s \in \SortSig$;
in that case we write
\ $F:\ \tuptimes{s}{1}{m}\to s$,
with \,$\dom{F} \eqdf \tuptimes{s}1m$.
\ (The case $m = 0$ corresponds to {\it constant symbols}.)

A 
\Sig-{\it product type\/}
has the form
\,$u = \tuptimes{s}{1}{m}$ \,($m\ge0$),
where \,\tup{s}{1}{m} \,are \Sig-sorts.
We use the notation
\,$u,v,w,\dots$ \,for \Sig-product types.

A \Sig-{\it algebra} $A$ has, for each sort $s$ of \Sig,
a non-empty {\it carrier set} \,\As\
\,of sort $s$,
and for each \Sig-function symbol
\,$F:\utos$,
\,a function
\,$\FA : \Au \to \As$
\,(where, for the \Sig-product type \ $u  = \tuptimes{s}{1}{m}$,
\,we write
\,$\Au \ \eqdf \ A_{s_1} \times\dots\times A_{s_m}$).

Given an algebra $A$, we sometimes write \,$\Sig(A)$ \,for its signature.

The algebra $A$ is {\it total} if 
\,\FA\ is total for each \Sig-function symbol $F$.
Without such a totality assumption,
$A$ is called {\it partial}.

In this paper we deal with total algebras,
except in \S8.4.

We will also consider classes \KK\ of \Sig-algebras.
A {\it \Sig-adt (abstract data type\/})
\,is defined to be any such class, closed under \Sig-isomorphism.
In particular, \AlgSig\ denotes the class of all \Sig-algebras.

\Examples
($a$)
\,The algebra of {\it booleans\/}
has the carrier \ $\BB = \{\ttt,\,\fff\}$
\ of sort \,\bools. It can be displayed as follows:
$$
\boxed{
\matrix \format\l&\quad\l\\
\algebras &\BBB\\
\carrierss &\BB\\
\functionss&\ttt, \fff:\ \ \to\BB,\\
&\ands^\BBB, \ors^\BBB:\BB^2\to\BB\\
&\nots^\BBB: \BB\to\BB\\
\ends&
\endmatrix
}
\quad
\tx{with signature}
\quad
\boxed{
\matrix \format\l&\quad\l\\
\signatures &\Sig(\BBB)\\
\sortss &\bools\\
\functionss&\trues, \falses:\ \ \to\bools,\\
&\ands, \ors:\bools^2\to\bools\\
&\nots: \bools\to\bools\\
\ends&
\endmatrix
}
$$
For notational simplicity, we will usually not distinguish between
function names in the signature (\trues, etc.)
and their intended interpretations ($\trues^\BBB = \ttt$, etc.)
\sn
($b$) \,The algebra \,\NNNo\ \,of naturals
has a carrier \,\NN\ \,of sort \,\nats,
\,together with the zero constant and successor function:
$$
\boxed{
\matrix \format\l&\quad\l\\
\algebras &\NNNo \\
\carrierss &\NN\\
\functionss&0:\ \ \to\NN,\\
&\Ss:\NN\to\NN\\
\ends&
\endmatrix
}
$$

($c$) \,The ring \,\RRRo\ \,of reals
has a carrier \RR\ of sort \reals:
$$
\boxed{
\matrix \format\l&\quad\l\\
\algebras &\RRRo \\
\carrierss &\RR\\
\functionss&0,1:\ \ \to\RR,\\
&+,\times:\RR^2\to\RR,\\
&-:\RR\to\RR\\
\ends&
\endmatrix
}
$$
\endpr

We make the following assumption about 
the signatures \Sig.

\pr{Instantiation Assumption}
{\sl For every sort $s$ of \Sig, there is a closed term of that sort,
called the default term \,\delbs\ \,of that sort.}
\endpr

This guarantees the presence of {\it default values} \,\delbsA\ \, 
in a \Sig-algebra $A$ at all sorts $s$, 
and {\it default tuples} \,\delbuA\ \,at all product types $u$.

\sheads{1.2}{Some definitions}
\Defn{1 \,(Subalgebra)}
Given \Sig-algebras $A$ and $B$,
we say that
$B$ is a \Sig-{\it subalgebra\/} of $A$
\,(written $B\subalg A$)
\,iff \,($i$) for all \Sig-sorts $s$, \,$B_s\sseq A_s$, 
\,and ($ii$) for every \Sig-function symbol $F$,
\,$F^B = \FA\rest B$.
\endpr

\Defn{2 \,(Expansions and reducts)}
Let \Sig\ and \Sigp\ be signatures with $\Sig \subset \Sig'$.
\sn
\ ($a$) If $A'$ is a \Sigp-algebra, \,then the \Sig-{\it reduct of} $A'$,
\ $A' \redSig$,
\ is the algebra
of signature \Sig, consisting of the carriers of $A'$
named by the sorts of \Sig\ and equipped with the functions
of $A'$ named by the function symbols of \Sig.
\sn
\ ($b$) If \,$A$ is a \,\Sig-algebra and $A'$ is a \Sigp-algebra,
\,then $A'$ is a {\it \Sigp-expansion} of $A$ iff $A$ is the
\Sig-reduct of $A'$.
\sn
($c$) If \,\KKp\ \,is a \,\Sigp-adt,
\,then \,$\KKp\redSig$ \,is the class of \Sig-reducts of algebras in \KKp.
\endpr

\shead{1.3}{Adding booleans: \,Standard signatures and algebras}
Recall the algebra \BBB\ of booleans (Example ($a$) in \S1.1).

A signature \Sig\ is called {\it standard} if
\,($i$) \,$\Sig(\BBB) \ \sseq \Sig$;
\,($ii$) \,the \Sig-function symbols include
a {\it conditional}
$$
\ifs_s :\bools\times s^2 \to s
$$
for all sorts $s$ of \Sig\ other than \bools; \,and
\,($iii$) \,the \Sig-function symbols include
an {\it equality  operation\/}
$$
\eqs_s: s^2 \to \bools
$$
for all $s\in\EqSortSig$, 
\,where \,$\EqSortSig\sseq\SortSig$
\,is the set of \Sig-{\it equality sorts\/}.

Given a standard signature \Sig,
a \Sig-algebra $A$ is {\it standard\/} if
\,($i$) \,it is an expansion of \BBB;
\,($ii$) \,the conditionals 
have their standard interpretation in $A$,
\ie, for $b\in\BB$ and $x,y \in \As$,
$$
\ifs_s(b,x,y) \ = \
\cases
   x \tif{$b = \ttt$}\\
   y \tif{$b = \fff$};
\endcases
$$
and \,($iii$)
\,the equality operator $\eqs_s$ is interpreted as {\it identity} on each
\Sig-equality sort $s$.

Note that any many-sorted signature \Sig\
can be {\it standardised} to a signature \,\SigBBB\
\,by adjoining the sort \,\bools\ \,together
with the standard boolean operations;
and, correspondingly, any algebra $A$ can be standardised
to an algebra \ABBB\
\,by adjoining the algebra \BBB\
and the conditional \,$\ifs_s$ \,at all \Sig-sorts $s$,
and the equality operator \,$\eqs_s$
\,at the specified equality sorts:
$$
\boxed{
\matrix \format\l&\quad\l&\quad\l\\
\algebras &\ABBB &\\
\imports &A,\,\BBB&\\
&\ifs_s:\BB\times A_s^2\to\As &(\sinSortSig),\\
&\eqs_s: A_s^2\to\BB &(s\in\EqSortSig)\\
\ends&&
\endmatrix
}
$$
Thus the standardisation of a \Sig-algebra $A$ depends on 
the specification of \,\EqSortSig.
These will be the sorts
for which an equality test is considered to be ``computable"
in some sense.

\Examples
($a$) \,The simplest standard algebra is the algebra \BBB\ of the booleans.
\smskipn
($b$) \,The standard algebra of naturals \NNN\
is formed by standardising the algebra \NNNo\ 
(Example ($b$) in \S1.1)
with \,\nats\ \,as an equality sort,
and, further, adjoining the order relation
\lsnat\ as a boolean-valued operation on \NN:
$$
\boxed{
\matrix \format\l&\quad\l\\
\algebras &\NNN \\
\imports &\NNNo, \,\BBB\\
\functionss
&\ifnat:\BB\times\NN^2\to\NN,\\
&\eqnat,\,\lsnat:\NN^2\to\BB\\
\ends&
\endmatrix
}
$$
($c$) \,The standard algebra \,\RRR\ \,of reals
is formed similarly by standardising the ring \RRRo\
(Example ($c$) in \S1.1), with \,\reals\ \,{\it not\/} an equality sort.
In fact, neither the equality nor the order relation on \RR\
is included as an operation 
on \reals. 
(The significance of this is discussed later; \cf\ Remark 3 in \S5.3.)

\StdAlgSig\ \,denotes the class of all standard \Sig-algebras.

\shead{1.4}{Adding counters: \ N-standard signatures and algebras}
A standard signature \Sig\ is called {\it N-standard\/}
if it includes (as well as \bools)
the {\it numerical sort\/} \nats,
and also function symbols for the
{\it standard operations\/} of {\it zero\/}
and {\it successor\/}, 
as well as
the {\it conditional\/} and {\it equality\/} and {\it order\/}
on the naturals:
$$
\align
0\: \ \ &\to\nats\\
\Ss\:\nats&\to\nats\\
\ifnat\:\bools\times\nats&\to\nats\\
\eqnat\:\nats^2&\to\bools\\
\lsnat\:\nats^2&\to\bools.
\endalign
$$
The corresponding \Sig-algebra $A$ is {\it N-standard} if the carrier
$A_\natss$ is the set of natural numbers
\ \NN = \{0,1,2,\dots\},
and the standard operations (listed above) have their
{\it standard interpretations\/} on \NN.
\endpr

Note that any standard signature \Sig\ can be
N-{\it standardised} to a signature \SigN\ 
by adjoining the sort \nats\ and the operations 0, \Ss,
\eqnat, \lsnat\ and \ifnat.
Correspondingly,
any standard \Sig-algebra $A$ can be N-{\it standardised\/} 
to an algebra \AN\ 
by adjoining the carrier \NN\
together with the corresponding standard functions:
$$
\boxed{
\matrix \format\l&\quad\l\\
\algebras &\AN \\
\imports &A,\,\NNN\\
\ends&
\endmatrix
}
$$

\Examples
($a$) \,The simplest N-standard algebra is \NNN\ (Example ($b$) in \S1.3).
\sn
($b$) \,The N-standard algebra \,\RRRN\ \,of reals is
formed by N-standardising the standard real algebra \,\RRR\
\,(Example ($c$) in \S1.3).
\endpr

\NStdAlgSig\ \,denotes the class of all N-standard \Sig-algebras.

\pr{N-standardness Assumption}\sl
We will assume throughout this paper
that the signatures and algebras are N-standard,
except where stated otherwise.
\endpr

\mn
We also consider a notion stricter than N-standardness.

\shead{1.5}{Strictly N-standard signatures and algebras}
An N-standard signature \Sig\ is
{\it strictly N-standard} 
if the {\it only operations of \Sig\/}
with range sort \nats\ or \bools\
are the {\it standard numerical operations\/}
\,$0,\,\Ss, \,\ifnat\,\eqnat, \,\lsnat$ (\S1.4)
\,and the {\it boolean operations\/}
\,$\trues,\,\falses, \,\ands,\,\ors,\,\nots$
\,(\S1.1).
An algebra is {\it strictly N-standard} if its signature is.

\Remarks
\itemm{(1)}
Any N-standardised signature and algebra
are automatically strictly N-standard.
\itemm{(2)}
A strictly N-standard signature
has no equality sorts other than \nats.
\itemm{(3)}
Any subterm of a term of sort \,\nats\ \,or \,\bools\
of a strictly standard signature is itself of sort \,\nats\ \,or \,\bools.
(Proved by structural induction on the term.)
\endpr

The notion of strict N-standardness will be used in Section 9.

\shead{1.6}{Adding arrays: \ Algebras \Ax\ of signature \Sigx}
The significance of arrays for computation
is that they provide
{\it finite but unbounded memory}.

Given a standard signature \Sig,
and standard \Sig-algebra $A$, we expand \Sig\ and $A$
in two stages:

\nin($1^\circ$) N-standardise these to form \SigN\ and \AN, as in \S1.3.

\nin($2^\circ$)
Define, for each sort $s$ of \Sig, the carrier \Asx\
to be the set of {\it finite sequences\/} or {\it arrays\/} \ax\
over \As, of ``starred sort" \sx.

The reason for introducing starred sorts
is the lack of effective coding
of finite sequences within abstract algebras in general.
(Note that, for simplicity, our definition excludes a starred sort \,\natsx,
\,which would be redundant.)

The resulting algebras \Ax\ have signature \Sigx,
which expands \SigN\
by including,
for each sort $s$ of \Sig, the new starred sort \sx,
and also the following new function symbols:

\nin($i$) the operator 
\ $\Lgths_s: \,s^*\to \nats$, \ 
where \,$\Lgths(\ax)$ \,is the length of the array \ax;

\smskipn
($ii$)  the application operator \ $\Aps_s: \,\sx\times\nats \to s$,
\ where
$$
\Aps_s^A  (\ax, k ) \
\ \cases
     \ax  [ k ] \tif{$k<\Lgths(\ax)$}\\
     \delbs \ow
  \endcases
$$
where \,\delbs\ \,is the default value at sort $s$ 
guaranteed by the Instantiation Assumption (\S1.1)\fn
{We assume that $\ax[k]$ is undefined for $k\ge\tx{\ssf Lgth}(\ax)$.};

\smskipn
($iii$)  the null array \ $\Nulls_s: \sx$ \ of zero length;

\smskipn
($iv$)  the operator \ $\Updates_s: \,\sx\times\nats\times s \to s^*$, 
\ where
\ $\Updates_s^A  (\ax,n ,x )$ \ is the array \ $\bx\in\Asx$  
of length $\Lgths(\bx) = \Lgths(\ax)$, \,such that
for all $k< \Lgths(\ax)$
$$
   \bx  [k] \ = \
      \cases
         \ax [k]        \tif{$k \neq n$} \\
            x           \tif{$k = n$}
      \endcases
$$
($v$)  the operator \ $\Newlengths_s: \ \sx\times\nats \to s^*$,
\ where  \ $\Newlengths_s^A   (\ax, m )$  \ is the array \bx\ of length $m$ 
such that for all $k<m$,
$$
\bx  [k] \ = \
    \cases
       \ax [k]        \ift{$k < \Lgths(\ax)$} \\
       \delbs         \ift{$\Lgths(\ax) \le k <m$}
    \endcases
$$
($vi$) the {\it conditional\/} on \Asx\
for each sort $s$; \ and
\smskipn
($vii$) the {\it equality} operator on \Asx\ for each equality sort $s$.

\nin
Note that
\Ax\ is an N-standard \Sigx-{\it expansion} of $A$.

The justification for ($vii$) is that if a sort $s$ has
``computable" equality, then clearly so has the sort \sx,
since it amounts to testing equality of finitely
many pairs of objects of sort $s$, up to a computable length.

\Shead2{Proof systems and theories for \Sig-algebras}
To reason about computations,
we choose a first-order language based on \Sig\ as a specification
language.

Note, in this connection, that the {\it operations\/} in \Sig\
are used for  computation.
In particular, boolean-valued operations
are used for tests in computations.
By contrast, for specification and reasoning about these algebras,
we may add {\it predicates\/} to the language, which are
not, in general, computable or testable.
For example, our specification language 
will include the equality predicate
at all sorts (as we will see),
whereas only the {\it equality sorts\/} 
$s$ have the ``computable" equality operator \,$\eqs_s$
(\S1.3).
In writing specifications on the reals we may also add the `$<$' predicate
(again, not computable, at least if defined totally),
as we will do later (\S5.3) for the specification of
approximable computability.  Note that 
these predicates added to the language
do not form part of the signature.
Intuitively, think of the equality operation as
a ``computable" boolean test, but the equality predicate
as a ``provable" assertion of equality between two terms.
\endpr

So let $\LangSig$ be the first order language
over the signature \Sig, with the {\it equality predicate\/} at all sorts.
The syntax of \LangSig\
is generated as follows.
For each \Sig-sort $s$
there are countably many {\it variables\/}
of sort $s$, denoted 
\,$\att,\,\btt, \,\dots, \,\xtt,\,\ytt,\,\dots$\,.
Next, for each \Sig-sort $s$,
there are {\it terms\/} of sort $s$,
generated from variables and the function
symbols of \Sig\ according to
the standard typing rules.
We write \,$t^s$ \,or \,$t:s$ \,if $t$ is a term of sort $s$,
and, for a product type $u = \tuptimes{s}1m$,
we write \,$t:u$ \,if $t$ is a $u$-tuple of terms,
\ie, an $m$-tuple of terms of type \,\tuptimes{s}1m.

The {\it atomic formulae} of \,\LangSig, \,then, are
equations 
\,($t_1^s = t_2^s$)
\,between terms of sort $s$,
for all \Sig-sorts $s$ (whether equality sorts or not),
and the propositional constants \,\trues\ \,and \,\falses.
{\it Formulae\/} of \LangSig\ are built up from these
by the logical connectives
\,$\land,\;\lor,\;\to,\;\neg$,
\,and the quantifiers \,$\all_s$ and $\ex_s$\,
\ for all sorts $s$ of \Sig.

\endpr

We will consider (in the following four subsections)
four formal systems in \LangSig, 
conveniently formulated as {\it sequent calculi\/}.
The first is our basic system \FOLSig,
full first order logic with equality over \Sig.
The next two are subsystems of this,
which will be used in Section 7.
The final system is a subsystem of \FOLSig,
extended by an infinitary proof rule.

Background information on sequent calculus proof systems
can be found in \cite{tak}.

Note that we do {\it not\/} assume (N-)standardness of \Sig\ 
in subsections 2.1 and 2.2 (only) below.

\shead{2.1}{\FOLSig: \ Full first order logic with equality over \Sig}
This can be formulated in a system
\,\LKeSig,
\,which is an adaptation to the many-sorted signature \Sig\
of the systems \LK\ and \LKe\ of \cite{gentzen,tak}.
The atomic formulae are equations at all \Sig-sorts.

A {\it sequent} of \LKeSig\ is a construct of the form \ \GamDel,
\ where \Gam\ and \Del\ are each finite sequences of formulae of \LangSig.

{\it Derivations\/} (of sequents) are then constructed
from certain specified {\it initial sequents\/} (``axioms")
by means of specified {\it inference rules\/}.

The system \LKe\ can be augmented in two ways:
\sn
($a$) Adding {\it axioms of a theory},
or rather all substitution instances of these,
as initial sequents;
\sn
($b$) Adding {\it induction\/} for a class \CCC\ of formulae
(in case \Sig\ is N-standard),
in the form of the inference rule
$$
\CCCIndSig:
\quad{
\GamDel, F(0)\qquad F(\att), \Pi \seqt \Lam, F(\Ss \att)
\over
\Gam,\Pi \seqt \Del,\Lam, F(t)}
$$
where the induction variable \att\ has sort \nats,
and the induction formula $F(\att)$ belongs to the class \CCC.
We write \,\IndSig\ \,for full \Sig-induction,
\ie, where \CCC\ is the set of all first-order \Sig-formulae.

We will also be interested in the ``intuitionistic"
version \,\CCCIndi\ \,of \,\CCCInd,
in which the sequences \Del\ and \Lam\ above are empty.

Analogous augmentations can be made for the other systems considered below.

In the next three subsections we will consider 
three further systems, the first two of which are subsystems of \FOLSig\
and the third of which is a subsystem of \FOLSig\
augmented by an infinitary \om-rule.
These subsystems are,
in fact, also subsystems of \LJeSig,
which is an adaptation to \Sig\ of the ``intuitionistic" system \LJe\
(\loccit),
in which the sequents have only one formula on the rhs.
(When we are working with these subsystems,
the scheme \,\CCCInd\ \,will 
consist of intuitionistic sequents, so that
the sequences \Del\ and \Lam\ above are empty.)

\shead{2.2}{\CondEqSig: \ conditional equational logic over \Sig}
A {\it conditional equation} is a formula of the form
$$
P_1 \con \dots \con P_n \to P
\tagx
$$
where $n \ge 0$ and $P_i$ and $P$ are equations.
A {\it conditional equational theory}
is a set of such formulae (or their universal closures).
An {\it equational sequent} is a sequent of the form
$$
P_1, \,\dots, \,P_n \ \seqt \ P
$$
where $n \ge 0$ and $P_i$ and $P$ are equations.
This sequent {\it corresponds to} the conditional equation ($*$).

The {\it initial sequents\/} are all substitution instances of 
the \Sig-equality axioms
(expressing that equality is a congruence relation with respect to \Sig),
and the {\it inferences\/} are
{\it structural inferences\/},
{\it atomic cuts\/}
and 
{\it substitution\/} of terms for free variables in sequents.

\shead{2.3}{\CondBUEqSig: \ Conditional BU equational logic over \Sig}
A {\it BU (bounded universal) quantifier\/} is a quantifier
of the form `$\all \ztt < t$',
where \,$\ztt:\nats$ \,and \,$t:\nats$.
(The most elegant approach is to think of this 
as a primitive construct, with its own
introduction rule: see below.)
A (\Sig-){\it BU equation} is formed by prefixing
an equation by a string of 0 or more
bounded universal quantifiers.
A {\it conditional BU equation} is a formula of the form
$$
Q_1 \con \dots \con Q_n \to Q
\tagxx
$$
\ where $n \ge 0$ and $Q_i$ and $Q$ are BU equations.
A {\it conditional BU equational theory}
is a set of such formulae (or their universal closures).
A {\it BU equational sequent} is a sequent of the form
$$
Q_1, \,\dots, \,Q_n \ \seqt \ Q
$$
where $n \ge 0$ and $Q_i$ and $Q$ are BU equations.
This sequent {\it corresponds to} the conditional BU equation
($**$).

The system \,\CondBUEqSig\
consists of {\it BU equational sequents\/}.
The initial sequents are the \Sig-equality axioms,
as before, plus the {\it boundedness axioms}
\TOL
$$
P(0),\dots,P(\ol{n - 1}) \ \seqt \ \all \ztt < \kbar P(\ztt)
\tag"\BddAxSig:"
$$
\TOR
for all \Sig-equations $P$
and all \,\ninNN,
\,where \nbar\ is the {\it numeral\/} for $n$,
\ie, the term \,$\Ss\dots\Ss 0$ \,($n$ times `\Ss').
The only inferences are {\it structural inferences},
{\it cut\/}, {\it substitution\/},
and the rules for the BU quantifiers:
$$
\all_b L :
\ {{\Gam \seqt s < t \qquad Q(s),\Del \seqt Q}
  \over
  {\all \ztt<t Q(\ztt), \Gam, \Del \seqt Q}}
\qqquad
\all_b R:
\ {{\att<t, \Gam \seqt Q(\att)}
  \over
  {\Gam \seqt \all \ztt<t Q(\ztt)}}
$$
where $s$ and $t$ are terms of sort \nats,
\ `$s<t$' \ stands for \ `$\lsnat(s,t) = \trues$',
\ and the variable \,$\att:\nats$
\,is the `{\it eigenvariable}' of the inference $\all_b R$,
which does not occur in the conclusion of that inference.

\Remarkn{\,(Boundedness axioms)}
The boundedness axioms \,\BddAxSig\ \,hold (of course)
in N-standard models.
We remark here that they are derivable in \,\FOLSig\ 
\,from the {\it N-standardness axioms\/}
\,\NStdAxoSig\
(a set of conditional equations
defined in \S7.2),
plus the single formula
$$
\ \ztt_1<\Ss \ztt_2 \ \longimp\  \ztt_1 < \ztt_2 \,\dis \,\ztt_1 = \ztt_2
$$
which is, however, {\it not a conditional
BU equation\/}.
This formula
is derivable, in turn, in
\,$\FOLSig +\QF\tx{-}\IndSig$
\,(induction for quantifier-free formulae),
from \,\NStdAxoSig.
It is not clear whether the boundedness axioms
are derivable in conditional BU equational logic alone
from \,\NStdAxoSig,
\,which is why we are adding them as axioms.
\endpr

\shead{2.4}{\CondSUEqomSig: \ Conditional SU equational logic over \Sig}
The final two systems that interest us,
in this and the next subsection, 
are not subsystems of \LKe,
but {\it infinitary} systems.
They will be used
for another illustration of a Malcev-type theorem
for N-standard algebras
(see Section 6, Theorem 4).
However they will not be used in the investigation
of the relationship between computability 
and algebraic specifiability in Section 8.

A (\Sig-){\it SU (standard universal) equation\/} is
formed by prefixing
an equation by a string of 0 or more
universal quantifiers of sort \nats.
A {\it conditional SU equation\/} is a formula of the form
$$
R_1 \con \dots \con R_n \to R
\tagxxx
$$
where $n \ge 0$ and $R_i$ and $R$ are SU equations.
A {\it conditional SU equational theory\/}
is a set of such formulae (or their universal closures).
An {\it SU equational sequent\/} is a sequent of the form
$$
R_1, \,\dots, \,R_n \ \seqt \ R
$$
where $n \ge 0$ and $R_i$ and $R$ are SU equations.
This sequent {\it corresponds to} the conditional SU equation
(\xxx).

The system \,\CondSUEqomSig\
contains SU equational sequents.
It contains the equality axioms and the following inferences:
the structural inferences, cut, and the following
rules for the universal number quantifier
(where \,$t:\nats$):
$$
\all L :
\ {{R(t),\Gam \seqt R}
  \over
  {\all \ztt R(\ztt), \Gam \seqt R}}
\qqquad
\allomR:
\ {{\dots \ \Gam \seqt R(\bar n) \ \dots \quad (\tx{all} \ n \in \NN)}
  \over
  {\Gam \seqt \all \ztt R(\ztt)}}
\tagx
$$
Note that the rule \,\allomR\ \,is actually an {\it infinitary \om-rule}.

\shead{2.5}{\FOLomSig: \ full first-order logic with equality 
and an \om-rule over \Sig}
This modifies the system \,\FOLSig\ \,(\S2.1)
by replacing the usual universal number quantifier rule \,\allR\
\,by the infinitary rule \,\allomR\ \,(\S2.4),
with also the corresponding rule \,\exomL\
\,dually.
We omit details, except to point out that
\,$\FOL + \IndSig$
can easily be interpreted in it.

We write \,\EqSig, \,\BUEqSig\ \,and \,\SUEqSig\
\,for the classes of equations, BU equations and SU equations (respectively)
over \Sig.

\shead{2.6}{Conservativity lemmas}
One reason for the importance of
(finite or infinite) conditional equational logic lies in the following
lemmas.
First we need a definition
which will be 
given again in context in Section 7.
Let \,\FFF\ \,be a formal system
(typically \,\CondEqSig\ \,or \,\CondEqomSig),
and let $T$ be a theory over \Sig\
(typically a conditional equational or
\om-conditional equational theory). 
We say that
$T$ {\it determines \nats} in \FFF\
if every closed term of sort \nats\ is, provably in \,\FFF\ \,from $T$,
equal to a numeral.

\nin
(1) \,(\FOL\ \,over \,\CondEq.)
\ Let $E$ be a \Sig-conditional equational theory,
\,and let \,\GamP\ \,be a \Sig-equational sequent.
Then \,\GamP\ \,is provable from $E$ in \,\FOLSig\
\,if, and only if, it is provable from $E$ in \,\CondEqSig.
\endpr

\nin
(2) \,($\FOL+\Ind$ \,over \,\CondEq.)
\ Let $E$ be a \Sig-conditional equational theory
\,which determines \,\nats\ \,in \,\CondEqSig,
\,and let \,\GamP\ \,be a {\it closed\/} \Sig-equational sequent.
Then \,\GamP\ \,is provable from $E$ in \,$\FOLSig+\IndSig$
\,if, and only if, it is provable from $E$ in \,\CondEqSig.
\endpr

\nin
(3) \,($\FOL+\Ind$ \,over \,\CondBUEq.)
\ Let $F$ be a \Sig-conditional BU equational theory
\,which determines \,\nats\ \,in \,\CondBUEqSig,
\,and let \,\GamQ\ \,be a {\it closed\/} \Sig-BU equational sequent.
Then \,\GamQ\ \,is provable from $F$ in \,$\FOLSig+\IndSig$
\,if, and only if, it is provable from $F$ in \,\CondBUEqSig.
\endpr

\nin
(4) \,(\FOLom\ \,over \,\CondSUEqom.)
\ Let $G$ be a \Sig-conditional SU equational theory over \Sig\
\,which determines \,\nats\ \,in \,\CondSUEqomSig,
\,and let \,\GamR\ \,be a {\it closed\/} \Sig-conditional SU
equational sequent.
Then \,\GamR\ \,is provable from $G$ in \,\FOLomSig\
\,if, and only if, it is provable from $G$ in \,\CondSUEqomSig.

All four lemmas can be proved
by cut elimination.
We omit proofs, except to note briefly that the
two conditions, that $E$ determines \,\nats\ \,
and that \,\GamP\ \,is closed,
are used in (2) and (3) to eliminate induction inferences,
and in (4) to eliminate cuts of formulae universally or existentially
quantified over \,\nats.

\Remarks
(1) \,These conservativity lemmas (at least for simple equations)
also follow from the Birkhoff-Mal'cev-type completeness theorems 1--4
in Section 7.
\sn
(2) \,Infinitary systems come into their own
when reasoning about infinite objects such as 
infinite streams of data.
Some applications in this direction,
using a related infinitary system (\CondEqom),
are given in \cite{tz:fef}.
\endpr

\Shead3{Specifiability of functions by theories}
\sshead{3.1}{Specifiability over algebras and over classes of algebras}
Recall from Section 2 that \,\LangSig\ \,is the first order 
language over \Sig, with equality as the only predicate at all sorts.

A \Sig-{\it theory} is just a set $T$ of formulae in \LangSig.
The {\it axioms} of $T$ are the formulae in $T$.
We will be particularly interested in theories $T$
satisfying certain syntactic conditions;
for example, $T$ might be a set of conditional equations.
This is considered more carefully in Section 7.

We are also interested (when \,\Sig\ \,is N-standard)
in classes \KK\
of the {\it N-standard models} of such \Sig-theories:
\,$\KK = \NStdAlgSigT \subseteq \NStdAlgSig$.
In this case we say also that \,\SigT\ \,is an
(N-standard) {\it specification} for 
the adt \,\KK.

{\it Assume, for the rest of this section, that \,\Sig,
\Sigp\ and \Sigpp\ are N-standard signatures with 
$\Sig \subset \Sig' \subset \Sig''$.
Also, $A$ is an N-standard \Sig-algebra and 
$A'$ is an N-standard \Sigp-algebra.
Also, $T$ is a \Sig-theory,
$T'$ is a \Sigp-theory and $T''$ is a \Sigpp-theory.}

Note that
any expansion of a standard algebra
is also standard, and
any expansion of an N-standard algebra
is also N-standard.

\Defn1
Let $A_1'$ and $A_2'$ be two \Sigp-algebras with
\,$A_1'\redSig = A_2'\redSig$.
Then $A_1'$ and $A_2'$ are 
{\it \Sigp/\Sig-isomorphic},
written 
\,$A_1' \cong_{\Sigp/\Sig} A_2'$,
\ if there is a \Sigp-isomorphism from $A_1'$ to $A_2'$
whose restriction to \Sig\ is the identity on $A_1'\redSig$.
\endpr

\Defn2
Suppose $A'$ is a \Sigp-expansion of $A$.
We say that \,\SigpTp\ \,{\it specifies $A'$ over $A$\/} \,iff 
$A'$ is the unique (up to \Sigp/\Sig-isomorphism) 
\Sigp-expansion of $A$
satisfying $T'$;
\ in other words:
\smskipn 
($i$) $A' \ttstile T'$; \ and 
\smskipn 
($ii$) for all \Sigp-expansions $B'$ of $A$,
\ if \,$B' \ttstile T'$ \, then \,$B' \cong_{\Sigp/\Sig} A'$.
\endpr

We will occasionally write: \,``$T'$ specifies $A'$ over $A$"
\,instead of \,``\SigpTp\ \,specifies $A'$ over $A$".

An important special case of Definition 2 is the following.

\Defn{${\bk 2}^{\bk f}$}
Suppose \,$\Sig' = \Sig\cup\curly{\fs}$.
We say that \,$\SigpTp$ \,{\it specifies $f$ over $A$\/} \,iff
$f$ is the unique 
(up to \Sigp/\Sig-isomorphism) 
function on $A$ (of the type of \fs\,)
such that \,$\Af\ttstile T'$.
\endpr

\Defn3
Suppose $A'$ is a \Sigp-expansion of $A$.
We say that \,\SigppTpp\ \,{\it specifies $A'$ over $A$} 
{\it with  hidden sorts and/or functions} 
\,iff $A'$ is the unique (up to \Sigp/\Sig-isomorphism) 
\Sigp-expansion of $A$
such that some \Sigpp-expansion of $A'$ satisfies $T''$;
in other words:
\itemm{($i$)}
$A'$ is a \Sigp-reduct of a \Sigpp-model of $T''$; \ and
\itemm{($ii$)}
for all \Sigp-expansions $B'$ of $A$,
\ if $B'$ is a \Sigp-reduct of a standard \Sigpp-model of $T''$,
then \,$B' \cong_{\Sigp/\Sig} A'$.
\endpr

Again, an
important special case:

\Defn{${\bk 3}^{\bk f}$}
Suppose \,$\Sig' = \Sig\cup\curly{\fs}$.
We say that \,$\SigppTpp$ \,{\it specifies $f$ over $A$\/}
{\it with  hidden sorts and/or functions} \,iff
$f$ is the unique function on $A$ (of the type of \fs\,)
such that some \Sigpp-expansion of \Af\ satisfies $T''$.
\endpr

\Defn4
An operator \ $\Phi : \NStdAlgSig \to \NStdAlg(\Sig')$ 
\ is {\it expanding (over \Sig)}
iff for all N-standard \Sig-algebras $A$, 
\,$\Phi(A)$ \,is a \Sigp-expansion of $A$,
\ie, $\Phi(A)\redSig = A$.
\endpr

\Example
The array construction \,$A \mapsto \Ax$ 
\,is an expanding operator.
\endpr

{\it Assume further, for the rest of this section, that
\ $\Phi : \NStdAlgSig \to \NStdAlg(\Sig')$ 
\, is an expanding operator over \Sig,
and that \,$\KK\sseq\NStdAlgSig$.}

\Notation
(1) We will write \,\APhi\ \,for \,$\Phi(A)$.

\itemm{(2)}
We write \KKPhi\ for
(the closure w.r.t\. \Sigp-isomorphism of) the class
$\{ \APhi \mid A \in \KK \} \subseteq \NStdAlg(\Sig')$.
\endpr

\Defn5
\itemm{($a$)\ }
\SigpTp\ {\it specifies \Ph\ uniformly over} \KK\ \,iff 
for all \AinKK,
\SigpTp\ specifies \APhi\ over $A$.
\itemm{($b$)\ }
\SigpTp\ {\it specifies \Ph\ uniformly over} \Sig\ \,iff 
\,\SigpTp\ specifies \Ph\ uniformly over 
\nl
\NStdAlgSig.
\endpr

\Propn1
Suppose \SigpTp\ specifies \Ph\ uniformly over \KK.
\sn
($i$) \ \ For \,\AinKK, 
\ $A \ttstile T \ \ \llongtofrom \ \ \APhi \ttstile T + T'$.
\sn
($ii$) \ If \ $\KK = \NStdAlgSigT$, \,then 
\ $\KKPhi = \NStdAlg(\Sig', \,T + T')$.
\endpr

\newpage

\Defn6
\itemm{($a$)\ }
\SigppTpp\ {\it specifies \Ph\ uniformly over \KK\ with 
hidden sorts and/or functions}
\,iff \,for all \AinKK,
\,\SigppTpp\ \,specifies \APhi\ over $A$ with hidden sorts and/or functions.
\itemm{($b$)\ }
\SigppTpp\ {\it specifies \Ph\ uniformly over \Sig\ 
with hidden sorts and/or functions} \,iff
\,\SigppTpp\ \,specifies \Ph\ uniformly over \NStdAlgSig\ 
with hidden sorts and/or 
functions.
\endpr

\Propn2
Suppose \,\SigppTpp\ \,specifies \Ph\ uniformly over \Sig\
with hidden sorts and/or functions.
\sn
($i$) \ \ \ $A \ttstile T \ \llongtofrom 
\ \tx{\APhi\ is a \Sigp-retract of a \Sigpp-model of \,$T+T''$}$.
\sn
($ii$) \ If \ $\KK = \NStdAlgSigT$, \,then 
\ $\KKPhi = \bigl(\NStdAlg(\Sig'', \,T + T'')\bigr) \,|\,_{\Sig'}$.
\endpr

Interesting special cases of the above notions,
in which the theories
$T$, $T'$ and $T''$ are subject to certain
syntactic conditions, are considered below (\S3.3) and in Section 7.
First we give an important example of a specification
of an expanding operator.

We write {\it conditional equational specification\/}
and {\it conditional BU equational specification\/} for specifications
in which the formulae are all conditional equations
and conditional BU equations, respectively.

\shead{3.2}{Conditional BU equational specification of the array construction}
Let \,\ArrAxSig\,\  be the following set of axioms in $A$
(dropping sort subscripts):
$$
\boxed{
\gathered
\Lgths(\Nulls) = 0, \\
\lsnat(\ztt,\Lgths(\att)) = \falses \ \ \to\ \ \Aps(\att,\ztt) = \,\delb, \\
\Lgths(\Updates(\att,\ztt,\xtt)) = \Lgths(\att), \\
\eqnat(\ztt,\ztt_0) = \falses \ \ \to\ \ \Aps(\Updates(\att,\ztt_0,\xtt),\ztt) 
  = \Aps(\att,\ztt), \\
\lsnat(\ztt, \Lgths(\att)) = \trues \ 
  \ \to \ \ \Aps(\Updates(\att,\ztt,\xtt),\ztt) = \xtt, \\
\Lgths(\Newlengths(\att,\ztt)) = \ztt, \\
\lsnat(\ztt,\ztt_1) = \trues \ \ \to\ \ \Aps(\Newlengths(\att,\ztt_1),\ztt) 
  = \Aps(\att,\ztt),\\
\Lgths(\att_1) = \Lgths(\att_2) \con 
\all \ztt<\Lgths(\att_1) \bigl[\Aps(\att_1,\ztt) 
= \Aps(\att_2,\ztt)\bigr]
\ \ \to\ \ \att_1 = \att_2.
\endgathered
}
$$
The last axiom relates equality on \sx\
to equality on $s$, for all equality sorts $s$
except \nats\ \,(since there is no starred sort \,\natsx, \,as 
explained in \S1.6).

Note that all the axioms of \,\ArrAxSig\,\ are conditional equations,
except for the last one, which is a conditional BU equation!

\Thmn1
The specification \,$(\Sigx,\,\ArrAxSig)$ \,specifies the array construction
\ $A \mapsto \Ax$ \ uniformly over \Sig.
\endpr

\n{\bf{Proof \,(outline):}}
\,Given an N-standard \Sig-algebra $A$, and a \Sig-sort $s$,
the axioms for `\Nulls', `\Newlengths' and `\Updates'
guarantee that {\it at least\/} all the ``standard" arrays over \As\
are present (or can be ``constructed").
On the other hand, the axiom for array equality
guarantees that there are no ``non-standard" arrays,
\ie, no elements of \Axs\ other than these.
\endpf

This array specification will be considered again, 
from the viewpoints of specification of \muPRx\ computations
(\S5.2),
and initial algebra specifications (\S8.2).

\sheads{3.3}{Reducing conditional BU to 
conditional equational specifications}
\Thmn{2 \,(BU elimination)}
Let \,$\Sig\sset\Sigp$,
\,let $A'$ be a \Sigp-expansion of $A$, 
\,and let $F$ be a conditional BU equational \Sigp-theory
which specifies $A'$ over $A$.
Then there is an expansion \Sigpp\ of \Sigp\ by function symbols,
and a conditional equational \Sigpp-theory $E$ which specifies 
$A'$ over $A$, with hidden functions.
If $F$ contains $q$ occurrences of BU quantifiers,
then \Sigpp\ expands \Sigp\ by $q$ new function symbols.
Moreover, if
$F$ is finite, with $e$ axioms (say),
then so is $E$, with $e+4q$ axioms.
\endpr

\Pf
The idea is to incorporate in the signature,
for each BU quantifier occurring in $F$,
a {\it characteristic function}
for that quantifier,
or (expressed differently) a function which computes that quantifier.
Consider (for notational simplicity)
the case of an equation with a single BU quantifier
$$
\all \ztt < s(\xtt) \,\bigl[ t_1(\ztt,\xtt) = t_2(\ztt,\xtt)\bigr].
\tagx
$$
with \,$\xtt:u$.
(In the general case, we ``eliminate" the quantifiers
successively, from the inside out.)
We adjoin, for each such BU quantifier ($*$) occurring in $F$,
a boolean-valued function symbol 
$$
\fs:\ \nats\times u \ \to \ \bools
$$
intended to satisfy in $A$ 
$$
\fs(n,x) = \trues\ \ \ \llongtofrom\ 
\ \all z<n \bigl[t_1(z,x)=t_2(z,x)\bigr].
$$
for all \,\ninNN, \,\xinAu.
This interpretation is imposed on \fs\ by
{\it adjoining\/} to $F$ the following axioms giving the
inductive definition for \fs:
$$
\gathered
\fs(0,\xtt) = \trues\\
\fs(\ztt,\xtt) = \trues\ \con t_1(\ztt,\xtt) = t_2(\ztt,\xtt) \ \ \to
  \ \ \fs(\Ss \ztt,\xtt) = \trues\\
\fs(\Ss \ztt,\xtt) = \trues \ \ \to\ \ \fs(\ztt,\xtt) = \trues\\
\fs(\Ss \ztt,\xtt) = \trues \ \ \to\ \ t_1(\ztt,\xtt) = t_2(\ztt,x)
\endgathered
\tagxxx
$$
and {\it replacing} ($*$) in $F$ by 
$$
\fs(s(\xtt),\xtt) = \trues.
\tagxxxx
$$
In this way we replace $F$ by a conditional
equational \Sigpp-theory $E$, with the stated properties.
\endpf

Note that if $F$ contains infinitely many occurences of BU quantifiers,
then \Sigpp\ contains, correspondingly, infinitely many new function symbols, 
which is (strictly speaking) not allowed by our definition of signature,
although it is harmless enough here.

We will return to this topic in the context
of initial algebra specifications in \S7.7.
\endpr

\Shead4{Computable functions}
In this section we consider various notions of 
computability over abstract algebras.
(An equivalent approach, using an 
imperative model of programming featuring the \qwhiles\ construct,
was developed in \cite{tz:book,tz:hb}
where the equivalence of these two approaches are explained.)
In \S4.1 two computability classes are introduced.
In \S4.2 two more classes are formed by adjoining 
the \muu\ operator to these.

\shead{4.1}{\PRSigb\ and \PRxSigb\ computable functions}
Given an N-standard signature \Sig,
we define \PR\  {\it schemes\/} over \Sig\
which generalise the schemes for primitive recursive functions 
over \NN\ in \cite{kleene:im}.
They
define (total) functions $f$ either outright
(as in the base cases ($i$)---($ii$) below)
or from other functions ($g,\dots,$ $h,\dots$) 
(as in the inductive cases ($iii$)---($v$)) as follows:

\itemm{\bi(a)}
{\bi Basic schemes: \,Initial functions\/}
\itemm{($i$)}
{\it Primitive \Sig-functions:}
$$
\align
f(x) \ &= \ F(x)\\
f(x) \ &= \ c
\endalign
$$
of type \,\utos,
\,for all the primitive function symbols \,$F\:\utos$ 
\,and constant symbols $c$ of \Sig, 
where $x:u$. 
\itemm{($ii$)}
{\it Projection:}
$$
      f (x) \ = \ x_i
$$
of type \,$u\to s_i$,  \,where \,$x = (\tup{x}1m)$ \,is of type
\,$u = \tuptimes{s}1m$.

\itemm{\bi(b)}
{\bi Inductive schemes:\/}
\itemm{($iii$)}
{\it Composition:}
$$
      f (x) \ = \ h ( g_1 (x) ,\ldots, g_m (x) )
$$
of type \utos,
where \,$g_i\:u\to s_i$ ($i=1,\dots,m$) \,and 
\,$h\:\tuptimes{s}1m\to s$.

\itemm{($iv$)}
{\it Definition by cases:}
$$
f (b, x, y) \ = \
\cases
x \tif{b = \ttt} \\
y \tif{b = \fff}
\endcases
$$
of type \,$\bools\times s^2 \to s$.

\itemm{($v$)}
{\it Simultaneous primitive recursion on \NN\/}:
\,This defines, on each \AinNStdAlgSig,
for fixed $m > 0$ (the degree of simultaneity),
$n \geq 0$ (the number of parameters),
and product types $u$
and $v = \tuptimes{s}1m$,
an $m$-tuple of functions
\,$f = ( f_1 ,\ldots, f_m )$ \,with
\,$f_i:\nats \times u \to s_i$, \,such that
for all  \,\xinAu\
\,and  $i = 1 ,\ldots, m$,
$$
\align
f_i (0,x ) & \ = \ g_i (x) 	\\
f_i ( z+1, x ) & \ = \
h_i ( z,x, f_1 (z,x) ,\ldots, f_m (z,x) )
\endalign
$$
where \,$g_i\:u\to s_i$ \,and \,$h_i\:\nats\times u \times v \to s_1$
\,($i=1,\dots,m$). 

\n
Note that
the last scheme 
uses the N-standardness of the algebras,
\ie\ the carrier \NN.

For details of the syntax and semantics of PR schemes,
see \cite[\S4.1.5]{tz:book},
from which it can be seen that
a scheme for a function
contains (hereditarily) the schemes for
all the auxiliary functions used to define it.

In the context of algebraic specification theory,
it often turns out to be more convenient to work with
PR {\it derivations\/} instead of PR schemes.
A PR derivation is, roughly, a ``linear version"
of a PR scheme, in which all the auxiliary functions
are displayed in a list.  More precisely:

\Defn{\,(PR derivation)}
A {\it \PRSig\ derivation\/} \al\
is a list of pairs 
$$
\al\ = \ ((f_0,\sig_0),\ (f_1,\sig_1),\ \dots \ (f_n,\sig_n))
\tagx
$$
of functions (actually function symbols) $f_i$ and PR schemes $\sig_i$
($i=1,\dots,n$)
where for each $i$, either $f_i$ is an initial function,
or $f_i$ is defined by $\al_i$ from
functions $f_j$, for certain $j<i$.
The derivation \al\ is said to be a {\it PR derivation of\/} $f_n$,
with {\it auxiliary functions\/} \tup{f}0{n-1}.
The {\it type\/} of \al\ is the type of $f_n$.
\endpr

\Notation
A \,\PRSigutos\ \,scheme (or derivation)
is a \,\PRSig\ \,scheme (or derivation) of type \,\utos.
\endpr

\Remarks
(1) \,The formalism of \PRSig\ derivations
is equivalent to that of \PRSig\ schemes:
from a PR scheme we can derive an  equivalent PR derivation
by ``linearising" the subschemes, and conversely,
given the derivation ($*$), the scheme $\sig_n$
is equivalent to it.
Below, we will usually work with derivations.
\sn
(2) \,A \,\PRSigutos\ 
\,derivation \,$\al\:\utos$
\,defines, or rather computes, a
function \,$\falA\:\Au\to\As$, or, more generally,
a family of functions  
\curly{\falA \mid \AinNStdAlgSig}
uniformly over \NStdAlgSig.
\sn
(3) \,We assume a standard G\"odel numbering of \PRSig\ derivations,
writing \cnr{\al} for the G\"odel number of derivation \al.
\endpr

It turns out that a broader class
of functions provides a better generalisation of the
notion of primitive recursiveness, namely
{\it \PRx\ computability}.
A function on $A$ is \PRxSig\ computable
if it is defined by a PR derivation over \Sigx,
interpreted on \Ax\
(\ie, using starred sorts for the auxiliary functions 
used in its definition).
 
\shead{4.2}{\muPRSigb\ and \muPRxSigb\ \,computable functions}
The \,\muPR\ {\it schemes\/} over \Sig\
are formed by adding to the PR schemes of \S4.1
the inductive scheme: 

\itemm{($vi$)}
{\it Least number {\rm or} \muu\ operator:}
$$
f (x) \ \simeq \ \mu  z [ g ( x , z ) = \ttt ]
$$
of type \,$u\to\nats$,  \,where
\,$g:u\times\nats\to\bools$
\,is \muPR.
Here \,$f (x) \da z$ \,if, and only if, \,$g (x , y ) \da \fff$ 
\,for each $y < z$ and $g (x ,z ) \da \ttt$.

\n
Note that this scheme also uses the N-standardness of the algebra.
Also, \muPR\ computable functions are, in general, {\it partial}.
We use the notation \,$f(x)\da y$ \,to mean that $f(x)$
is defined and equal to $y$.
The notation \,`$\simeq$'
\,means that the two sides are either 
both defined and equal,
or both undefined. The schemes for composition and simultaneous
primitive recursion are correspondingly re-interpreted
to allow for partial functions.

These schemes generalise the schemes given in \cite{kleene:im}
for partial recursive functions
over \NN.

As before, we can define
the concepts of
\muPRSig\ {\it derivations\/} and \muPRSig\ {\it computability\/}.

Again, a broader class turns out to be more useful, namely
{\it \muPRx\ computability}.
This is just \PRx\ computability with \,\muu.
 
\Notation
\PRA\ \,is the class of functions \,\PR\ \,computable
on $A$,
\,and
\,\PRAutos\ \,is the subclass of \,\PRA\ \,of functions of type \,\utos.
Similarly for \,\PRxA, \,\muPRA, \,etc.
\endpr

There are many other models of computability, due to 
Moschovakis, Friedman, Shepherdson and others, 
which turn out to be equivalent to \muPRx\ computability:
see \cite[\S7]{tz:hb}.
All these equivalences have led to the postulation of a 
{\it generalised Church-Turing Thesis
for deterministic computation of functions},
which 
can be roughly formulated as follows:

{\displaytext
Computability of functions on many-sorted algebras 
by deterministic algorithms
can be formalised by \,\muPRx\ computability.

}

\shead{4.3}{Equivalent imperative programming models of computation}
In \cite{tz:hb} we investigate computation on many-sorted \Sig-algebras,
using imperative programming models:
\,\WhileSig, \,based on the \,\qwhiles\ \,loop construct over \Sig,
\,\ForSig, \,based similarly on the \,\qfors\ \,loop,
and \WhilexSig\ \,and \,\ForxSig,
\,which use arrays, \ie, auxiliary variables of starred sort over \Sig.

Writing \WhileA\ \,for the class of functions \While-computable on $A$, etc.,
we can list the equivalences
between the ``schematic" 
and ``imperative" computational models as follows.

\Thm
\itemm{($i$)}
$\PRA \ = \ \ForA$
\itemm{($ii$)}
$\PRxA \ = \ \ForxA$
\itemm{($iii$)}
$\muPRA \ = \ \WhileA$
\itemm{($iv$)}
$\muPRxA \ = \ \WhilexA$,

\n
in all cases, uniformly for \,\AinNStdAlgSig.
\endpr

These results are all stated in \cite{tz:hb}, and 
can be proved by the methods of \cite{tz:book}.

\newpage

\shead{4.4}{Universal Function Theorem for \muPRxb}
The following is a uniform version of a theorem proved in
\cite[\S4.9]{tz:hb}
(using the equivalent formalism of \,\Whilex\ \,programs):

\Thm
For any \Sig-function type \utos,
there is a \,\muPRxSig\ \,derivation \,$\ups:\natxutos$
\,which is
universal for \,\muPRxSig\ \,derivations of type \,\utos.
\endpr

In other words, we can enumerate all the \muPRx\ derivations 
of type \utos:
$$
\al_0,\ \al_1,\ \al_2,\ \dots
$$
so that, putting
$$
\varphi^A_i \ \eqdf \ \fs^A_{\al_i}\: \Au\ \to\ \As
$$
and
$$
\UnivusA \ = \ \fs^A_\ups: \NN\times\Au\to\As
$$
we have
$$
\UnivusA(i,a) \ = \ \varphi^A_i(a)
$$
for all \,\AinNStdAlgSig\ \,and \,$i=0,1,2,\dots$.

\Remarksn{ \,(Canonical forms of \,\muPRxb\ \,derivations)}
(1) \,From the construction of the universal \,\muPRxSigutos\ \,derivation
\ups\
\cite[\S4]{tz:hb},
it can be seen that
\ups\ uses the \muu-operator exactly once.
\sn
(2) \,For any \,\muPRxSigutos\ \,derivation \al,
the universal derivation \,$\ups\:\natxutos$
\,provides an equivalent
{\it canonical  {\rm or} normal form derivation\/} \,\alhat, such that
\,$\falhatA \,= \,\falA$
\,for all N-standard \Sig-algebras $A$.
This canonical derivation is formed 
in a simple way from \,\ups, \,essentially
by substituting the G\"odel number \cnr{\al} of \al\ 
for the distinguished \,\nats\ \,variable of \ups,
\,so that for all N-standard $A$,
$$
\fs_{\alhat}^A \ = \ \varphi_{\cnr{\al}}^A \ = \ \fs_\al^A.
$$
This is, in fact, a generalisation to \,\NStdAlgSig\
\,of Kleene's Normal Form Theorem for (essentially) \,\muPR(\NNN)
\,\cite{kleene:im}.
\sn
(3) \,From the constructions in (1) and (2) it follows that \,\alhat\
\,also uses the \muu-operator exactly once, 
and in such a way that for any N-standard $A$,

{\displaytext
\falhatA\ \,is total if, and only if,
this application of the \muu-operator is total on $A$.

}

\Shead5
{Algebraic specifications for computable functions}
We will consider functions $f$ {\it computable\/} on a \Sig-algebra,
by \,\PR\ \,and \,\muPRx\ \,derivations,
and show that they are 
algebraically {\it specifiable\/} by
conditional equational, and conditional BU equational, theories.

We will also consider,
in the context of {\it metric algebras\/}
(\ie, algebras with metrics  such that 
the functions in the signature are continuous)
a broader class of functions than \muPRx\ computable,
namely those functions
{\it uniformly approximable\/} by \muPRx\ computable functions,
and show that such functions
are specifiable by {\it conditional equations and inequalities\/},
which are conditional formulae built up from inequalities ($t_1<t_2$)
on the reals
as well as equations ($t_1=t_2$) between terms of the same sort.

\shead{5.1}{Algebraic specifications for \,\PR\
\,computable functions} 
Let \Sig\ be an N-standard signature.  
For each \PRSig\ derivation \al, there is a finite
set \Eal\ of ``specifying equations" for the function $f$,
as well as the auxiliary functions
\,$g=(\tup{g}1{k_\al})$,
\,defined by \al.

The set \Eal\ consists of {\it equations\/}
in an expanded signature \,$\Sigal = \Sig \cup \{ \gal, \fal \}$,
\,where \,$\gal \ident \gs_{\al,1},\dots,\gs_{\al,k_\al}$.
It is defined by course of values induction on the length of the 
derivation \al,
with cases ($i$)---($v$) (\S4.1)
according to the last scheme in \al.
In fact, \Eal\ is formed by adjoining, in each case,
specifying equation(s) like those shown for that case in \S4.1.
These are simple (\ie, not conditional) equations;
for example, in the case ($iv$) {\it definition by cases\/},
there are two equations:
$$
\align
\fs(\trues,\,\xtt,\,\ytt) \ &= \ \xtt \\
\fs(\falses,\,\xtt,\,\ytt) \ &= \ \ytt
\endalign
$$
and in the case ($v$) {\it simultaneous primitive recursion\/},
there are \,$2m$ \,equations (where $m$ is the degree of simultaneity):
$$
\align
\fs_i (0,\,\xtt ) & \ = \ \gs_i (\xtt)      \\
\fs_i ( \ztt+1, \,\xtt ) & \ = \
        \hs_i ( \ztt,\,\xtt, \,\fs_1 (\ztt,\xtt) ,\ldots, \fs_m (\ztt,\xtt) )
\endalign
$$
for \,$i=1,\dots,m$.

Thus we have an effective map 
\ $\al \mapsto (\Sigal, \Eal)$
\ from \PRSig\ derivations to (simple) equational specifications.

Now for each \PR\ derivation \al\ and N-standard \Sig-algebra $A$, let
\,\falA\ \,be the function on $A$ computed by \al,
and let \,\galA\ \,be the corresponding auxiliary functions on $A$.
Consider the operators
$$
A \longmapsto \AfalA
\tagx
$$
and
$$
A \ \longmapsto \ \AgalAfalA.
\tagxx
$$

Recall the definition of uniform specification of an operator 
over a class of \Sig-algebras (\S3.1, Definitons 5 and 6).

\Thmn{1 \,(Equational specification of PR functions)}
For each \PRSig\ derivation \al,
the equational specification \,$(\Sigal, \Eal)$ 
\, specifies the operator ($**$)
uniformly over 
$A \in \NStdAlgSig$.
Hence it specifies the operator
($*$) uniformly over all N-standard \Sig-algebras $A$,
with hidden functions.
\endpr

\Pf
By course of values induction on the length of \al.
\endpf

In other words, the equations \Eal\ specify not only \falA,
but also the auxiliary functions \galA, uniformly over 
all N-standard \Sig-algebras $A$.

Similarly with
\PRx\ computability:
for a \PRxSig\ derivation \al, 
let \Eal\ be the set of
specifying equations for the function \,\ \fal\
and the auxiliary functions \gal\
defined by \al, 
in the signature \,$\Sigalx = \Sigx \cup \{ \gal, \fal \}$.

\Cor
For each \PRxSig\ derivation \al,
the equational specification \,\SigalxEal\
specifies the operator
($*$) uniformly over \Sig,
with hidden sorts and functions.
\endpr

\shead{5.2}{Algebraic specifications for \muPRxb\ computable functions}
We now consider \,\muPRxSig\ \,derivations \al.
For each such derivation there is again
a finite set \Fal\ of ``specifying equations"
for the function $f$ defined by \al\
and its auxiliary functions $g$.
This set is constructed like \Eal\ (\S5.1),
by structural induction on \al.
Now, however,
\Fal\ consists of 
{\it conditional BU equations}
in a signature \ $\Sigalx = \Sigx \cup \{ \gal, \fal \}$,
\ because of the new case, \ie,
scheme ($vi$) for the \muu-operator
(\S4.2),
which results in the addition to \Fal\ of the conditional BU equation
\TOL
$$
\all \ztt < \ytt \,(\gs_0(\xtt,\,\ztt) \,= \,\falses) \ 
\con \ (\gs_0(\xtt,\,\ytt) \,= \,\trues) 
\ \ \longimp \ \ \fs(\xtt) = \ytt .
\tag{\Fmu}
$$
\TOR

Again we have an effective map 
\ $\al \mapsto \SigalxFal$
\ from \,\muPRxSig\ \,derivations to conditional BU equational specifications.

Now there are complications in the theory, 
since \muPRx\ computable functions are, in general, partial.
We intend to study specification theory for partial algebras
systematically in a future paper.
Here we limit ourselves to 
the case where the \muPRx\ computable function
is, in fact, total.

As before, for a \muPRx\ derivation \al\
and an N-standard \Sig-algebra $A$, let
\,\falA\ \,be the function on $A$ defined by \al,
and let \,\galA\ \,be corresponding auxiliary functions on \Ax.
A further problem is that, even if \,\falA\ \,is total,
the functions \,\galA\ \,might not be.
We will now show that we can, without loss of generality, 
restrict attention to the case that the
\,\galA\ \,are also total.
We accomplish this
by the use of the uniform derivations
provided by the Universal Function Theorem for \muPRx\ (\S4.4),
as we now explain.

\Def
A \muPRx\ derivation \al\ is {\it total on\/} $A$ iff 
the auxiliary functions \galA, as well as \falA, are all total on \Ax.
\endpr

\pr{Totality Lemma}
Given any \muPRxSig\ derivation \,$\al\:\utos$,
\,we can effectively find a \muPRxSig\ derivation \,$\alhat\:\utos$
\,such that for any N-standard \Sig-algebra $A$,
\itemm{$(i)$}
$\falhatA \,= \,\falA$;
\itemm{$(ii)$}
if \,\falA\ \,is total, then \alhat\ is total on $A$.
\endpr

\Pf
This follows from the Universal Function Theorem
and the three remarks following it (\S4.4).
\endpf

Now consider the operators ($*$) (\S5.1 above)
and
$$
A \ \longmapsto \ (\Ax,\,\galhatA,\,\falhatA)
\tagxxx
$$
where \alhat\ is constructed from \al\ as in the totality lemma.
Let \,$\Sigalx = \Sigx \cup \curly{\galhat,\falhat}$.
Recall the definition of 
the array specification \,\ArrAxSig\ \,in \S3.2,
and the definition of 
the conditional BU specification
\,\Falhat\ \,of \,\falhatA\ ($= \falA$).

\Thmn{2 \,(Conditional BU equational specification of \muPRxb\ functions)}
\nl
For each 
\muPRxSig\ derivation \al, let
$$
\Falx \ \eqdf \ \ArrAx(\Sig) \,+ \,\Falhat
$$
where \alhat\ is constructed from \al\ 
as in the totality lemma.
Then the conditional BU equational specification
\,$(\Sigalx,\,\Falx)$\,\ specifies the operator (\xxx)
in the following sense:
for any $A$ on which \falA\ is total,
$$
(\Ax,\,\galhatA,\,\falhatA)\ \ttstile \ \Falx.
$$
Hence
\,$(\Sigalx,\,\Falx)$
\,specifies the operator $(*)$ 
uniformly over all N-standard \Sig-algebras $A$
\,on which \,\falA\ \,is total,
with hidden sorts and functions.
\endpr

\Pf
As with Theorem 1, by course of values induction on the length of \al.
\endpf

Note that the specification given in Theorem 2 is uniform
over all N-standard \Sig-algebras $A$
on which \,\al\ \,is total.
In fact, there is a stronger form of uniformity
for \muPRx\ computability,
following from the Universal Function Theorem for \muPRx.
(Actually, this is already implicit in the construction of
the derivation \alhat\ in the
totality lemma, which is really a normal form lemma for \muPRx\ derivations.)

\Thmn{3 \,(Universal conditional BU equational specification)}
For each \Sig-function type \,\utos\
\,we can effectively find a 
signature \,\Sigxus\ \,which expands \Sigx\ 
by function symbols, and a
finite conditional BU equational specification
\,$(\Sigxus,\,\FUus(\ztt))$
\,which is 
universal for specifications of total \,\muPRxSig-computable functions
of that type,
in the following sense:
it contains a distinguished number variable \,\ztt\
\,such that 
for each \,\muPRxSig\ \,derivation \,$\al:\utos$,
\,and each N-standard \Sig-algebra $A$,
if \,\falA\ \,is total on $A$,
then 
\,$(\Sigxus,\,\FUus(\kbar))$,
\,where \,$k=\cnr{\al}$,
\,specifies \,\falA\ \,on $A$, with hidden sorts and functions.
\endpr

(Here \,$\FUus(\kbar)$
\,is the result of substituting the numeral \kbar\ for \ztt\ in 
\,$\FUus(\ztt)$.)

Next, by the BU Reduction Theorem (Theorem 2 in Section 3),
we derive as a corollary to Theorem 3:

\Thmn{4 \,(Universal conditional equational specification)}
For each \Sig-function type \,\utos\
\,we can effectively find a 
signature \,\Sigxusp\ \,which expands \,\Sigxus\  
\,(of Theorem 3)
by function symbols, and a
finite conditional specification
\,$(\Sigxusp,\,\EUus(\ztt))$
\,which is 
universal for specifications of total \,\muPRxSig-computable functions
of that type,
in the following sense:
it contains a distinguished number variable \,\ztt\
\,such that 
for each \,\muPRxSig\ \,derivation \,$\al:\utos$,
\,and each N-standard \Sig-algebra $A$, 
if \falA\ is total on $A$,
\,then 
\,$(\Sigxusp,\,\EUus(\kbar))$,
\,where \,$k=\cnr{\al}$,
\,specifies \,\falA\ \,on $A$, with hidden sorts and functions.
\endpr

From the above uniformity theorems it follows trivially that 
for a given \Sig-function type \,\utos\
\,there is a uniform bound to the lengths of conditional
BU \Sigx-specifications,
{\it or\/} conditional
equational \Sigx-specifications respectively,
for total \,\muPRx-computable functions 
on N-standard \Sig-algebras.
\endpr


\Shead6{Algebraic specifications for computably approximable functions}
We have shown that 
$$
\tx{\sl computability $\implies$ algebraic specifiability}
$$
where (for example) if ``computability" means \,\muPRx\
\,(or, equivalently, \,\Whilex) \,compu\-tability,
then ``algebraic specifiability" means specifiability
by conditional BU equations.

It is natural to ask in what sense a converse holds.
We will see (below) that a full converse to the above cannot be expected,
since algebraic specifiability is more powerful,
in some sense, than \muPRx-computability.
(In Section 7 we will investigate partial converses.)
We show here in fact that,
on metric algebras,
$$
\tx{\sl computable approximability\ $\implies$ \ algebraic specifiability.}
$$
``Computable approximability", to be defined shortly,
is a strong extension of the notion
of computability;
while ``algebraic specifiability" will be (re-)defined 
so as to permit the order relation (as well as equality) 
between pairs of terms of sort \reals.

\shead{6.1}{Metric algebras}
We refer to \cite{tz:top} and \cite[\S6]{tz:hb} for definitions of
(total) {\it metric algebra\/} and related concepts.
We review some definitions and results from these references.
(Note that in these references the subject is discussed
in the broader context of {\it partial algebras\/}.)

A {\it metric \Sig-algebra \Ad, based on a \Sig-algebra\/} $A$,
is an algebra of the form
$$
\boxed{
\matrix \format\l&\quad\l\\
\algebras &\Ad \\
\imports &A \\
\carrierss & \RR \\
\functionss&d_s:A_s^2\to\RR \quad (\sinSortSig)\\
\ends&
\endmatrix
}
$$
where $d$ is a family \,\ang{d_s\mid \sinSortSig}
\,of metrics $d_s$ on the carriers \As, where
(in the case that $A$ is standard or N-standard)
\,$d_\boolss$ \,and \,$d_\natss$ \,are the discrete metrics
on \,\BB\ \,and \,\NN\ \,
respectively,
and such that the primitive functions on $A$
are all continuous under these metrics.

We will often write `$d$' for the metric $d_s$, and
`$A$' for the metric algebra \Ad.

\Examples
($a$)
\,The metric algebra \,\RRRd\
\,on the reals (``$d$" for ``distance") is defined by
$$
\boxed{
\matrix \format\l&\quad\l\\
\algebras &\RRRd \\
\imports &\RRRN\\
\functionss&\divN:\RR\times\NN\to\RR,\\
&d_\realss:\RR^2\to\RR,\\
&d_\natss:\NN^2\to\RR,\\
&d_\boolss:\BB^2\to\RR\\
\ends&
\endmatrix
}
$$
where \,\RRRN\ \,is the N-standard algebra of reals (\S1.4, Example ($b$)),
\divN\ is division of reals by naturals
(where division by zero is defined as zero),
the metric on \,\RR\ \,is the standard one,
and the metrics on \,\NN\ \,and \,\BB\ \,are discrete.

Note that \,\RRRd\ \,does not contain the (total) boolean-valued
functions \eqreal\ or \lsreal,
since they are not continuous with respect
to this metric.

\smskipn ($b$)
\,The interval metric algebra \,\IIId:
\ Here the unit interval
\,$I = [0,1]$
\,is included as a separate carrier of
sort `\intvls', again with the usual metric.
This is useful for studying real continuous functions
with compact domain.
(We could also choose \,$I = [-1,1]$, \,etc.)
The algebra \,\IIId\ \,is defined by
$$
\boxed{
\matrix \format\l&\quad\l\\
\algebras &\IIId \\
\imports &\RRRd\\
\carrierss & I\\
\functionss &\iI:I\to\RR,\\
&d_\intvlss:I^2\to\RR\\
\ends&
\endmatrix
}
$$
where \iI\ is the embedding of $I$ into \,\RR.
Because of the importance of the metric algebra \IIId\ as
in our computation theory, let us review 
its construction.
It contains \RR\ with its standard ring operations,
\NN\ and \BB\ with their standard operations,
functions for definition by cases on \RR, \NN\ and \BB,
division of reals by naturals,
the unit line interval \II\ and its embedding in \RR,
and the standard metrics on all four carriers.
\endpr

\shead{6.2}{Definitions and theorems}
Now let $A$ be an N-standard metric \Sig-algebra with metric $d$.

\Defn{1 \,(\muPRxb\ computably approximable functions)}
A total function \,$f:\Au \to \As$ \,on $A$
is \,\muPRx\ {\it computably approximable,
uniformly on\/} $A$,
if there is a total \,\muPRx\ computable function
$$
G:\NN\times \Au\ \ \to \ \As
$$
and a total computable function \,$g:\NN\to\NN$
on $A$ such that, putting
\,$G_n \eqdf G(n,\,\cdot\,)$,
\,the sequence $G_n$ {\it approximates $f$ uniformly\/} on \Au\
with modulus of approximation $g$,
\ie, for all $n$, $k$ and all $x \in \Au$,
$$
k\ge g(n) \ \implies\ d(G_k(x),f(x)) \ < \ 2^{-n}.
$$
\endpr

\Defn{2 \,(Fast \,\muPRxb\ computably approximable functions)}
A total function \,$f:\Au \to \As$ \,on $A$
is {\it fast \,\muPRx\ computably approximable,
uniformly on\/} $A$,
if there is a total \,\muPRx\ computable function
\ $G:\NN\times \Au\ \,\to \,\As$
\ on $A$ such that, putting
\,$G_n \eqdf G(n,\,\cdot\,)$,
\,the sequence $G_n$ {\it approximates $f$ uniformly fast\/} on \Au,
\ie, for all $n$ and all $x \in \Au$,
$$
d(G_n(x),f(x)) \ < \ 2^{-n}.
\tagx
$$
\Remarkn1
It is easy to see that Definitions 1 and 2 are equivalent;
for given a (computable) approximating sequence $G_n$
with modulus of approximation $g$,
we can effectively replace it by the fast (computable)
approximating sequence \,$G_n' \eqdf G_n\circ g$.
We will therefore usually tacitly assume w.l.o.g. that
our approximating sequences are fast,
and work with the (simpler) Definition 2.
\endpr

\Defn{3 \,(Fast \muPRxb\ approximating derivations)}
Let $A$ be a metric \Sig-algebra.
A derivation 
\,$\gam\:\natxutos$
\,is an {\it approximating derivation for\/} 
a total function
\,$f\:\Au\to\As$ 
\,if 
\,($i$) \,the function \,$G:\NN\times\Au\to\As$ \,computed by \,\gam\ \,on $A$
is total on $A$; and ($ii$) \,$G$ and $f$
satisfy ($*$) above.
\endpr
Note that {\it at most one\/} function is
\,\muPRx\ \,approximable by a given derivation on any metric algebra.

\Defn{4 \,(Conditional equation or inequality)}
\nl
($a$) \,A {\it conditional equation or inequality \/} is defined
like a conditional equation, except that the atomic
statements may be {\it either\/} equations ($t_1 = t_2$)
between terms of the same sort,
{\it or\/}
order 
($t_1<t_2$)
between terms of sort \reals.
\sn
($b$) A {\it conditional BU equation or inequality\/} 
is defined like a conditional equation, except that the atomic
statements may be {\it either\/} equations ($t_1=t_2$)
or BU equations 
\linebreak
($\all \ztt<t\,[\,t_1=t_2\,]$)
between terms of the same sort,
or inequalities
($t_1<t_2$)
between terms of sort \reals.
\endpr

\Remarkn2
Here we are treating the order relation on the reals
as a new atomic predicate of \,\LangSigx\
\,(like equality), {\it not\/} as a boolean-valued operation
$$
\lsreal:\reals^2\to\bools.
$$
This predicate (unlike such an operation)
does not form part of the signature \,\Sig.
(See the analogous Remark concerning equality
at the beginning of Section 2.)
\endpr

Note also that ($*$) is a conditional inequality
(actually a simple inequality, without an antecedent).

\Notation
We write \,\muPRxApproxA\ \,for the class of \muPRx\ computably
approximable functions on $A$, and
\,\muPRxApproxAutos\ \,for those of type \,\utos.
\endpr

In preparation for the next theorem,
we note that 
a ``Universal Function Theorem"
holds for \,\muPRxApproxA,
\,in the following sense. 
For any \Sig-function type \utos, let
$$
H^\us\ \eqdf\ \Univ^A_\natxutoss:
\ \NN\times\NN\times \Au\ \to\ \As
$$
be the universal function for
\,$\muPRxA_\natxutoss$
\,given by the Universal Function Theorem (\S4.4).
Then for each \,$f\in\muPRxApproxAutos$,
there is a number $k$ such that
(writing \,$H^\us_{k,n} = H^\us(k,n,\,\cdot\,)$)
\,the sequence of functions
\,$H^\us_{k,0},\,H^\us_{k,1},\,H^\us_{k,2},\,\dots$
\,uniformly approximates $f$.
The number $k$ can be chosen as the G\"odel number of an
approximating derivation for $f$,
\ie, a derivation \,$\gam\:\natxutos$
\,of the function \,$H^\us(k,\,\cdot\,)$. 
Combining this with Theorem 3 of Section 5, we obtain:

\Thmn{1 \,(Universal conditional BU specification 
of \,\muPRxb\ approximable functions)}
For each \Sig-function type \,\utos\
\,we can effectively find a
signature \,\Sigxus\ \,which expands \Sigx\
by function symbols, and a
finite conditional 
BU specification \,$(\Sigxus,$ $\FVus(\ztt))$
\,consisting of conditional BU equations and inequalities,
\,which is
universal for specifications of 
\,\muPRxSig-computably approximable functions of that type,
in the following sense:
it contains a distinguished number variable \,\ztt\
\,such that
for each \,\muPRxSig\ \,derivation 
\,$\gam\:\natxutos$
\,and each metric \Sig-algebra $A$
and total function \,$f\:\Au\to\As$,
\,if \,\gam\ \,is an approximating derivation for $f$ on $A$,
then
\,$(\Sigxus,\,\FVus(\kbar))$,
\,where \,$k=\cnr{\gam}$,
\,specifies \,$f$ \,on $A$, with hidden sorts and functions.
\endpr

\Pf
Define
$$
\FVus(\ztt) \ \eqdf\ \FUus(\ztt) + E_{\invexpss} + E_*(\ztt)
$$
where \,$\FUus(\ztt)$ \,is the conditional BU equational specification
constructed as in Theorem 3 in Section 5 for 
the universal function $H$ for \,\muPRx\ \,computable functions
of type 
\linebreak
\,$\nats \times u \to s$,
\,$E_{\invexpss}$ \,is the set of specifying equations for
the computable real-valued function
\,$\invexps(n) = 2^{-n}$,
\,\ie, its recursive definition: 
$$
\invexps(0) \ = \ 1,\qquad \invexps(\Ss \ntt) \ = \ \divN(\invexps(\ntt),\,2),
$$
and \,$E_*(\ztt)$ \,is the inequality ($*$) above ---
or rather, its formal version
$$
\ds(\Hs(\ztt, \,\ntt,\,\xtt), \,\fs(\xtt)) \ < \ \invexps(\ntt).
\tagxx
$$
(Note that every \,\muPRx\ \,function \,$G\:\NN\times\Au\to\As$
\,is obtainable from $H$ by substituting the G\"odel number
of its derivation for the first argument \,\ztt\ \,of $H$.)
Let \,\Sigxus\ \,be the signature formed by
expanding \,\Sigx\ \,by symbols for \,\Hs\ \,and 
\,\invexps, \,as well as the auxiliary functions 
used in their computations.
Then for any \muPRx\ derivation \,$\gam\:\natxutos$,
\,metric \Sig-algebra $A$
and function \,$f\:\NN\times\Au\to\As$,
\,if \gam\ is an
approximating derivation for $f$
then \,$(\Sigxus,\,\FVus(\kbar))$
\,(where \,$k=\cnr{\gam}$)
\,is a conditional BU specification of $f$ on $A$,
with hidden sorts and functions,
consisting of conditional BU equations and inequalities
\endpf

Now, by adapting the BU Reduction Theorem (\S3.3)
to specifications with inequalities,
we derive as a corollary to Theorem 1:

\newpage

\Thmn{2 \,(Universal conditional specification 
of \,\muPRxb\ approximable functions)}
For each \Sig-function type \,\utos\
\,we can effectively find a
signature \,\Sigxusp\ \,which expands \,\Sigxus\  
\,(of Theorem 1)
by function symbols, and a
finite conditional 
specification \,$(\Sigxusp,\,\EVus(\ztt))$,
\,consisting of conditional equations and inequalities,
\,which is
universal for specifications of approximably
\,\muPRxSig-computable functions of that type,
in the following sense:
it contains a distinguished number variable \,\ztt\
\,such that
for each \,\muPRxSig\ \,derivation 
\,$\gam\:\natxutos$
\,and each metric \Sig-algebra $A$
and total function 
\,$f\:\Au\to\As$,
\,if \,\gam\ \,is an approximating derivation for $f$ on $A$,
then
\,$(\Sigxusp,\,\EVus(\kbar))$,
\,where \,$k=\cnr{\gam}$,
\,specifies \,$f$ \,on $A$, with hidden sorts and functions.
\endpr

\Remarkn{3 \,(Replacing  the order predicate by a boolean-valued operation)}
The order relation in the above specification is used in one place only: 
in the (conditional) relation ($*$) (or ($**$)).
In fact (Remark 2 above notwithstanding)
($*$) {\it could\/} be interpreted as a conditional {\it equation\/}
(so that $f$ is {\it conditionally equationally definable\/}
with hidden sorts and functions)
by interpreting \,`$t_1<t_2$' \,as \,`$\lsreal(t_1,t_2)=\trues$',
\,where the boolean-valued operator
$$
\lsreal:\reals^2\to\bools
$$
is included in the signature of the metric algebra over \RR.
The problem here is that (as discussed in \cite{tz:top})
whereas all functions in the signature of metric algebras
(and hence all functions computable over these)
are {\it continuous\/}, the (total) function \,\lsreal\ \,is discontinuous.
The only way to restore continuity is to consider
a {\it partial\/} continuous \,\lsreal\ \,operator,
which leads to a study of {\it topological partial algebras\/}.
This can be done, and the whole of the present theory
could be re-cast in such a context,
but that would take us too far afield from the present study.
\endpr

Let us apply Theorem 2 to 
the classical notion of
{\it Gzegorczyk-Lacombe (GL) computability\/}
on the  unit interval $I=[0,1]$.
This includes all the well-know functions of real analysis
(sin, exp, log, etc.) restricted to $I$.

\Notation
We write
\,\GLtmI\ for the class of GL-computable total functions 
\,$f\:I^m\to \RR$.
\endpr

Many concrete models of computation on \,\II\ 
\,are equivalent to this class \cite{pourel-richards,weih:book}.
It has been shown that (recall the definition of \IIId\
in Example ($b$)):
$$
\GLtmI \ = \ \muPRxApprox(\IIId)_{\intvlss^m\to\realss}.
$$
(For details, see 
\cite[\S9]{tz:top}, \cite[\S5.9]{tz:hb}.)
Hence, again, a kind of
``Universal Function Theorem"
holds for \GLtmI,
in the following sense. 
For $m=1,2,\dots$, \,let
$$
H^m\ \eqdf\ \Univ^{\IIId}_{\natss\times\intvlss^m\to\realss}:
\ \NN\times\NN\times I^m\ \to\ \RR
$$
be the universal function for
\,$\muPRx(\IIId)_{\natss\times\intvlss^m\to\realss}$
\,given by the Universal Function Theorem (\S4.4).
Then for each \,$f\in\GLtmI$,
there is a number $k$, 
effectively obtainable from the GL-code for $f$,
such that (writing \,$H^m_{k,n} = H^m(k,n,\,\cdot\,)$)
\,the sequence of functions
\,$H^m_{k,0},\,H^m_{k,1},\,H^m_{k,2},\,\dots$
\,uniformly approximates $f$ on $I$.

So by
Theorem 2 applied to \IIId:
\endpr

\Thmn{3 \,(Universal specification of GL computable functions)}
For each $m>0$
there is a signature
\,\Sigxm\ \,which is an expansion of \,\SigIIIdx\ \,by finitely many
function symbols,
and a finite conditional specification 
\,$(\Sigxm,\,\EWm(\ztt))$
\,consisting of conditional equations and inequalities,
\,which is 
universal for specifications of \GLtmI,
in the following sense:
it contains a distinguished natural number variable \ztt\
such that 
each function
$f\in\GLtmI$ 
is specified (with hidden sorts and functions)
by a suitable substitution instance
\,$(\Sigxm,\,\EWm(\kbar))$,
\,where $k$ can be found effectively from a GL-code for $f$.
\endpr

\Remarkn{4 \,(Description of the signature \,\Sigxm\ \,of Theorem 3)}
The signature \Sigxm\ is an expansion of \,\SigIIId\
\,(for a description of which see the remark at the end of \S6.1)
by the following sorts and functions:

\itemm{($i$)}
the sorts and functions of the array structure over \IIId\ (\S1.6);

\itemm{($ii$)}
the \muPRx\ ``universal function" $F^m$ for \GLtmI\
(as described in the above discussion)
together with the auxiliary functions in its derivation;

\itemm{($iii$)}
the function $2^{-n}$, used for assertions about 
computable approximations, as explained in the proof of 
Theorem 1;

\itemm{($iv$)}
the characteristic function for BU quantification,
as described in the proof of the
BU elimination theorem (\S3.3).

\sn
Note that there is only one function of type ($iv$) in \Sigxm,
namely that obtained by eliminating
the conditional BU equation \Fmu\ (\S5.2) specifying the (single!)
\muu-operator occurrence in the 
\muPRx\ derivation for $H^m$ in ($ii$)
(see Remark 1 in \S4.4).
(There are no conditional BU equality axioms for arrays (\S3.2)
to eliminate here, since \,\reals\ \,and \,\intvls\ 
\,are not equality sorts.)
\endpr

\shead{6.3}{Illustration: \ Specification of dynamical system}
We illustrate the connection between algebraic specification methods
and models of physical systems.

A {\it deterministic dynamical system\/} with finite dimensional state space 
\,$S\sseq\RR^n$ \,and \,time \,$T\sseq\RR$
\,is represented in a model by a function
$$
\phi: \ T\times S\ \to\ S
$$
where for \,$t\in T$, $s\in S$, 
\,$\phi(t,s)$ \,is the state of the system at time $t$ with 
initial state $s$.
For example, the state of a particle in motion
is represented by position and velocity.
Thus, for a system of $n$ particles in 3-dimensional space,
the state space has $6n$ dimensions.

In practice, the model is specified by
ordinary differential equations (ODEs) whose complete solution is \,$\phi$.
Specifically, in the modern qualitative theory of ODEs
\cite{arnold}, \,$\phi$ \,is differentiable,  and the function
\,$\phi_t:S\to S$ \,defined by
$$
\phi_t(s) \ = \ \phi(t,s) \quad \tx{for} \ \ t\in T, s\in S,
$$
is a 1-parameter group of diffeomorphisms of $S$;
the action of this group on $S$ is 
called the {\it flow\/} on the {\it phase space\/} $S$.
This flow can be specified by a vector field on $S$.

In modelling a physical system, one aim is to compute
values of the function \,$\phi$ \,on some time interval
and subspace of the space of initial conditions.
Many methods exist to derive algorithms for $\phi$
\,from the equations that define it.
Indeed, various fields of applied mathematics
exist in order to design such equations,
and the field of numerical analysis exists
to design such solution methods.

Conversely, we suppose that \,$\phi$ can be simulated on a
digital computer, \ie, \,$\phi$ \,is a classically 
computable (\eg, GL-computable) function.
{\it Assume also that the state space $S$ is the unit $n$-cube $I^n$,
and the time dimension $T$ is the unit interval $I$.\/}
Thus
$$
\phi\: I\times I^n\ \to \ I.
$$
We can now apply Theorem 3 to show that 
the dynamical system has a finite algebraic specification.

\Thmn{4 \,(Universal specification of computable dynamical systems)}
For each $n>0$
there is a signature \,\Sigtilxn\ \,which
extends \,\SigIIIdx\ \,by function symbols,
and a finite conditional specification
\,$(\Sigtilxn,\,\EtilWn(\ztt))$
\,consisting of conditional equations and inequalities,
which is universal for all 
classically computable dynamical systems
on the unit $n$-cube $I^n$ over the unit time interval $I$.
\endpr
Note that \Sigtilxn\ is essentially the signature \Sigxm\ 
of Theorem 3, with $n=m+1$.

We have shown above how powerful algebraic specifications are,
even for topological data types.
More research needs to be done to determine the extent of its power,
especially on metric algebras.
Here topological notions such as continuity can play a part,
as we see from the following example.

\shead{6.4}{Example: \,Specification of function assuming continuity}
Consider the two equations for a (total) function $f$ on the real line:
$$
f(x+y) \ = \ f(x)\times f(y), \qqquad f(1) \ = \ c
\tagx
$$
for some constant $c>0$.
If we assume that $f$ is continuous, even at one point,
then it is easy to see that these equations are satisfied uniquely
by the function 
$$
f(x) \ = \ c^x.
$$
However, in the absence of any such
continuity assumption,
it can be shown that (for any $c>0$) there are
$2^{2^{\aleph_0}}$ non-constructive solutions to ($*$).
Here ``non-constructive" means both that these solutions 
are non-computable, and that their existence is proved 
by non-constructive means, using Zorn's Lemma
to show the existence of a Hamel basis on \,\RR, \,\ie,
a maximal linear independent subset of \,\RR\ \,over \,\QQ.

Note that any solution $f$ of ($*$) is a homomorphism 
from the additive group of reals to the multiplicative group
of positive reals.

This example suggests the following 

\Question
On metric algebras, does conditional equational
specifiability,
together with a topological condition
such as continuity,
imply computability?
\endpr

Specifically, is there a continuous function on $I$ which
is definable by equations but not approximably computable?

Note, in this connection, that there are other ``equational specifications"
for the exponential function \,$e^x$:
\itemm{(1)}
the differential equation \ $f'(x) = f(x)$ \ with 
initial condition \ $f(0) = 1$;
\itemm{(2)}
from the polynomial approximations
given by the partial sums of the Maclaurin expansion 
\ $\sum_{i=0}^\infty {x^i}\slash{i!}$,
\ a specification consisting of conditional equations and inequalities
can be derived 
by the methods of this section for approximating computations;
\itemm{(3)}
similarly, from the polynomial approximations,\ $(1+x\slash n)^n$,
\,a specification consisting of
conditional equations and inequalities
can be derived.

Note that in the first of these specifications, differentiability
of $f$ is (of course) implicitly assumed,
and uniqueness of the solution follows by
the Lipschitz condition;
however no assumptions of continuity are needed in (2) or (3).

The above question points to an open field of research.
The investigation of computable solutions of recursive equations
in \cite{gaertner-hotz} would be relevant here.

\bn

\itemm{\bbf 7 \ \ }
{\bbf Initial algebra specifications with
conditional equations and conditional BU equations}

\mn
In this section
we will consider theories $T$, which we assume to be 
formalised in logical formalisms \FFF\
of the kind described in Section 2;
for example, \,$\FFF= \CondBUEqSig$.
\shead{7.1}{Pre-initial and initial models}
In this subsection (only), we make no assumptions 
concerning the (N-)standardness of
signatures or algebras.
Let \Sig\ be a signature and
let \KK\ be a \Sig-adt.

A formalism \FFF\ is said to be {\it valid for\/} \KK\
if the axioms and inference rules of \FFF\
hold for all algebras in \KK.
Note, for example, that \CondBUEqSig\
is valid for \NStdAlgSig,
but not, in general, for \AlgSig.

A \Sig-algebra $A$ is {\it pre-initial for} \KK\
if there is a unique \Sig-homomorphism from $A$ to
every algebra in \KK;
\,{\it pre}-initial in that
it might not itself belong to \KK.
(The notion of \Sig-{\it homomorphism\/} 
between \Sig-algebras is defined as usual \cite{meinke-jvt}.)

Note that the {\it closed term algebra} \ \TSig\
\ is pre-initial for \KK.

An {\it initial algebra of\/} \KK\
is a pre-initial algebra 
which belongs to \KK.
As is easily seen,
any two initial algebras of \KK\
must be \Sig-isomorphic.
We denote any initial algebra of \KK\ by \ISigKK.

We will be interested in the case that 
$$
\KK \ = \ \AlgSigT,
$$
the class of models of a first-order \Sig-theory $T$,
where $T$ may have certain syntactic restrictions.
We will assume:

\bull 
in this subsection that $T$ is a conditional
equational theory; 

\bull
in \S7.2 
likewise, but restrict attention to N-standard models of $T$;

\bull
in \S7.3 that $T$ is a conditional BU equational theory
(again with N-standard models);

\bull
and in \S7.4 that it is a conditional SU equational theory
(ditto).

\n
(Recall the formal systems defined in Section 2.)
Finally in \S7.5 we will show how conditional BU equational initial algebra
specifications can be ``reduced" to conditional equational initial algebra
specifications.

Let $T$ be a \Sig-theory.
We write \,\ISigT \,for the initial algebra \,
$\Init\bigl(\Sig,\AlgSigT\bigr)$
\,(if it exists), and call it the {\it initial model of $T$}.

Consider the {\it closed term algebra\/} \,\TSigTF\
\,formed from \,\TSig\ \,by identifying closed terms
provably equal from $T$, in some formalism \FFF, \ie, 
$$
\TSigTF \ \eqdf\ \TSig/\approxTF
$$
where
$$
t_1 \approxTF t_2 \ \ \ \llongtofrom_{df}\ \ \ 
  \tx{$t_1=t_2$ is provable from $T$ in \FFF}.
$$

\Lemma
If \FFF\ is valid for \AlgSigT,
then \TSigTF\ is pre-initial for \AlgSigT.
\endpr

We will investigate whether \,\TSigTF\
is, further, {\it initial\/} for \,\AlgSigT, \,\ie, whether
$$
\TSigTF \ = \ \ISigT.
$$

\pr{Initiality Lemma}\sl
Suppose \FFF\ is valid for \AlgSigT.
If $\TSigTF \in \AlgSigT$,
\,then it is (\Sig-isomorphic to) \ISigT.
\endpr

\Defs
Let $A$ be a \Sig-algebra.
\itemm{(1)}
$A$ has an {\it initial algebra specification} \SigT\
if \,$A \,\cong\, \ISigT$.
\itemm{(2)}
$A$ has an {\it initial algebra specification with
hidden sorts and/or functions} \,$(\Sigp,$ $ T')$
\,if \Sigp\ is an expansion of \Sig\ by sorts and/or functions, 
$T'$ is a \Sigp-theory and 
$$
A \ \cong \ \Init\bigl(\Sig,\,\Alg(\Sigp,T')\redSig\,\bigr).
$$
\endpr

\newpage

\Thmn{1 \,\cite{malcev}}
Let $E$ be a conditional equational theory over \Sig.
Let \,$I \eqdf \Tbi(\Sig,E,\CondEqSig)$.
Then $I$ is an initial model of $E$.
\,Furthermore, if
\,$t_1, t_2$ \,are two closed \Sig-terms of the same sort,
then the following are equivalent:
\itemm{($i$)}
$t_1$ and $t_2$ have the same value in $I$,
\itemm{($ii$)}
$t_1$ and $t_2$ have the same value in all models of $E$,
\itemm{($iii$)}
$t_1 = t_2$ \ is provable from $E$ in \,\CondEqSig,
\itemm{($iv$)}
$t_1 = t_2$ \ is provable from $E$ in \,\FOLSig.
\endpr

\Pf
The main thing here is to show that 
\,$I\ttstile E$,
from which \,$(ii)\tto(i)$ \,will follow.
Since $I$ is a (closed) term model,
it is sufficient 
to show that $I$ satisfies all {\it closed substitution instances}
of the axioms of $E$.
So consider any closed instance
\ $P_1 \con \dots \con P_n \to P$ 
\ of an axiom of $E$,
where $P_i$ and $P$ are closed equations.
Note that the corresponding sequent
$$
P_1, \dots, P_n \seqt P
\tagx
$$
is derivable from $E$ in \,\CondEqSig,
\,by the substitution rule.
Suppose \ $I \ttstile P_i$ \ for $i = 1,\dots,n$.
Then, by the definition of $I$, $P_i$ is provable from $E$ in \CondEqSig.
But then $P$ is also provable, 
by repeated (atomic) cuts
of the sequent ($*$)
with the sequents \ $\seqt P_i$,
and so $I \ttstile P$.

Hence \,$I\ttstile E$.
It follows, by the Initiality Lemma,
that $I$ is an {\it initial\/} model of $E$.
Hence also \,$(ii)\tto(i)$.
The further implications
\,$(i)\tto(iii)\tto(iv)\tto(ii)$ 
\,are all trivial.
\endpf

\Remarkn{\,(Completeness and conservativity)}
Mal'cev's Theorem \cite{malcev}, in the form given above, can be viewed
as expressing both
\ ($a$) {\it completeness} of \,\CondEqSig, \,given by the implication
\,$(ii)\impp(iii)$, \,and
\ ($b$) {\it conservativity\/} of first order logic with equality
over \,\CondEqSig, \,given by the implication \,$(iv)\impp(iii)$.
(\Cf\ conservativity lemma (1) and the remark in \S2.6.)
\endpr

Necessary and sufficient conditions for the existence of initial models
of theories are given in \cite{mahr-makowsky}.

\shead{7.2}{Initial N-standard models}
Assume, from now on, that 
\Sig\ is N-standard, and that \KK\ 
consists of {\it N-standard} \Sig-algebras;
for example,
\ $\KK = \NStdAlgSigT$,
\ for some \Sig-theory $T$.
Then 
\ \TSigTF,
\ although it is pre-initial for \KK,
might fail to be initial for \KK\
for two reasons:
it might not satisfy $T$,
and it might not even be N-standard! 
(We return to the second point below.)

An {\it initial N-standard model of\/} $T$
is an initial algebra of \,\NStdAlgSigT.
Any two initial N-standard models of $T$
are \Sig-isomorphic.
We denote any such model by 
$$
\INSSigT \ \eqdf\ \Init(\Sig,\,\NStdAlgSigT).
$$
\sn
{\bf N-Standard Initiality Lemma\/}.\sl
\ \ Suppose \FFF\ is valid for \NStdAlgSigT.
\nl
If \,$\TSigTF \in \NStdAlgSigT$
\,then it is (\Sig-iso\-mor\-phic to) 
\INSSigT.
\endpr

\Defs
Let $A$ be an N-standard \Sig-algebra.
\itemm{(1)}
$A$ has an {\it initial N-standard algebra specification} \SigT\
\,if \ $A \,\cong\, \INSSigT$.
\itemm{(2)}
$A$ has an {\it initial N-standard algebra specification with
hidden sorts and/or functions} \ $(\Sigp, T')$
\ if \Sigp\ is an expansion of \Sig\ by sorts and/or functions, 
$T'$ is a \Sigp-theory and 
$$
A \ \cong\ \Init\bigl(\Sig, \, \NStdAlg(\Sigp,T')\redSig \,\bigr).
$$

Note that \INSSigT\ (if it exists)
might not be an initial model of $T$,
\ie, $T$ might have another, non-N-standard, initial model,
as the following example demonstrates.

\Examplen{\,(Initial N-standard model of a theory
which is not an initial model of that theory)}
Let \Sig\ contain (in addition to the standard operations on \nats\ and \bools)
a constant \,$\uuu:\bools$, \,and let $T$ contain the single axiom
\ `$\uuu \ne \trues$'.
Then the term algebra \ \TSig\
trivially satisfies $T$,
and is hence (by the Initiality Lemma of \S7.1)
an {\it initial} model of $T$.
It is {\it not N-standard}, since
it has a 3-element carrier of sort \bools,
with distinct denotations of \trues, \falses\ and \uuu.
There is, however, also an {\it initial N-standard} model of $T$
with an N-standard (2-element) carrier of sort \bools,
formed by identifying \uuu\ and \falses.
\endpr

Now \TSigTF\ \ may fail to be N-standard for two reasons:
that $T$ proves ``too little" or ``too much",
roughly speaking.
The first reason
is connected with non-N-standard interpretations of the sorts \nats\
and \bools.  
Thus,
there may be a function symbol $f$ in \Sig\
with range sort \nats,
without corresponding axioms in $T$ capable of ``reducing"
\,$f(t)$, for some closed term $t$, to a numeral.
Similarly (as in the above example), 
not all closed boolean terms (\ie, terms of sort \bools)
may be (provably in $T$) equal to \trues\ or \falses.
(In the terminology of \cite{guttag-horning} the specification \SigT\ is not
``sufficiently complete".)
The second reason is that $T$ may be {\it inconsistent},
in the sense that it proves
`$\trues = \falses$' (or, equivalently in a suitable 
weak background theory, `$0 = 1$').
This motivates the following definitions.
Note that we must (to begin with) 
speak of provability relative to some formal system
\,\FFF, \,which will typically be 
one of the system \,\CondEqSig\ \,or \,\CondBUEqSig\
\,of Section 2.

\Defn3
$T$ is {\it consistent} in \,\FFF\ \,if the equation `$\trues = \falses$'
is not provable in \FFF\ from $T$.
\endpr

\Defn4
$T$ {\it determines \nats} in \FFF\
if every closed term of sort \nats\ is, provably in \FFF\ from $T$,
equal to a numeral;
and $T$ {\it determines \bools} in \FFF\
if every closed term of sort \bools\ is, provably in \FFF\ from $T$,
equal to \trues\ or \falses.
\endpr

\newpage

\Defn{5 \,(N-standardness axioms)}
\sn
($a$) \,\NStdAxSig\ \, is the following set of conditional equations:
$$
\boxed{
\gathered
\ands(\trues,\trues) = \trues, \qquad
\ands(\trues,\falses) = \ands(\falses,\trues)
  = \ands(\falses,\falses) = \falses, \\
\ors(\falses,\falses) = \falses, \qquad
\ors(\trues,\trues) = \ors(\trues,\falses)
  = \ors(\falses,\trues) = \trues, \\
\nots(\trues) = \falses, \qquad \nots(\falses) = \trues, \\
\ifs_s(\trues, \,\xtt_1^s, \xtt_2^s) \ = \ \xtt_1^s,
\qquad \ifs_s(\falses, \,\xtt_1^s, \xtt_2^s) \ = \ \xtt_2^s, \\
\eqnat(0,0) = \trues, \qquad \eqnat (\Ss \ztt,0) = 
\eqnat(0,\Ss \ztt) = \falses, \\
\eqnat(\Ss \ztt_1,\Ss \ztt_2) = \eqnat(\ztt_1,\ztt_2), \\
\lsnat(0,\Ss \ztt) = \trues, \qquad \lsnat (\ztt,0) = \falses, \\
\lsnat(\Ss \ztt_1,\Ss \ztt_2) = \lsnat(\ztt_1,\ztt_2), \\
\eqs_s(\xtt^s,\xtt^s) = \trues, \\
\eqs_s(\xtt^s_1, \xtt^s_2) = \trues \ \to \ t^s_1 = t^s_2.
\endgathered
}
$$
where, in the axioms for \,$\ifs_s$, 
\,$s$ ranges over all \Sig-sorts other than \,\bools;
\,and in the axioms for \,$\eqs_s$,
\,$s$ ranges over all \Sig-{\it equality sorts} other than \,\nats,

\sn
($b$) \,$\NStdAxoSig$ 
\,is the set of all {\it closed \Sig-substitution instances\/} of 
\,\NStdAxSig.
\endpr


Note that
\,$\NStdAxSig + \IndSig$ \,holds in any N-standard \Sig-algebra.

We use the terminology:
$T$ {\it proves \NStdAxoSig\ in\/} \FFF\
\,to mean: \,\NStdAxoSig\ is derivable from $T$ in \FFF.

We now state some lemmas which give sufficient conditions for 
a term model \,\TSigTF\ \,to be N-standard.

\Lemman{1 \,(N-standardness lemma)}
Suppose that in \,\FFF\
\itemm{($i$)}
$T$ is consistent,
\itemm{($ii$)}
$T$ determines \,\nats\ \,and \,\bools, \,and
\itemm{($iii$)}
$T$ proves \NStdAxoSig.
\sn
Then \,\TSigTF\ \,is N-standard.
\endpr

\Lemman2
If \Sig\ is strictly N-standard
then \,\NStdAxoSig\ \,determines \,\nats\ \,and \,\bools\
\,in \,\CondEqSig.
\endpr

\Pf
By structural induction on all closed \Sig-terms of sort
\,\nats\ \,and \,\bools\ \,(simultaneous\-ly).
\endpf

The following is 
an immediate consequence
of Lemmas 1 and 2.

\newpage

\Lemman{3 \,(Strict N-standardness lemma)}
Suppose \Sig\ is {\it strictly\/} N-standard, 
\,\FFF\ \,is at least as strong as \,\CondEqSig,
\,and in \,\FFF\
\itemm{($i$)}
$T$ is consistent, \,and
\itemm{($ii$)}
$T$ proves \NStdAxoSig\ \,(or \,\NStdAxSig).
\sn
Then \,\TSigTF\ \,is N-standard.
\endpr

\shead{7.3}{Conditional equational theories}
We now give the analogue of Mal'cev's Theorem (\S7.1)
for N-standard models of conditional equational theories.

\Thmn2
Let $E$ be a conditional equational theory over \Sig.
Suppose that in 
\linebreak
\CondEqSig, 
\,$E$ is consistent, determines \nats\ and \bools,
and proves \,\NStdAxoSig.
Then 
\linebreak
$I \eqdf \Tbi(\Sig,\,E,\,\CondEqSig)$
\,is an initial N-standard model of $E$.
Furthermore, if \ $t_1, t_2$ \ are two closed \Sig-terms of the same sort,
then the following are equivalent:
\itemm{($i$)}
$t_1$ and $t_2$ have the same value in $I$,
\itemm{($ii$)}
$t_1$ and $t_2$ have the same value in all N-standard models of $E$,
\itemm{($iii$)}
$t_1 = t_2$ \ is provable from $E$ \ in \,\CondEqSig,
\itemm{($iv$)}
$t_1 = t_2$ \ is provable from $E$ in \,$\FOLSig+\IndSig$.
\endpr

\Pf
By the N-standardness Lemma (\S7.2),
$I$ is an N-standard algebra.
As in Theorem 1, the main thing is to show
that \,$I\ttstile E$.
This is done exactly as in the proof of Theorem 1.
It follows, by the N-standard Initiality Lemma (\S7.2),
that $I$ is an {\it initial\/} N-standard model of $E$.
The rest of the
proof is similar to that for Theorem 1.
Note for the implication \,$(iv)\tto(ii)$,
\,we use the fact that the rule \,\IndSig\ \,is valid
for N-standard \Sig-algebras.
\endpf

\Remarks
(1) \,By Lemma 2 in \S7.2,
the assumption in the theorem that $E$ determines \,\nats\ 
\,and \,\bools\ \,can be replaced by the assumption that 
\Sig\ is strictly N-standard.
\sn
(2) \,({\it Completeness and conservativity.\/})
\,Here again, the implication \,$(ii)\tto(iii)$ 
\,can be construed as a 
completeness theorem,
\,and \,$(iv)\tto(iii)$ \,
as a conservativity theorem.
(See the Remark in \S2.6 and the Remark following Theorem 1.)
\sn
(3) \,({\it The N-standardness axioms\/}.)
We have ``incorporated" the N-standard\-ness axioms \,\NStdAxoSig\
\,in the theory $E$, so to speak,
by assuming that $E$ proves them.
Another feasible approach would be to incorporate these axioms
in the logics \,\CondEq, \,\CondBUEq\
\,and \,\FOL,
\,by adding them as axioms
(as we did with the boundedness axioms \,\BddAx\ \,in \,\CondBUEq).
This would entail some minor re-wording of the theorems.
\endpr

We turn our attention to theories with
syntactic structure more complicated than conditional equations.

\newpage

\shead{7.4}{Conditional BU equational theories}
We give the analogue of Mal'cev's Theorem for
N-standard models of
BU conditional equational theories.

\Thmn3
Let $F$ be a conditional BU equational theory over \Sig.
Suppose that in \,\CondBUEqSig, 
\,$F$ 
is consistent, determines \nats\ and \bools\ and
proves \,\NStdAxoSig.
Then \ $I \eqdf \Tbi(\Sig,\,F,\,\CondBUEqSig )$
\ is an initial N-standard model of $F$.
\ Furthermore, if $t_1, t_2$ are two closed \Sig-terms of the same sort,
then the following are equivalent:
\itemm{($i$)}
$t_1$ and $t_2$ have the same value in $I$,
\itemm{($ii$)}
$t_1$ and $t_2$ have the same value in all N-standard models of $F$,
\itemm{($iii$)}
$t_1 = t_2$ \,is provable from \,$F$
\,in \,\CondBUEqSig,
\itemm{($iv$)}
$t_1 = t_2$ \,is provable from \,$F$ 
\,in \,$\FOLSig+\IndSig$.
\endpr

\Pf
By the N-standardness Lemma, $I$ is N-standard.
As in Theorems 1 and 2, the main thing is to show that \,$I\ttstile F$.
Again, since $I$ is a term model, 
it is sufficient to show that $I$
satisfies the set of closed substitution instances of $F$.
First note that, by definition,
$I$ satisfies precisely all closed equations provable
from $F$ in CondBUEq, \ie, for any 
{\it closed equation\/} $P$:
$$
I \ttstile P \ \ \llongtofrom \ \ F \tstile P
\tagx
$$
where `$\tstile$' here means provability in \CondBUEq.
Further, by use of the {\it boundedness axioms\/} \,\BddAx\ 
\,of \CondBUEq\ (\S2.3),
the same holds
for any {\it closed BU equation} $Q$:
$$
I \ttstile Q \ \ \llongtofrom \ \ F \tstile Q.
\tagxx
$$
For suppose
\,$Q\,\ident \,\all \ztt < t P(\ztt)$,
\ where $P(\ztt)$ is an equation. 
Since $I$ is N-standard, 
$$
I\ttstile t \ = \ \nbar
\tagxxx
$$
for some (unique) $n$.
Then
$$
\align
\ I \ttstile \all \ztt < t P(\ztt) \ \
   &\llongtofrom \ \ \tx{for all $k<n$,}\ I \ttstile P(\kbar)\\
   &\llongtofrom \ \ \tx{for all $k<n$,}\ F \tstile P(\bar k) 
     \qquad \tx{by} \ \ (*)\\
   &\llongtofrom \ \ F \tstile \all \ztt < t P(\ztt) 
     \qqquad\quad\, \tx{by \,\BddAx\ \,and (\xxx)}.
\endalign
$$
Now consider any closed instance
\ $f \ \ident \ Q_1 \con \dots \con Q_m \to Q$ 
\ of an axiom of $F$ 
(where $Q_i$ and $Q$ are closed SU equations).
Suppose \,$I \ttstile Q_i$ \,for $i = 1,\dots,m$.
Then by ($**$) $Q_i$ is provable from $F$ in \CondBUEq.
But then so is $Q$, 
by repeated cuts
of the sequent
\ $Q_1, \dots, Q_m \seqt Q$ 
\ corresponding to $f$
with the sequents \ $\seqt Q_i$,
and so $I \ttstile Q$.
\endpf

\Remarks
(1) \,As before,
the assumption in the theorem that $F$ determines \,\nats\ 
\,and \,\bools\ \,can be replaced by the assumption that 
\Sig\ is strictly N-standard.
\sn
(2) \,({\it Completeness and conservativity.\/})
\,Again, the implication \,$(ii)\tto(iii)$
\,can be construed as a completeness theorem,
and \,$(ii)\tto(iii)$ \,as a conservativity theorem.
\endpr

\shead{7.5}{Conditional SU equational theories}
Now we turn to the 
infinitary conditional SU equational logic (\S2.4).
Although it will not be used further in the paper,
it is interesting in its own right.

Remember that the infinitary \om-rule \,$\all_\om R$
\,obviates the need for an induction rule.

\Thmn4
Let $G$ be a conditional SU equational theory over \Sig.
Suppose that in \,\CondSUEqomSig, 
\,$G$ 
is consistent, determines \nats\ and \bools\ and
proves \,\NStdAxoSig.
Then \ $I \eqdf  \Tbi(\Sig,\,G,\,\CondSUEqomSig)$
\ is an initial N-standard model of $G$.
\ Furthermore, if $t_1, t_2$ are two closed \Sig-terms of the same sort,
then the following are equivalent:
\itemm{($i$)}
$t_1$ and $t_2$ have the same value in $I$,
\itemm{($ii$)}
$t_1$ and $t_2$ have the same value in all N-standard models of $G$,
\itemm{($iii$)}
$t_1 = t_2$ \,is provable from \,$G$
\,in \,\CondSUEqomSig,
\itemm{($iv$)}
$t_1 = t_2$ \,is provable from \,$G$ 
\,in \,\FOLomSig.
\endpr

\Pf
By the N-standardness Lemma, $I$ is N-standard.
Again, the main thing is to show that 
$I$ satisfies closed substitution instances of axioms of $G$.
By definition, 
for any {\it closed equation} $P$:
$$
I \ttstile P \ \ \llongtofrom \ \ G \tstile P
\tagx
$$
where `$\tstile$' here means provability in \,\CondSUEqom.
Further, by use of the $\all_\om R$ rule,
the same holds
for any {\it closed SU equation} $R$:
$$
I \ttstile R \ \ \llongtofrom \ \ G \tstile R.
$$
For suppose
\,$R \ident \all \ztt P(\ztt)$,
\ where $P(\ztt)$ is an equation. Then
$$
\align
\ I \ttstile \all \ztt P(\ztt) \ \
   &\llongtofrom \ \ \tx{for all $n$,}\ I \ttstile P(\bar n)\\
   &\llongtofrom \ \ \tx{for all $n$,}\ G \tstile P(\bar n) 
     \qquad \,\tx{by} \ \ (*)\\
   &\llongtofrom \ \ G \ \tstile \all \ztt P(\ztt) 
     \qqquad\quad \tx{by} \ \ \all_\om R
\endalign
$$
The rest of the proof follows the pattern of Theorems 1, 2 and 3.
\endpf

\Remarks
(1) \,As before,
the assumption in the theorem that $G$ determines \,\nats\ 
\,and \,\bools\ \,can be replaced by the assumption that 
\Sig\ is strictly N-standard.
\sn
(2) \,({\it Completeness and conservativity.\/})
\,Once again, the implication \,$(ii)\tto(iii)$
\,can be viewed as a completeness theorem,
and \,$(iv)\tto(iii)$
\,as a conservativity theorem.
\endpr

\shead{7.6}{Open term algebras}
So far (Theorems 1, 2, 3 and 4)
we have concentrated on closed term algebras.
We could also formulate our results in a more general setting,
namely, with term algebras constructed from
open terms, \ie, terms containing free variables
(from a given set $X$).

The problem here is that with open terms (an analogy of) the
N-Standardness Lemma (\S7.2) will fail in general.
However, under a certain syntactic condition 
(the ``N-term condition" below),
a version of this Lemma can still be formulated.

First we need some definitions and notation.
Given a signature \Sig, and a set
\,$X\sseq \VarSig$,
\,let \,\TSigX\ \,be the set of \Sig-terms in $X$, 
\ie, \Sig-terms containing variables from $X$ only.
In particular, for $X=\nil$, \,we have 
the set of closed \Sig-terms
\,$\TSig = \Tbi(\Sig,\nil)$.

Given a first-order \Sig-theory $T$
\,and formalism \,\FFF\
\,which is valid for \,\AlgSigT,
\,let \,\TSigXTF\ \,be the \Sig-{\it term algebra\/} formed 
from \,\TSigX\
\,by identifying terms provably equal from $T$ in \,\FFF.
(The closed term algebra \,\TSigTF\ \,considered above
corresponds to the special case \,$X=\nil$).

The algebra \,$I\eqdf\TSigXTF$ \,is 
{\it free for \,$T$ over $X$\/}.
This means that for every model $A$ of $T$,
and every assignment \,$\rho:X\to A$
\,of elements of $A$ to variables in $X$ (of the same sort),
\,there is a {\it unique \Sig-homomorphism\/}
\ $h:I\to A$ 
\ such that \,$h\rest X \rho$.
(This reduces to initiality in \,\AlgSigT\ \,when \,$X=\nil$.)

Note that $I$ need not itself be a model of $T$.
However, this will be the case,
provided $T$ satisfies certain syntactic conditions
(\eg, if $T$ is a conditional equational theory;
\cf\ Theorem 1 above).

Again, assuming that \Sig\ is N-standard, 
we are interested in the question whether 
$I$ is N-standard.  A useful criterion in this connection
is the following syntactic condition on \Sig\ and $X$:

\pr
{N-term Condition for $(\Sig, X)$}
{\sl No \Sig-term of sort \,\nats\ \,or \,\bools\
\,contains any variables from $X$.
\endpr

\Remarks
(1) \,The N-term condition for $(\Sig,X)$ 
is trivially satisfied when \,$X=\nil$. 
\sn
(2) \,When \Sig\ is strictly N-standard,
it is equivalent to the condition:
\sn

\ce{\sl there are no variables in $X$ of sort \,\nats\ \,or \,\bools.}

\sn
This follows from Remark 3 in \S1.5.

Now the theory given above, and specifically Theorems 1 to 4,
can be generalised to the case of open term models
\,\TSigXTF\ , where \,$(\Sig,X)$ \,satisfies the N-term condition.
First, the N-standardness lemma becomes:

\pr{N-Standardness Lemma$^{\bk X}$}\sl
Suppose that \,$(\Sig,X)$ \,satisfies the N-term condition.
\nl
Suppose further that in \,\FFF
\itemm{($i$)}
$T$ is consistent,
\itemm{($ii$)}
$T$ determines \nats\ and \bools, \,and
\itemm{($iii$)}
$T$ proves \NStdAxoSig.
\sn
Then \,\TSigXTF\ \,is N-standard.
\endpr

Next, the strict N-standardness lemma becomes
(using Remark 2 above):

\pr{Strict N-Standardness Lemma$^{\bk X}$}\sl
Suppose \Sig\ is {\it strictly\/} N-standard, 
and there are no variables in $X$ of sort \,\nats\ \,or \,\bools.
Suppose also \,\FFF\ \,is at least as strong as \,\CondEqSig,
\,and in \,\FFF\
\itemm{($i$)}
$T$ is consistent, \,and
\itemm{($ii$)}
$T$ proves \NStdAxSig.
\sn
Then \,\TSigXTF\ \,is N-standard.
\endpr

Consider next, for example, Theorem 2.
This can be reformulated as follows.

\Thmn{$\bk 2^{\bk X}$}
Suppose \,$(\Sig,X)$ \,satisfies the N-term condition.
Let $E$ be a conditional equational theory over \Sig.
Suppose that in \CondEqSig,
\,$E$ is consistent, determines \nats\ and \bools,
and proves \,\NStdAxSig.
Then \ $I \eqdf \Tbi(\Sig,X,\,E,\,\CondEqSig)$
\ is an N-standard model of $E$,
which is free for $E$ over $X$.
\ Furthermore, if \ $t_1, t_2$ \ are two terms in \,\TSigX\ \,of the same sort,
then the following are equivalent:
\itemm{($i$)}
$t_1$ and $t_2$ have the same value in $I$,
\itemm{($ii$)}
$t_1$ and $t_2$ have the same value in all N-standard models of $E$,
\itemm{($iii$)}
$t_1 = t_2$ \ is provable from $E$ \ in \,\CondEqSig,
\itemm{($iv$)}
$t_1 = t_2$ \ is provable from $E$ in \,$\FOLSig+\IndSig$.
\endpr

The strict N-standardness Lemma$^X$,
and Theorem 2$^X$, will be used in Section 9.

\shead{7.7}{Reducing conditional BU to conditional equational specifications}
We re-consider the work of \S3.3 from the viewpoint of initial algebra 
specifications.

\Thmn{5 \ (BU elimination for initial algebra specifications)}
Let $F$ be a conditional BU equational theory over \Sig.
Then there is an expansion \Sigp\ of \Sig\ and a 
conditional equational theory $E'$ over \Sigp\
which is equivalent to $F$ (relative to N-standard models) in the sense that:
\itemm{($i$)}
if $A$ is an N-standard \Sig-model of $F$, 
then it has a \Sigp-expansion which is
a N-standard model of $E'$;
\itemm{($ii$)}
if \ $A \cong \INSSigF$ \ then
it has a unique (up to \Sigp/\Sig-isomorphism) \Sigp-expansion $A'$ such that
\ $A' \cong \INS(\Sig', E')$;
\itemm{($iii$)}
if $A'$ is an N-standard \Sigp-model of $E'$, then its \Sig-reduct $A$
is an N-standard model of $F$;
and if 
\ $A' \cong \INS(\Sig', E')$
\ then \ $A \cong \INSSigF$.
\sn
If $F$ contains $q$ occurrences of BU quantifiers, 
then \Sigp\ \,expands \,\Sig\ by one new sort and $q$ new function symbols.
Moreover, if
$F$ is finite, with $e$ axioms (say),
then so is $E'$, with $e+4q$ axioms.
\endpr

\Pf
The idea, again, is to incorporate in the signature,
for each BU quantifier occurring in $F$,
a {\it characteristic function}
for that quantifier.
The problem with adjoining a boolean-valued function symbol
\ $\fs:\nats\times u \to \bools$
\ satisfying ($**$) in 
the BU elimination theorem in \S3.3,
is in the case that $A$ is an initial N-standard model of $F$.
In order that its \Sigp-expansion $A'$ be {\it N-standard},
the value of $\fs(n,x)$ must be either \trues\ or \falses\
for every value of the arguments $n,x$.
Furthermore, in order that $A'$ also be {\it initial},
the \Sig-homomorphism $h$ from $A$ to every N-standard model $B$ of $F$
must be extendible to a
\Sigp-homomorphism $h'$ from $A'$ to the \Sigp-expansion $B'$ of $B$.
However, the rhs of ($**$) in \S3.3 will hold ``more often" in $B$ than in $A$
(since $B$ is a homomorphic image of $A$),
with a corresponding {\it change} in the value of $\fs(n,x)$
from \falses\ to \trues!
Hence $h$ cannot, in general, be extended as desired.
(Making \fs\ a 0,1-valued function will cause exactly the same problem.)
 
We therefore adjoin a special sort \Ds\ for the range of
such functions \fs,
with a constant \ds\ which takes the place of `\trues' in ($**$) in \S3.3.
(The point is that when the condition on the rhs of ($**$) fails,
$\fs(n,x)$ is not ``forced" to equal anything else at all.)
Now for each BU quantifier as in ($*$) of \S3.3,
{\it adjoin} to the signature the function symbol 
$$
\fs: \nats\times u \ \to \ \Ds,
$$
and adjoin the axioms formed from (\xxx) and (\xxxx) in \S3.3 
by replacing `\trues' by `\ds' throughout.
In this way we {\it replace} $F$ by a conditional
equational theory $E'$ in \Sigp, with the stated properties.
\endpf

\Remark
If $A$ is an N-standard model of $F$, 
then its N-standard \Sigp-expansion $A'$
modelling $E'$,
given by part ($i$) of the theorem,
is not (in general) uniquely determined.
However, the added condition of initiality (on $A$ and $A'$)
determines $A'$ uniquely.
\endpr

\Shead8{Initiality-preserving operators on N-standard algebras}
In this section we combine the theory of Section 5
(``computability $\implies$ algebraic specifiability")
with the initial algebra theory of Section 7.

\shead{8.1}{Initiality preserving operators and the HEP}
Assume now (as in \S3.1) that
\Sigp\ and \Sigpp\ are N-standard signatures with 
\,$\Sig \subset \Sig' \subset \Sigpp$,
\,and \,$\Phi : \NStdAlgSig \to \NStdAlg(\Sig')$
\, is an expanding operator over \Sig.
Recall Definitions 5 and 7 in \S3.1.

\Defn1
\Ph\ is {\it initiality preserving\/}
({\it w.r.t\. \Sig\ and\/} \Sigp)
iff for all 
\,$\KK \subseteq \NStdAlgSig$
\,and \,$A \in \NStdAlgSig$,
\ $A$ is initial in \KK\ iff \APhi\ is initial in \KKPhi.
\endpr

\Lemman1
Suppose \Ph\ is initiality preserving,
and \,\SigpTp\ \,specifies \Ph\ uniformly over \Sig.
Then for any \Sig-theory $T$ and N-standard \Sig-algebra $A$,
$$
A \cong \INSSigT \ \ifff
\ \APhi \cong \INS(\Sigp, \, T + T').
$$
\endpr

\newpage

\Lemman2
Suppose 
\,$\Phi(A) = \Psi(A)\redSigp$
\,for all \,$A \in \NStdAlgSig$,
\,where 
$$
\Psi : \NStdAlgSig \to \NStdAlg(\Sig'')
$$
is an expanding operator 
which is initiality preserving w.r.t\. \Sig\ and \Sigpp.
Then 
\Ph\ is initiality preserving, and
for any \Sigpp-theory \,$T''$ \,and N-standard \Sig-algebra $A$,
if \,\SigppTpp\ \,specifies \Ps\ uniformly over 
\Sig, \,then
\,\SigppTpp\ \,specifies \Ph\ uniformly over \Sig\
with hidden sorts and/or functions; and
for any \Sig-theory T\ \,and N-standard \Sig-algebra $A$,
$$
\align
A \cong \INSSigT \
&\ifff
\ \APsi \cong \INS(\Sigpp, T+T'') \\
&\ifff
\ \APhi \cong \INS(\Sigpp, T+T'')\redSigp \\
&\ifff
\ \APhi \cong \Init\bigl(\Sigp, \,\NStdAlg(\Sigpp, T + T'') \redSigp \,\bigr).
\endalign
$$
\endpr

\Pf
From Lemma 1.
\endpf

\Defn2
\Ph\ has the {\it homomorphism extension property (HEP) 
(w.r.t\. \Sig\ and \Sigp)}
iff every homomorphism \ $h: A \to B$ \ between N-standard \Sig-algebras
can be {\it extended uniquely} to a homomorphism
\ $h^\Phi : \APhi \to B^\Phi$
\ between their images under \Ph.
\endpr

\Lemman3
If \Ph\ has the HEP,
then \Ph\ is initiality preserving.
\endpr

We will apply the above theory to three cases:
array specifications in \S8.2, and specifications
for PR and \muPRx\ computable functions in \S8.3 and \S8.4 respectively.

\shead{8.2}{Initial algebra specification of array algebras}
Recall the array specification 
\,$(\Sigx,\,\ArrAxSig)$ \,defined in \S3.2.

\Lemman1
The array construction \,$A \mapsto \Ax$ 
\,(\S1.6) has the HEP, and (hence) is initiality preserving.
\endpr

\Lemman2
For any N-standard \Sig-algebra $A$ and \Sig-theory $T$:
$$
A \cong \INSSigT \ \ifff \  \Ax \cong \INS(\Sigx, \, T + \ArrAxSig).
$$
\endpr

\Pf
By \S8.1, Lemma 1, and \S3.2, Theorem 1.
\endpf

Of particular interest is the case that $T$ is a {\it conditional BU
equational\/} theory:

\Thmn1
If a \Sig-algebra $A$ has an initial N-standard algebra specification 
by a set of conditional BU equations,
then so does \Ax.
Moreover, if the specification for $A$ is finite, 
with $e$ axioms (say),
then so is that for \Ax, with at most
\,$e + 8s$\,\ axioms, 
where $s$ is the number of sorts in \Sig.
\endpr

Next, from the BU elimination theorem
for initial algebras (Theorem 5 in Section 7)
we can reduce such a specification for \Ax\ further to one with
{\it conditional equations} only.

\Thmn2
If a \Sig-algebra $A$ has an initial N-standard algebra specification 
by a set of conditional equations,
then so does \Ax\
(with hidden sorts and functions).
Moreover, if the specification for $A$ is finite, 
with $e$ axioms (say),
then so is that for \Ax, with at most
\,$e + 12s$\,\ axioms, where $s$ is the number of sorts in \Sig.
\endpr

\Pf
First apply Theorem 1
(or Lemma 2) above.
Then replace the equality axiom for \sx\ in \ArrAxSig,
which is a conditional BU \Sigx-equation (\S3.2), by a conditional 
\Sigx-equation, for each \Sig-equality sort $s$ other than \nats,
by BU elimination (Theorem 5 in \S7.7, applied to \Sigx).
\endpf

\shead{8.3}{Initial algebra specifications for PR computable functions}
Now we apply the above theory to the results in \S5.1.

\Lemman1
For each \,\PRSig\ \,derivation \al,
the operator $(**)$ (\S5.1) has the HEP, and is (therefore)
initiality preserving.
Hence the operator $(*)$ is initiality preserving.
\endpr

\Pf
By structural induction on \al.
\endpf

Hence, by Theorem 1 in Section 5 and Lemma 2 in \S8.1:

\Lemman2
For each \PRSig\ derivation \al,
and for each N-standard \Sig-algebra $A$ and \Sig-theory $T$:
$$
\align
A \cong \INSSigT \
&\ifff
\ (A, \galA, \falA) \,\cong \,\INS(\Sigal, \,T+\Eal) \\
&\ifff
\ \AfalA \,\cong \,\INS(\Sigal, \,T+\Eal)\redSigf \\
&\ifff
\ \AfalA \,\cong 
\,\Init\bigl(\Sigf, \,\NStdAlg(\Sigal, \, T+ \Eal)\redSigf\,\bigr).
\endalign
$$
\rm
Here \,$\Sigf = \Sig \cup \{\fal \}$.
(Remember, \,$\Sigal = \Sig \cup \{ \gal, \fal \}$,
\,where \,\gal\ is the list of auxiliary functions of \al.)
Of particular interest is the case that $T$ is a 
{\it conditional equational} theory:

\Thmn3
Let $f$ be a PR function on a \Sig-algebra $A$.
If $A$ has an initial N-standard algebra specification 
by a set of conditional equations,
then so does \Af\
(with hidden functions).
\endpr

\shead{8.4}{Initial algebra specifications for \muPRx\ computable functions}
We turn to \,\muPRx\ \,computability (\S5.2).
The problem here (as noted in \S5.2)
is that even if the computed function is total,
the auxiliary functions need not be.
However, by applying the totality lemma (\S5.2),
we are able restrict our attention to total derivations.

\Lemman1
For each \,\muPRxSig\ \,derivation \al\ 
and each N-standard \Sig-algebra $A$ on which \falA\ is total,
the operator (\xxx) (\S5.2) has the HEP, and is (therefore)
initiality preserving.
Hence the operator $(*)$ (\S5.1) is initiality preserving.
\endpr

\Pf
By structural induction on \al.
\endpf

Hence, by Theorem 2 in Section 5 and 
Lemma 2 in \S8.1, we have:

\Lemman2
For each \muPRxSig\ derivation \al,
each N-standard \Sig-algebra $A$ on which
\falA\ is total, and each \Sig-theory $T$:
$$
\multline
A \isom \INSSigT \ \ifff \\
\AfalA\ \cong \ \Init\bigl
(\Sigf, \,\NStdAlg(\Sigalx, \, T+ \ArrAxSig +\Falhat)\redSigf\,\bigr).
\endmultline
$$
where \alhat\ is the total derivation for 
\,\fal\ \,given by the totality lemma,
and \,\Falhat\ \,is the conditional BU specification for \alhat.
\endpr
Here, as before,  \,$\Sigf = \Sig \cup \{\fal \}$.
Of particular interest are the two cases that $T$ is a 
{\it conditional BU equational} theory,
and a {\it conditional equational} theory.
First, assuming the former:

\Thmn4
Let $f$ be a total \muPRx\ function on a \Sig-algebra $A$.
If $A$ has an initial N-standard algebra specification \SigF,
where $F$ is a set of conditional BU equations,
then likewise \Af\
has such a specification $(\Sigf,\Ff)$
with hidden sorts and functions,
where \Ff\ is also a set of conditional BU equations.
Moreover, \Ff\
can be obtained by adjoining to $F$
an instantiation \,$\FU(\kbar)$
\,of some universal conditional BU equational specification 
\,$\FU(\ztt)$,
\,which depends only on \Sig\ and the type of $f$.
\endpr

The universal specification \,$\FU(\ztt)$
\,in this theorem is obtained as in Theorem 3 in Section 5.

Finally, by assuming $T$ in Lemma 2 
is a conditional equational theory,
and applying Theorem 4 above and then 
BU elimination for initial algebras (Theorem 5 in Section 7):

\Thmn5
Let $f$ be a total \muPRx\ function on a \Sig-algebra $A$.
If $A$ has an initial N-standard algebra specification \SigE,
where $E$ is a set of conditional equations,
then likewise \Af\
has such a specification $(\Sigf,\Ef)$
with hidden sorts and functions,
where \Ef\ is also a set of conditional equations.
Moreover, \Ef\
can be obtained by adjoining to $E$
an instantiation \,$\EU(\kbar)$
\,of some universal conditional equational specification 
\,$\EU(\ztt)$,
\,which depends only on \Sig\ and the type of $f$.
\endpr

\Shead9{Computability of algebraically specifiable functions}
In this section we prove (partial) converses to the results of 
Section 5.
First we need a definition.

\Defn{(Strong specifiability)}
Let \KK\ be a \Sig-class, 
let \,$\Sig' \supseteq \Sig \cup \{ \fs \}$
\,and let $T$ be a \Sigp-theory.
We say that $T$ {\it strongly specifies\/} a family
\,\curly{\fA \mid\AinKK}
\,(possibly with hidden sorts and/or functions)
\,iff
\nl
($i$) $T$ specifies \,\curly{\fA \mid\AinKK},
\,and further
\nl
($ii$) for every \,$A,B\in \KK$ \,with \,$B\subalg A$,
\,$f^B = \fA\rest B$.
\endpr
\n
(Here \,$\fA\rest B$ \,denotes the restriction of \fA\ to $B$.)

The significance of this concept is seen by rephrasing it 
in either of the following two ways.

\Lemman1
Let \KK\ be a \Sig-class, 
let \,$\Sig' \supseteq \Sig \cup \{ \fs \}$
\,and let $T$ be a \Sigp-theory.
$T$ {\it strongly specifies\/} a family
\,\curly{\fA \mid\AinKK}
\,(possibly with hidden sorts and/or functions)
\,iff
\nl
($i$) $T$ specifies \,\curly{\fA \mid\AinKK},
\,and further \,
\nl
($ii'$) for every \,$A,B\in \KK$ \,with \,$B\subalg A$,
\,$B$ is closed under \fA.
\endpr

\Lemman2
Let \KK\ be a \Sig-class which is closed under finitely generated subalgebras,
let \,$\Sig' \supseteq \Sig \cup \{ \fs \}$
\,and let $T$ be a \Sigp-theory.
$T$ {\it strongly specifies\/} a family
\,\curly{\fA \mid\AinKK}
\,(possibly with hidden sorts and/or functions)
\,iff
\nl
($i$) $T$ specifies \,\curly{\fA \mid\AinKK},
\,and further \,
\nl
($ii''$) for every \,\AinKK\ 
\,and every finitely generated \,$B\subalg A$,
\,$B$ is closed under \fA.
\endpr

We consider algebras and functions specified by
conditional equational theories.
We have to assume now that these theories have
{\it effective axiomatisations\/}:
that the axioms are finite, for example, or 
at least recursively enumerable.

We will also make use of Theorem $2^X$ in \S7.6.
Recall the remarks preceding the theorem there,
that the N-term condition for $(\Sig,X)$ follows from either
\,($i$) $X = \nil$;
\,or
\,($ii$) strict N-standardness of \,\Sig,
together with $X$ containing no variables of sort \,\nats\ \,or \,\bools.

We will prove two theorems, making each of these assumptions 
in turn.

\shead{9.1}{Computability of specifiable function on minimal algebras}
We first consider a partial converse,
using Remark 1 on the N-term condition (\S7.6),
that is, restricting our attention
to minimal models (\ie, models in which every element 
is named by a closed term).
We use the notation \,\MinNStdAlgSigT\
\,for the set of minimal N-standard \Sig-models of a theory $T$.

\Thmn1
Suppose \Sig\ is N-standard.
Let $E$ be an r.e\. conditional equational \Sig-theory
which in \,\CondEqSig\ is consistent, determines
\,\nats\ \,and \,\bools\ \,and proves 
\NStdAxoSig.
Suppose $\Sig' \supseteq \Sig \cup \{ \fs \}$,
\,and let $E'$ be an r.e\. conditional equational \Sigp-theory 
which strongly specifies \,\curly{f^A\mid \AinMinNStdAlgSigE}
\,(possibly with hidden sorts and/or functions).
Assume also that \,$E+E'$ 
\,determines \,\nats\ \,and \,\bools,  
\,and is conservative over $E$,
\,in \,\CondEqSig,
and also that 
all sorts of \,\dom{f} \,other than \,\bools\ \,are equality sorts.
Then \fA\ is uniformly \,\muPRx\ computable over 
\,\AinMinNStdAlgSigE.
\endpr

\Pf
We will describe a pseudo-\WhilexSig\ algorithm
for computing \fA\ uniformly over minimal N-standard \Sig-models $A$ of $E$.
Suppose \,$\fs:\utos$,
\,where \,$u = \tuptimes{s}1n$.
In general, some of the $s_i$ are \nats\ or \bools,
and the others not.
Suppose (w.l.o.g.) that for some $m<n$,
sorts \,$s_{m+1},\dots,s_n$ \,are all either \,\nats\ \,or \,\bools,
\,and sorts \,\tup{s}1m \,are not.
Write \,$u=v\times w$
\,where \,$v=\tuptimes{s}1m$
\,and
\,$w=\tuptimes{s}{m+1}n$.
By assumption, sorts \,\tup{s}1m\ \,are equality sorts.

For any \,\AinMinNStdAlgSigE,
we will show how to compute 
$$
f^A: \ A^u \ \to \ A_s.
$$
Choose a tuple 
\,$k= (\tup{k}1{n-m})\in\Aw$
\,(of naturals and truth values), and
consider the function
$$
f_k^A \eqdf f(\ \cdot\ ,\,k): \ \Av \ \to \ \As.
$$
We will show how to compute $f_k^A$
uniformly in the (numerical and boolean) parameters $k$.

Let \,$I = \Tbi(\Sig,\,E, \,\CondEqSig)$
\,and \,$J = \Tbi(\Sig',\,E+E', \,\CondEq(\Sigp))$
\,(recall the definitions in \S7.1).
By the N-Standardness Lemma (and the conservativity assumption
for \,$E+E'$ \,over $E$),
both $I$ and $J$ are N-standard.
(Below we denote elements of these
algebras by `$[t]$', \ie, suitable equivalence classes of terms $t$,
or tuples thereof.
We also write \,\kbar\ \,for the tuple of numerals and/or truth constants
corresponding to $k$.)

Note that the identity mapping on \,\TSig\ \,induces a \Sig-homomorphism
$$
\iI:\ I \ \to \ \JSig.
$$
By conservativity of \,$E+E'$ \,over $E$, \,\iI\ \,is injective.
Hence \,$I\subalg \JSig$.

Further, the function $f^J$ specified by $E'$ on \,\JSig\ \,is clearly
the same as that defined ``naturally" on $J$ by
\ $f^J([t]) = [\fs(t)]$.
By the strong specification assumption,
$$
f^I = f^J\rest I.
$$
Hence for any closed \Sig-term $t_0$,
$$
f_k^J ([t_0]) \ = \ f_k^I([t_0]) \ = \ [t]
$$
for some closed \Sig-term $t$.
By definition of $J$, this means that the equation
$$
\fs(t_0, \,\kbar) \ = \ t
\tagx
$$
is {\it provable\/} from \,$E+E'$ \,in \,\CondEq(\Sigp).

Now take any \,\AinMinNStdAlgSigE,
\,and any \,\ainAv.
Since $A$ is minimal, there is a tuple of closed \Sig-terms \,$t_0:v$ 
\,such that 
\,$t_0^A = a$.
By Theorem 2 of Section 6 applied to \Sigp, there is a \Sigp-homomorphism 
$$
h \,: \ J \ \to \ (A,\,f^A, \,\dots\,)
$$
with \,$h([t_0]) = a$.
Hence, since ($*$) holds in $J$, it also holds in 
(the \Sigp-expansion of) $A$, with \,`\fs' \,interpreted as \fA.

This suggests the following algorithm for \,$f_k^A$
with $A$ minimal.
\ With {\it inputs} \ainAv:
first generate all (G\"odel numbers of)
tuples of closed \Sig-terms of type $v$, until you find a tuple $t_0$
with \,$t_0^A=a$. 
(This is where we use computability of equality on type $v$.)
Then
{\it generate} all G\"odel numbers of theorems of $E+E'$
until you find one of the form
\,$\cnr{\fs(t_0) = t}$,
\,for some closed \Sig-term $t$.
Then the {\it output} is \,$t^A$.

The search is effective in the 
{\it term evaluation\/} function for closed \Sig-terms in $A$,
by recursive enumerability of $E$ and $E'$.
Further, since term evaluation is \PRx\ computable (\cite[\S4]{tz:hb}),
this algorithm can be formalised as a \,\muPRxSig\ \,derivation 
for \fA,
as desired.
\endpf

\Remark
The assumption that the sorts of \,\dom{f} \,are equality sorts
can clearly be weakened to the assumption that
equality is
(uniformly over \,\MinNStdAlgSigE)
computable on these sorts.
\endpr

\shead{9.2}{Computability of specifiable function in strictly N-standard
algebras}
We consider a second partial converse,
using Remark 2 
on the N-term condition,
\ie, no free variables of sort \,\nats\ \,or \,\bools,
plus strict N-standardness.
\Thmn2
Suppose \Sig\ is strictly N-standard.
Let $E$ be an r.e\. conditional equational \Sig-theory
which in \,\CondEqSig\ is consistent
and proves \,\NStdAxSig.
Suppose $\Sig' \supseteq \Sig \cup \{ \fs \}$
\,is also strictly N-standard
and proves \,\NStdAx(\Sigp).
Let $E'$ be an r.e\. conditional equational \Sigp-theory 
which strongly specifies \,\curly{f^A\mid \AinNStdAlgSigE}
\,(possibly with hidden sorts and/or functions).
Assume also that \,$E+E'$ \,is conservative over $E$  in \,\CondEq(\Sigp).
Then \,\fA\ \,is uniformly \,\muPRx\ computable over \,\AinNStdAlgSigE.
\endpr

\Pf
We will describe a pseudo-\WhilexSig\ algorithm
for computing \fA\ uniformly over \,\AinNStdAlgSigE.
Suppose \,$\fs:\utos$,
\,where \,$u = \tuptimes{s}1n$.
In general, some of the $s_i$ are \nats\ or \bools,
and the others not.
Suppose (w.l.o.g.) that for some $m<n$,
sorts \,$s_{m+1},\dots,s_n$ \,are all either \,\nats\ \,or \,\bools,
\,and sorts \,\tup{s}1m \,are not.
Write \,$u=v\times w$
\,where \,$v=\tuptimes{s}1m$
\,and
\,$w=\tuptimes{s}{m+1}n$.

For any \,\AinNStdAlgSigE,
we will show how to compute 
$$
f^A: \ A^u \ \to \ A_s.
$$
Choose a tuple 
\,$k= (\tup{k}1{n-m})\in\Aw$
\,(of naturals and truth values), and
consider the function
$$
f_k^A \eqdf f(\ \cdot\ ,\,k): \ \Av \ \to \ \As.
$$
We will show how to compute $f_k^A$
uniformly in the (numerical and boolean) parameters $k$.

Choose a tuple of variables \,$\xtt:v$
(\ie, of the same product type as $a$).
Let \,$I = \Tbi(\Sig,\,\xtt,\,E, \,\CondEqSig)$
\,and \,$J = \Tbi(\Sig',\,\xtt,\,E+E', \,\CondEq(\Sigp))$
\,(recall the definitions in \S7.6).
By the strict N-standardness Lemma$^X$ (\S7.6),
both $I$ and $J$ are N-standard.

Note that the identity mapping on \,\TSigx\ \,induces a \Sig-homomorphism
$$
\iI:\ I \ \to \ \JSig.
$$
By conservativity of \,$E+E'$ \,over $E$, \,\iI\ \,is injective.
Hence \,$I\subalg \JSig$.

Further, the function $f^J$ specified by $E'$ on \,\JSig\ \,is clearly
the same as that defined naturally on $J$ by
\ $f^J([t]) = [\fs(t)]$.
By the strong specification assumption,
$$
f^I \ = \ f^J\rest I.
$$
Hence
$$
f_k^J (\xtt) \ = \ f_k^I(\xtt) \ = \ [t]
$$
for some \,\tinTSigx.
By definition of $J$, this means that the equation
$$
\fs(\xtt, \,\kbar) \ = \ t
\tagx
$$
is {\it provable\/} from \,$E+E'$ \,in \,\CondEq(\Sigp).

Now take any \,\AinNStdAlgSigE,
\,and any \,\ainAv.
By Theorem $2^X$ applied to \Sigp, there is a \Sigp-homomorphism 
$$
h \,: \ J \ \to \ (A,\,f^A, \,\dots\,)
$$
where \,$h(\xtt) = a$.
Hence, since ($*$) holds in $J$, it also holds in 
(the \Sigp-expansion of) $A$, with \,`\fs' \,interpreted as \fA\
and $a$ assigned to \,\xtt.

This suggests the following algorithm for \,$f_k^A$.
\ With {\it inputs} \ainAv:
{\it generate} all G\"odel numbers of theorems of $E+E'$
until you find one of the form
\,$\cnr{\fs(\xtt) = t}$,
\,for some \Sig-term $t$ (in \xtt).
This search is effective, by recursive enumerability of $E$ and $E'$.
Then the {\it output} is 
the {\it evaluation of the term\/} $t$ in $A$ with $a$ assigned to \,\xtt.

Since term evaluation is \PRx\ computable \cite[\S4]{tz:hb},
this algorithm can be formalised as a \,\muPRxSig\ \,derivation for \fA,
as desired.
\endpf

\Remarks
\itemm{(1)}
The above algorithm gives,
for each tuple of numerical and boolean arguments $k$,
a {\it fixed\/} term \,\tinTSigx\ 
\,as the value of \,$f_k^A(a)$
\,for all \,\AinNStdAlgSigE\
\,and all \,\ainAv.
\itemm{(2)}
Theorems similar to Theorems 1 and 2 can be formulated for conditional 
BU equational theories and specifications,
using a variation of Theorem 3 (instead of Theorem 2) in Section 7.
\endpr

\mn

\itemm{\bf 9.3}
{\bf Significance of strong specifiability;
\ Equivalence of specifiability and 
\linebreak
computability}
\sn
We want to combine some of the above results into an
equivalence result between computability and specifiability.

Note that by the Locality Theorem for \While\ \,computations
\cite[\S2.8]{tz:hb}, if $f$ is \,\muPRx\
computable on an algebra $A$,
then any subalgebra of $A$ is closed under $f$.
This suggests the following formulations for equivalence theorems,
which are simple consequences of the above theorems
and the lemmas on strong specifiability at the beginning of this section.

We give one formulation (Theorem 3) for minimal algebras (\cf\ Theorem 1),
and another (Theorem 4) for strictly N-standard algebras (\cf\ Theorem 2).

\Thmn3
Suppose \Sig\ is N-standard.
Let $E$ be an r.e\.   conditional equational \Sig-theory,
which in \,\CondEqSig\  
\,is consistent, determines \,\nats\ \,and \,\bools\
\,and proves \,\NStdAxoSig.
Let \,$\fbi = \ang{\fA\mid\AinMinNStdAlgSigE}$
\,be a family of functions on 
\MinNStdAlgSigE.
Assume that all sorts of \,\dom{\fbi} 
\,other than \,\bools\ \,are equality sorts.
Then the following are equivalent:
\itemm{($i$)}
\fbi\ \,is \,\muPRx\ computable uniformly on \,\MinNStdAlgSigE;
\itemm{($ii$)}
\fbi\ \,is strongly specifiable uniformly on \,\MinNStdAlgSigE,
\,with hidden sorts and functions,
by a finite set of conditional equations
which (together with $E$) is conservative over $E$ in \CondEqSig.
\endpr

\Thmn4
Suppose \Sig\ is strictly N-standard.
Let $E$ be an r.e\.   conditional equational \Sig-theory,
which in \,\CondEqSig\  
\,is consistent
and proves \,\NStdAxoSig.
Let \,$\fbi = \ang{\fA\mid\AinNStdAlgSigE}$
\,be a family of functions on 
\,\NStdAlgSigE.
Then the following are equivalent:
\itemm{($i$)}
\fbi\ \,is \,\muPRx\ computable uniformly on \,\NStdAlgSigE;
\itemm{($ii$)}
\fbi\ \,is strongly specifiable uniformly on \,\NStdAlgSigE,
\,with hidden sorts and functions,
by a finite set of conditional equations
which (together with $E$) is conservative over $E$ in \CondEqSig,
and such that the signature of these equations is also strictly N-standard.
\endpr

\Remarkn{\,(Herbrand-G\"odel computability on \NN)}
The above theorem 
generalises the classical equivalence result on \,\NNN\ \cite{kleene:im}.
\endpr

\Shead{10}{Concluding remarks and future directions}
\sshead{10.1}{Computation on Topological Data Types}
We have extended the theory of algebraic specifications
from the world of countable computable algebras to that of
all algebras, and especially metric algebras, by means of
abstract computability theory. Topological data types and
algebraic specifications play a fundamental role in many areas of
computing, including semantics and scientific computation.

Our main theorems concern the transformation of abstract
algorithms to algebraic specifications and provide some basic
techniques for the theory of specifying and verifying
abstract computations. An obvious question is:

{\displaytext
Under what circumstances can the conditional equations be
replaced by equations in our theory?

}
However, the converse results on the derivation of
algorithms from algebraic specifications need strengthening
to provide completeness or equivalence theorems. Improving
results in the reverse direction
is an important problem,
as stated in the Introduction.
There is much more to this topic than the results 
in Section 9.
A key technical problem in this area is:

{\displaytext
To develop general techniques for solving equations, conditional
equations and other algebraic formulae in topological algebras.

}

In semantics, for example, special cases of the problem are
common.  Semantic modelling makes heavy use of fixed-point
equations. One thinks of the introduction of metric methods
into semantics by M. Nivat 
(see \cite{nivat,arnold-nivat:tcs,arnold-nivat:fi}), or their use in
concurrency theory by De Bakker and others
\cite{db-z,db-rutten,db-devink}.
Studies of the methods of equation
solving in ultrametric algebras, including equivalence
between metric, algebraic and domain-theoretic techniques,
are in Stoltenberg-Hansen and Tucker 
\cite{stolt-jvt88,stolt-jvt91,stolt-jvt93};
see also \cite{stolt94}.

In scientific computation, numerical methods are concerned
with obtaining computable solutions from differential and
integral equations. Mathematical models of systems in the
world are specified by sets of equations, from which
algorithms are sought to compute their solutions and hence
to simulate the system. Our main theorems and examples in
Section 5 show the opposite: if a system can be
approximately simulated on a computer then there exist
algebraic specifications that capture the system's behaviour. Such
results seem to be new and, in our view, draw loci that
help delimit the computability theory of physical
systems. We conjecture that it is possible

{\sl
To show that certain parts of the theory of numerical
approximation of differential and integral equations are special
instances of a general theory of algebraic specifications.

}
This is an exciting and difficult problem with many obvious applications.

Given the wealth of algorithms and theory in numerical
methods, it seems to us that relatively little is known
about the computational and logical scope and limits of
equations, the classical mathematical methods of science.
Progress in the area has awaited the creation of stable
computation theories for topological data types. Over the
past decade, computability theory for topological spaces
and algebras has developed dramatically. Several general
approaches have produced deep results and have been shown
to be equivalent. Some approaches are

\bull
metric spaces \cite{moscho64},
\bull
axiomatic computation structures \cite{pourel-richards},
\bull
type two effectivity \cite{weih:book},
\bull
algebraic domain representability \cite{stolt-jvt88,stolt-jvt95,blanck97},
\bull
continuous domain representability \cite{edalat97},
\bull
abstract computability \cite{tz:top,tz:hb,bss}

The equivalence of the first five approaches is proved in
\cite{stolt-jvt99:tcs}.
The equivalence of all these with the last one is proved in
\cite{brattka97,brattka:thesis,tz:top,tz:abs-conc}.

However, this computability theory needs to be complemented
by a logical theory which includes equation solving in topological
algebras.

\shead{10.2}{Theory of computable data types}
The theory of algebraic specifications of computable
(semicomputable, and cosemicomputable) data types contains
many techniques for proving special properties of algebraic
specifications, and showing the equivalence or
non-equivalence of specification methods.  Can some of
these results be generalised? We believe the answer is yes,
but not without much further work. Many results depend on
special techniques of classical computability theory on the
natural numbers. The theory for computable algebras uses
representations by recursive algebras of numbers. It is
possible to make a representation theory for topological
algebras based on Baire space \,$\NN^\NN$ \,using the {\it type two
effectivity\/} methods of Klaus Weihrauch \cite{weih:book}.
The use of the Diophantine Theorem  for r.e\. sets is more
difficult: the theory of r.e\. sets in abstract
computability differs from the classical case, and no
Diophantine Theorem is known (even for minimal algebras).

Since abstract computability theory is uniform over classes
of algebras, our results on specifications are uniform,
yielding parameterised specifications. As we have seen,
this process is surprisingly delicate because it leads to
questions about standardness. In abstract computations it
is natural to augment an algebra by basic data types such
as booleans, naturals and finite arrays. These have an
effect on the axiomatisations.  There are other important
additional types, of both theoretical and practical
interest, that may be used to augment a given data type
and are in need of a standard algebraic
specification theory, including:
\itemm{$(i)$}
infinite streams (necessary for developing the theory of interactive systems);
\itemm{$(ii)$}
real numbers (necessary for developing the theory of metric algebras
and normed linear spaces).

\n
An attempt to extend the specification methods
of this paper to both these data types,
using infinitary equational specifications,
is made in \cite{tz:fef}

Finally, we note there are several other basic properties
of specifications in need of investigation, especially term
rewriting properties.

\bigskip
\bigskip
\goodbreak

\cbb{References}\bigskip

\bibliographystyle{alpha}
\bibliography{abbrev,bib}

\enddocument